\documentclass[a4paper,12pt]{article}
\usepackage[utf8]{in
    putenc}
\usepackage{cancel}
\usepackage{ulem}
\usepackage{amsfonts}
\usepackage{amssymb}
\usepackage{graphicx}
\usepackage{amsmath}
\usepackage{enumerate}
\usepackage{mathtools}
\usepackage{subfloat}
\usepackage{color}
\usepackage{float}
\usepackage{here}
\usepackage{cite}
\usepackage{mathrsfs}
\usepackage{float,epsfig}
\usepackage{dcolumn}

\usepackage{tabularx, ragged2e, booktabs, caption}
\usepackage{blindtext}
\usepackage{caption} 
\usepackage{subcaption}
\usepackage{tablefootnote}
\usepackage{booktabs} 
\usepackage[export]{adjustbox}
\usepackage{authblk}

\usepackage{graphicx}
\usepackage{bm}
\usepackage{amsmath,amssymb,amsthm}
\usepackage[colorlinks=true,linkcolor=blue,citecolor=red]{hyperref}
\newcommand{\be}{\begin{equation}}
\newcommand{\ee}{\end{equation}}
\newcommand{\bea}{\setlength\arraycolsep{2pt} \begin{eqnarray}}
\newcommand{\eea}{\end{eqnarray}}

\def\0{{\sst{(0)}}}
\def\1{{\sst{(1)}}}
\def\2{{\sst{(2)}}}
\def\3{{\sst{(3)}}}
\def\4{{\sst{(4)}}}
\def\5{{\sst{(5)}}}
\def\6{{\sst{(6)}}}
\def\7{{\sst{(7)}}}
\def\8{{\sst{(8)}}}
\def\sst#1{{\scriptscriptstyle #1}}

\title{\bf On  Einstein-non linear-Maxwell-Yukawa de-Sitter black hole thermodynamics}

\author{ M. Chabab$^{1}$\footnote{Email: mchabab@uca.ac.ma (Corresponding author) }, H. El Moumni$^{2}$,  J. Khalloufi$^{1}$
	\\ 
	{\small $^{1}$ High Energy and Astrophysics Laboratory, Physics Department, FSSM, Cadi Ayyad University,
	}\\
	{\small  P.O.B. 2390 Marrakech, Morocco.
	}\\
	{\small $^{2}$ EPTHE, Physics Department, Faculty of Science,  Ibn Zohr University, Agadir, Morocco. }
}

\vspace{-3.6em}

\date{}
\begin{document} 
\maketitle
\begin{abstract}
We investigate the thermodynamics of the Einstein-non linear-Maxwell-Yukawa black hole solution in de-Sitter spacetime. After presenting the black hole solution and its horizons, we use different thermodynamics points of view to probe the phase structure of such a black hole, ranging from local, global thermodynamics to the effective thermodynamics formalism with two horizons. Our analysis shows that the black hole can undergo a thermal phase transition between an unstable phase and stable phase. These results are consolidated by our proposal of generalizing  effective thermodynamics to three horizons black holes. As by product, we find that the effective temperatures behave as an association of ohmic resistances in an electric circuit, which may reveal a possible analogy between the electric and  thermodynamic systems.

{\noindent}

\end{abstract}
\newpage
\tableofcontents



\section{Introduction}

Black holes which are the solutions of Einstein’s equation are the most important and intriguing objects in physics with many fascinating physical properties. A special attention is devoted to their thermodynamics proprieties which have been studied over years in several different contexts since the seminal works of Hawking, Page, Bardeen and Bekenstein \cite{Hawking:1982dh,Hawking:1974sw,Bardeen:1973gs,Bekenstein:1972tm,Bekenstein:1973mi,Bekenstein:1974ax}. In particular  the
 study of
phases transitions of the black holes in AdS space \cite{Chamblin:1999tk,KM,our} and the analogy between the critical  van der Waals gas behavior and the charged AdS black hole one 
 \cite {Chabab:2018lzf,Chabab:2019mlu, our3,Chabab:2017knz,our8,our6,our7} leading to so-called extended phase space. In this context  the negative cosmological constant $\Lambda$ can be the thermodynamic pressure defined as  $P=-\frac{\Lambda}{8\pi}$. However, when a positive cosmological constant (de Sitter spacetime), the situation is quite different.

The current cosmological observations indicate that the universe is expanding with an accelerated rate, and may tend to de Sitter space asymptotically then in order to construct the whole history of the evolution of our universe, we need to understand the
classical and quantum nature of the de Sitter spacetime. This comprehension triggered a deep investigation of the thermodynamic properties of de Sitter background \cite{chin1,chin2,Urano:2009xn,Sekiwa:2006qj,Dolan:2013ft,Li:2016zdi,Zhao:2014raa,Zhang:2014jfa,Zhang:2016yek, Simovic:2019zgb,Mbarek:2018bau,Zhang:2016nws,Liu:2019qxt,Ma:2019pya,Kubiznak:2015bya}, knowing that black holes in such spacetime generally possess an event horizon and a cosmological one, which radiate thermally \cite{chin1,chin2}. Consequently, there are different thermodynamics quantities  corresponding to each of the two horizons that satisfy the first law of thermodynamics while the corresponding entropies verify the area formula \cite{Sheykhi,Dolan:2013ft}. 


 Born and Infeld have first introduced the nonlinear electrodynamics in order to remove the central singularity of the point-like charges and obtained finite energy solutions for particles by extending Maxwell's theory \cite{ref1}.
 Hereafter, Plebanski and al. have extended the model and presented other kinds of nonlinear electrodynamics Lagrangians \cite{ref2}. Lately,  a variety of nonlinear electrodynamics models have been studied extensively, one of the main reason of this intensive investigation is that such theories appear as effective theories at low energy limits of heterotic string theory \cite{ref3}. In addition, the nonlinear electrodynamics has also been used to get solutions describing baryon configurations which are consistent with confinement in the  AdS/CFT conjecture context \cite{ref4}. In other contexts, the ability on nonlinear electrodynamics to remove
  curvature singularities showed a very promising  results.  As an example, in cosmology, one can call upon nonlinear electrodynamics to explain the inflationary epoch and the late-time accelerated expansion of the universe \cite{ref5,ref6}.  Different black hole solutions  in general relativity coupled to nonlinear electrodynamics have appeared in black hole field theory literature, where 
  the nonlinear electrodynamics is a source of field equations satisfying the weak energy condition, hence recovering the Maxwell theory in the weak field limit \cite{ref7,ref8,ref9,ref10,ref11,ref12,fan}.

In addition to Born-Infeld model,   other  classes  of nonlinear electrodynamics have been proposed, ranging from 
power Maxwell invariant (PMI)\cite{Hassaine:2007py,Hassaine:2008pw},  $\arcsin$ \cite{Kruglov:2014iwa}, logarithmic  \cite{Soleng:1995kn} and exponential \cite{Hendi:2012zz,Hendi:2013mka} to models inspired from string theory with very complex form \cite{Fradkin:1985qd,Karlhede:1987bg,Hamada:1987ph,Tseytlin:1997csa}.
In \cite{base}, the authors consider gravity coupled with a Yukawa-like electric potential, described by the scalar potential, $\phi(r)=\frac{q}{r} e^{-\alpha r}$, where $q$ stands the electric charge and $\alpha$ is a positive
constant. By using this ansatz, the theory becomes highly nonlinear and involves two degrees of freedom, namely the charge $q$ and
Yukawa charge $\alpha$. The consideration of Yukawa-like electric potential is lent from the nuclear physics\cite{Yukawa:1935xg,Birrell:1982ix} and some studies of gravitational potential  in the framework of $f(R)$ modified theory of gravity \cite{DeMartino:2018yqf,DeLaurentis:2018ahr,Ribas:2016ulz,Berezhiani:2009kv}. In our framework, the parameter $\alpha$ is employed  to confine the electromagnetic force to a shorter range. In light of all these motivations, we perform in the present work a natural generalization of the paper \cite{base} to de-Sitter spacetime and  present a comprehensive analysis of the thermodynamic properties of such black hole from various points of view.

The organization of the paper is as follow: First, we 
 extended the  static and spherically symmetric  Maxwell-Yukawa solution found in \cite{base} to de-Sitter spacetime, and also discuss its horizons and extremal case. Then we perform a thermodynamics analysis to probe the thermal phase picture via the continuity of heat capacity, its sign and the swallowtail structure of the Gibbs free energy.
In section $4$, we use a local approach, which means that each horizon is treated independently. Whereas in section $5$ the event, cosmological and inner horizons are treated simultaneously. This treatment  is dubbed the thermodynamics from a global point of view.
After that, we recall the notion of effective thermodynamics of the black hole, which has been emerged recently years to investigate the de Sitter black hole thermodynamics. In section $7$, we discuss the Strong-Weak electric interaction transitions and explain how it can be related to the phase transitions. Then we extend the formalism of the effective thermodynamics from two to three horizons. The last section is devoted to our conclusion.

\section{Einstein-nonlinear-Maxwell-Yukawa dS black hole solution }
Here, we start by a concise review of the static spherical symmetry black hole solution in Einstein-Yukawa-Maxwell gravity and our extension to de-Sitter spacetime. Yukawa potential in the spherical coordinate system is given by
\begin{equation}\label{1}
\phi(r)=\dfrac{q}{r}e^{-\alpha r},
\end{equation}
in which $q$ denotes the electric charge located at the origin and $\alpha$ is a positive constant. We consider the electric potential and Maxwell's field  as,
 \begin{equation}\label{2}
\text{A} = \phi(r) dt,\quad  \text{F} = E dt \wedge dr,
\end{equation}

The electromagnetic field  and the Maxwell invariant stand for,
\begin{equation}\label{4}
E = - \phi'(r) = \dfrac{q\left( 1+\alpha r\right) }{r^2}e^{-\alpha r},\quad 
\mathcal{F}= F_{\mu \nu}F^{\mu \nu}=-2E^2.
\end{equation}
Then, we introduce the static and spherically symmetric line element  ansatz,
\begin{equation}\label{5}
ds^2 = - f(r) dt^2 + \dfrac{dr^2}{f(r)}+ r^2 d\Omega^2,
\end{equation}

We examine the Einstein-Hilbert action  within  the nonlinear electrodynamics  Lagrangian  $\mathcal{L}=\mathcal{L}(\mathcal{F})$ associated with  the Maxwell-Yukawa potential in de-Sitter spacetime

\begin{equation}\label{7}
S =  \dfrac{1}{16\pi G}\int d^4x\sqrt{-g}\left[\left( R-\dfrac{\Lambda}{2}\right) +\mathcal{L}(\mathcal{F}) \right] 
\end{equation}
in which $R$ is the Ricci scalar and $\Lambda>0$ is the cosmological constant that characterizes the de Sitter space.  The Lagrangian $\mathcal{L}$ is given by \cite{base}
\begin{equation}\label{17}
\mathcal{L} = \dfrac{ 4 C_0 q }{r^4 }\left[ \left( \alpha^3 r^3 - 1 - \left( 1+\alpha r\right) ^2\right)  e^{-\alpha r} - \alpha^4 r^4 \mathcal{E}_1 (\alpha r)\right]  + C_1
\end{equation}
where $C_0$ and $C_1$ are  integration constants, $\mathcal{E}_1(x)$ is the exponential integral defined by the formula \cite{bookintegral}
\begin{equation}\label{18}
 \mathcal{E}_1(x) = \int_{1}^{\infty}\dfrac{e^{-x t}}{t} dt .
\end{equation}

It is worth to notice that the   $C_1$ may be interpreted as an effective cosmological constant which we will neglect in the rest  of this study since we are already considering a dS space. Thus, one obtains the metric function,
\begin{equation}\label{19}
f(r) =  1 - \dfrac{2m}{r} + \dfrac{qC_0 \left(\alpha^3 r^3 - \alpha^2 r^2+ 2\alpha r - 6 \right) e^{-\alpha r} } {6 r^2} - \left( \dfrac{\Lambda}{3} + \dfrac{\alpha^4 q C_0}{6} \mathcal{E}_1 (\alpha r)\right) r^2 .
\end{equation}
The parameter $m$ is an integration constant regarded as mass. To determine $C_0$, 
we consider the limit of vanishing $\alpha$ $(\alpha \to 0)$ for the metric function 
\begin{equation}\label{20}
f(r) =  1 - \dfrac{2m}{r} - \dfrac{qC_0 } {r^2} -  \dfrac{\Lambda}{3} r^2.
\end{equation}
hence recovering the Reissner-Nordstrom dS black hole solution. This implies  $C_0 = -q$. Consequently, Eq.~\eqref{19} becomes, 
\begin{equation}\label{21}
f(r) =  1 - \dfrac{2m}{r} - \dfrac{q^2 \left(\alpha^3 r^3 - \alpha^2 r^2+ 2\alpha r - 6 \right) e^{-\alpha r} } {6 r^2} - \left( \dfrac{\Lambda}{3} - \dfrac{\alpha^4 q^2}{6} \mathcal{E}_1 (\alpha r)\right) r^2; 
\end{equation}
which is the final form of the ENLMY-dS black hole solution. 
By rescaling the variables : $\rho = \alpha r$, $M = \alpha m$, $Q = \alpha q$ and $\lambda =  \Lambda/\alpha$; the metric function reads 
\begin{equation}\label{22}
f(\rho) =  1 - \dfrac{2M}{\rho} - \dfrac{Q^2 \left(\rho^3 - \rho^2+ 2\rho - 6 \right) e^{-\rho} } {6 \rho^2} - \left( \dfrac{\lambda}{3} - \dfrac{Q^2}{6} \mathcal{E}_1 (\rho)\right) \rho^2 ,
\end{equation}
Eq.~\eqref{22} shows that $f(\rho)$ is not explicitly dependent on  $\alpha$ which acts as a scale factor, so it can be chosen as $\alpha = 1$. Hence, irrespective of the value of $\alpha$, one can study the global properties of the resulting solution. Nevertheless, we should keep in mind that $\alpha$ is a "charge" that corresponds to a potential which play a part in black hole thermodynamics. Having obtained the essential of the ENLMY-dS black hole solution,
 we will investigate its  phase structure  within different thermodynamics points of view.

\section{Black hole horizons and extremal cases }
\begin{figure}[h!]
	\centering \includegraphics[scale=0.7]{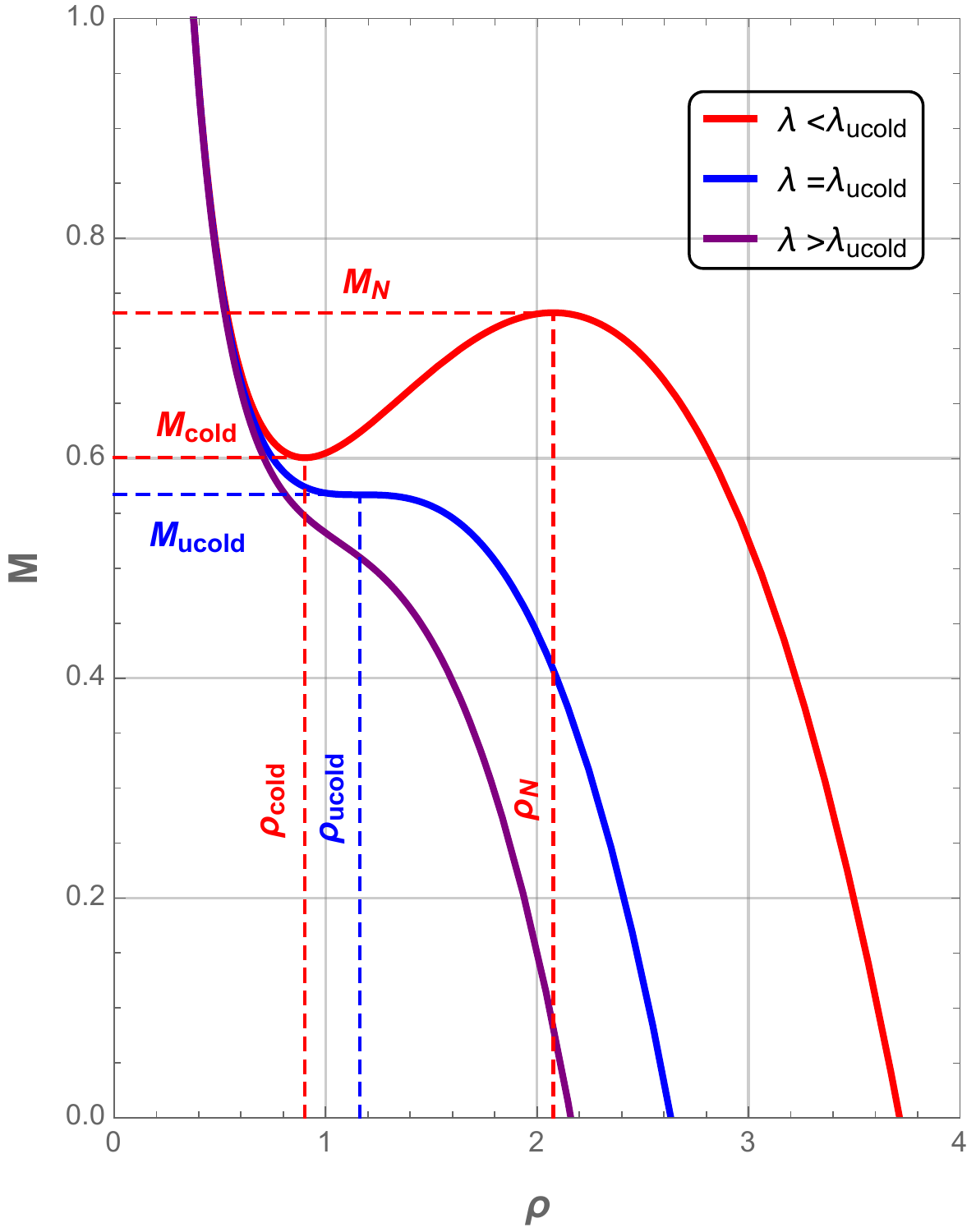}
	\caption{\footnotesize Black hole mass as a function of radius for different values of cosmological constant $\lambda$ with $Q = 1$. }
	\label{f2}
\end{figure}

\begin{figure}[h!]
 	\centering
 	\begin{subfigure}[h]{0.45\textwidth}
 		\centering \includegraphics[scale=0.5]{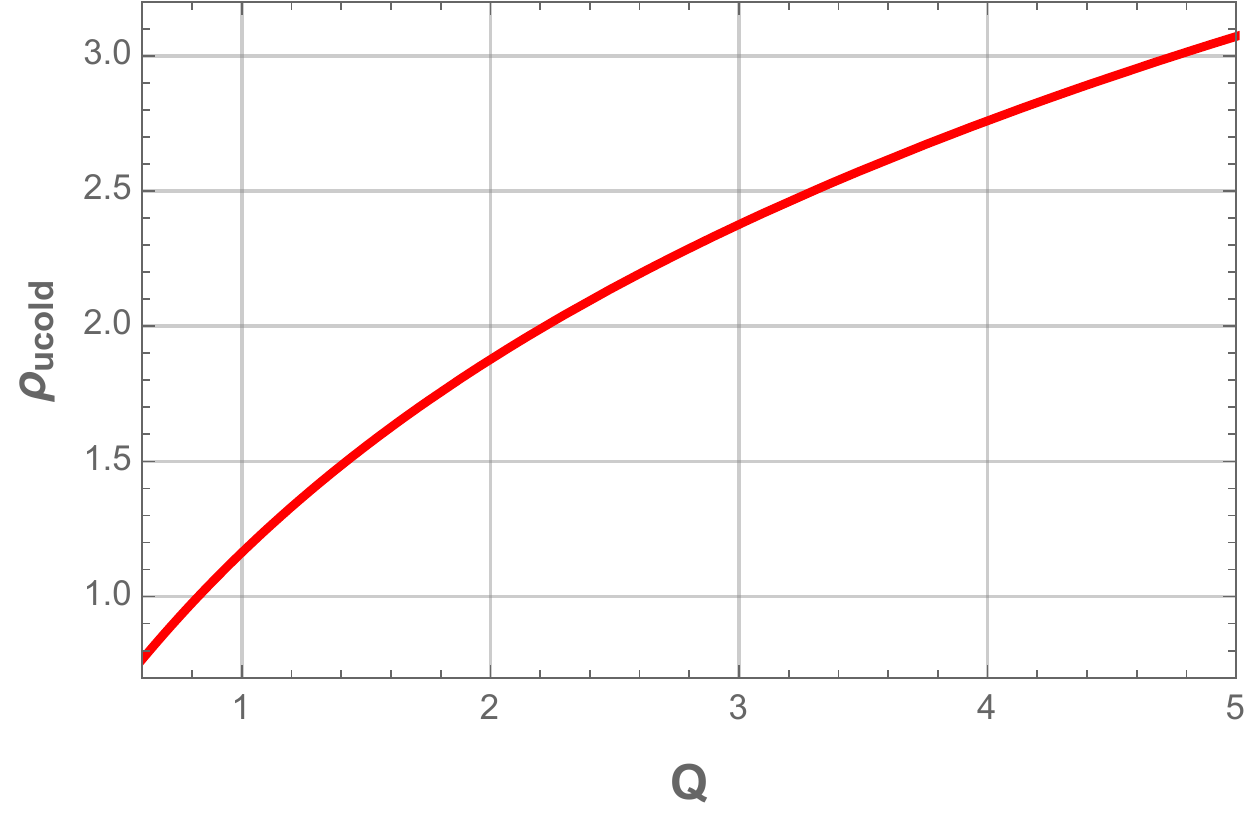}
 		\caption{\footnotesize Ultra-cold horizon radius $\rho_{ucold}$ as a function of electric charge $Q$.}
 		\label{f3_1}
 	\end{subfigure}
 	\hspace{1pt}	
 	\begin{subfigure}[h]{0.45\textwidth}
 		\centering \includegraphics[scale=0.5]{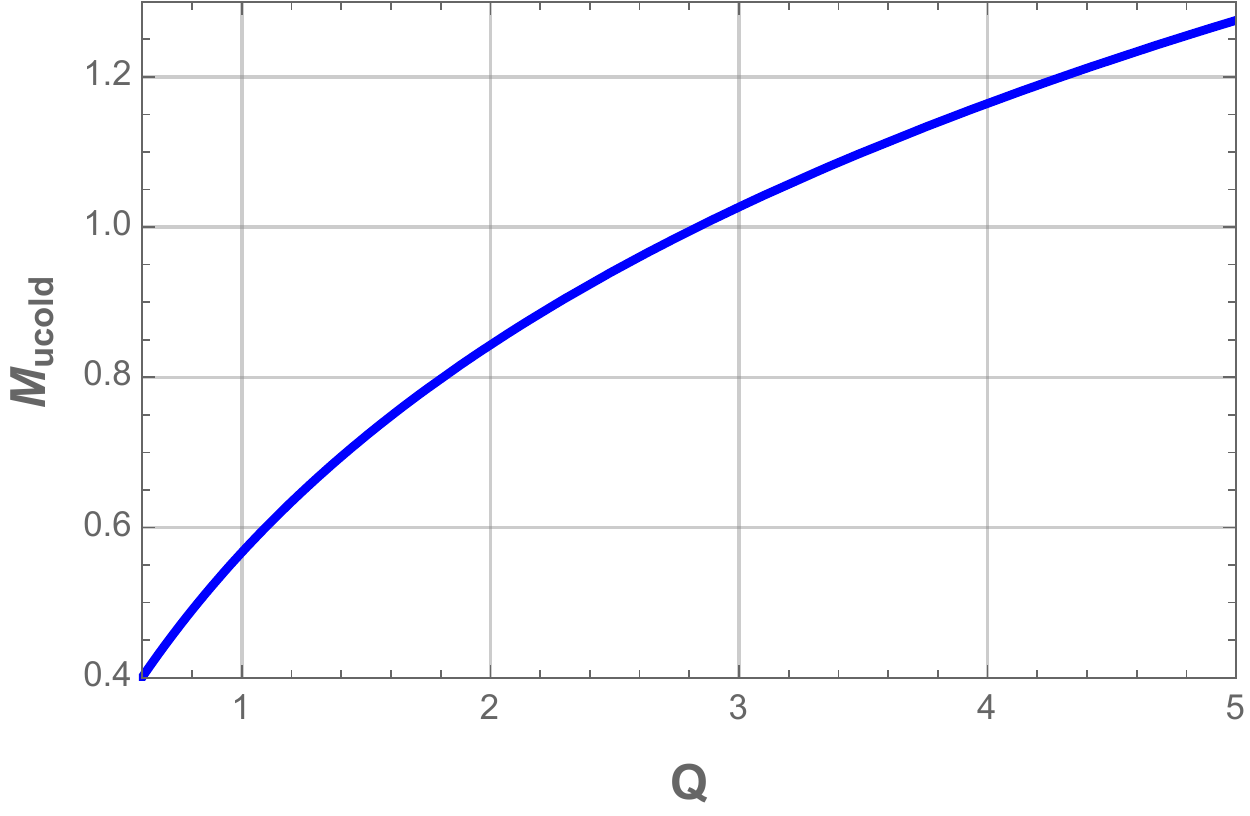}
 		\caption{\footnotesize Ultra-cold horizon mass $M_{ucold}$ as a function of electric charge $Q$.}
 		\label{f3_2}	
 	\end{subfigure}
  	\begin{subfigure}[h]{0.45\textwidth}
 	\centering \includegraphics[scale=0.5]{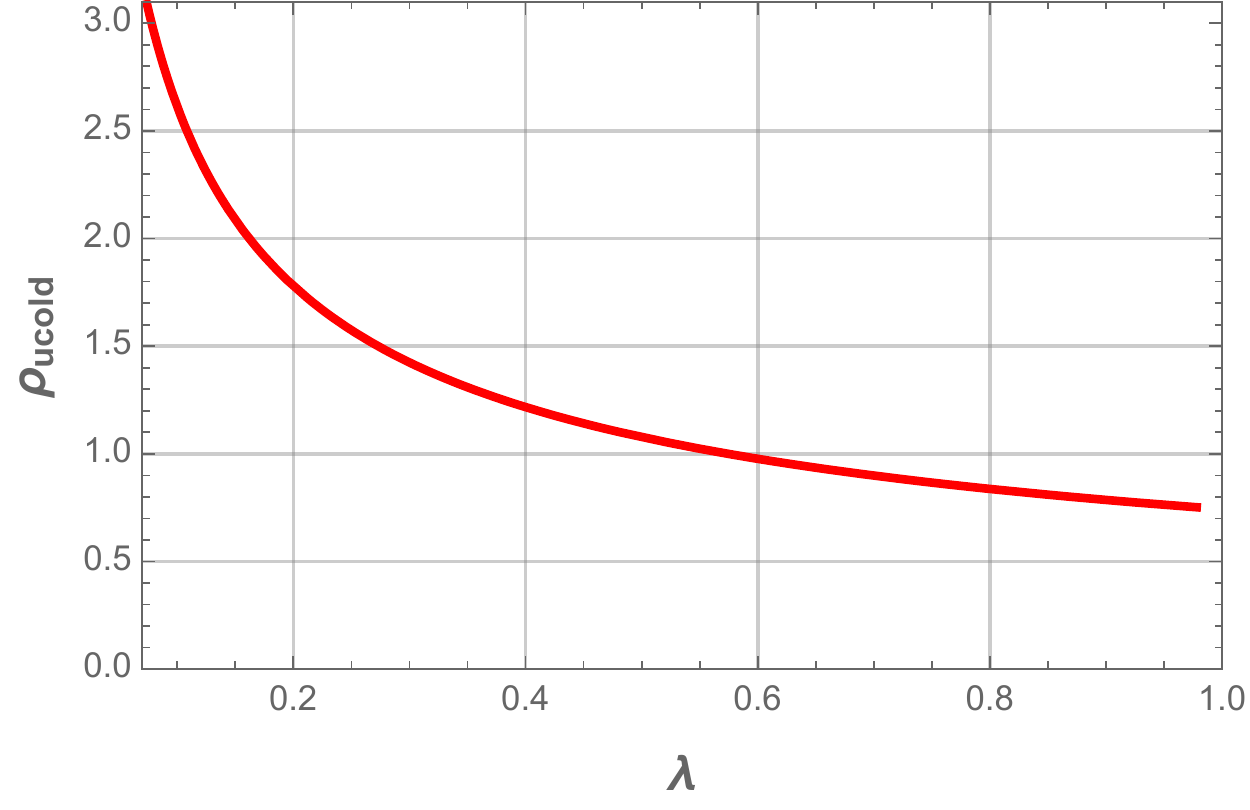}
 	\caption{\footnotesize Ultra-cold horizon radius $\rho_{ucold}$ as a function of cosmological constant $\lambda$.}
 	\label{f3_3}
 \end{subfigure}
 \hspace{1pt}	
 \begin{subfigure}[h]{0.45\textwidth}
 	\centering \includegraphics[scale=0.5]{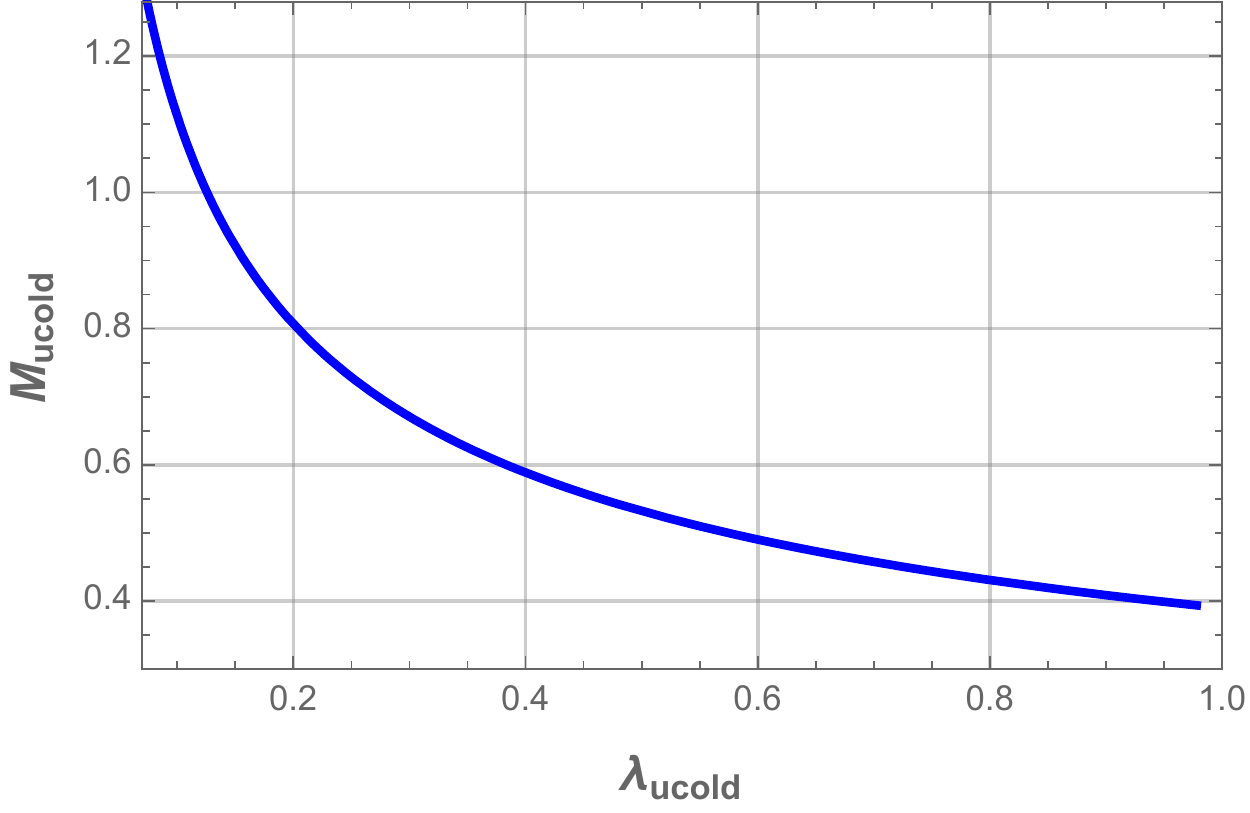}
 	\caption{\footnotesize Ultra-cold horizon mass $M_{ucold}$ as a function of cosmological constant $\lambda$.}
 	\label{f3_4}
 \end{subfigure}	
 	\caption{\footnotesize Ultra-cold horizon radius $\rho_{ucold}$ (red) and mass $Mass_{ucold}$ (blue) behavior as a function of cosmological constant $\lambda$ and electric charge $Q$ .}
 	\label{f3}
 \end{figure}

Before probing the phase picture, an overview of the black hole horizon and their extremal cases is necessary. The relation between the black hole mass $M$ and its horizon radius is naturally established from the equation of horizons $f (\rho) = 0$,
\begin{equation}\label{24}
M = \dfrac{\rho}{2} - \dfrac{Q^2 \left(\rho^3 - \rho^2+ 2\rho - 6 \right) e^{-\rho} } {12 \rho} - \left( \dfrac{\lambda}{6} - \dfrac{Q^2}{12} \mathcal{E}_1 (\rho)\right) \rho^3.
\end{equation}
We depict  in Fig.~\ref{f2} the behavior of the black hole mass $M$ as function  of the radius $\rho$ for different values of cosmological constant $\lambda$. 
For the appropriate values of the electric charge $Q$ and the cosmological constant $\lambda < \lambda_{ucold}$, the black hole mass plot presents a local minimum followed by a local maximum that are both located above the horizontal axis. These local extrema imply that the black hole has three horizons: the inner horizon $\rho_-$, the event horizon $\rho_+ (>\rho_-)$ and the cosmological horizon $\rho_c (>\rho_+)$. When the cosmological constant reaches the critical value $\lambda=\lambda_{ucold}$, the mass curve shows an inflection point where the three horizons coincide. In this case, the black hole is dubbed ultra-cold black hole. From Fig.~\ref{f3}, we can see that ultra-cold black hole with larger electric charge has larger horizon radius and larger mass,  whereas the critical value for the cosmological constant is smaller one.

In the case where $\lambda>\lambda_{ucold}$,  no black hole exists whatever its mass phase. The critical value of the cosmological constant for the existence of the ENLMY charged dS black hole is larger than that for the usual charged dS black hole. This can be explained by the electric repulsion which is much more weaker in the nonlinear Maxwell-Yukawa electrodynamics than in the usual electrodynamics.

\begin{figure}[h!]
	\centering
	\begin{subfigure}[h]{0.45\textwidth}
		\centering \includegraphics[scale=0.5]{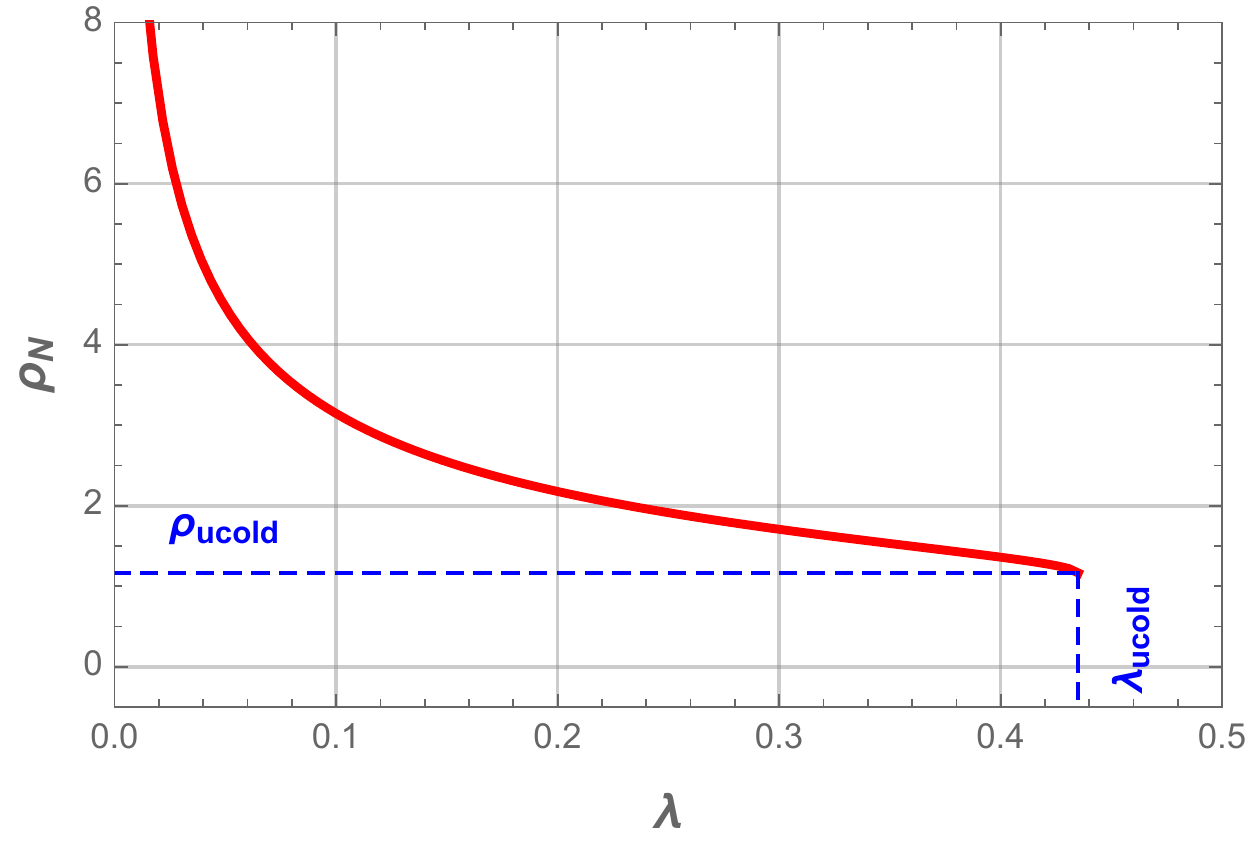}
		\caption{\footnotesize Nariai black hole horizon radius $\rho_{N}$ as a function of cosmological constant $\lambda$.}
		\label{f4_1}
	\end{subfigure}
	\hspace{1pt}	
	\begin{subfigure}[h]{0.45\textwidth}
		\centering \includegraphics[scale=0.5]{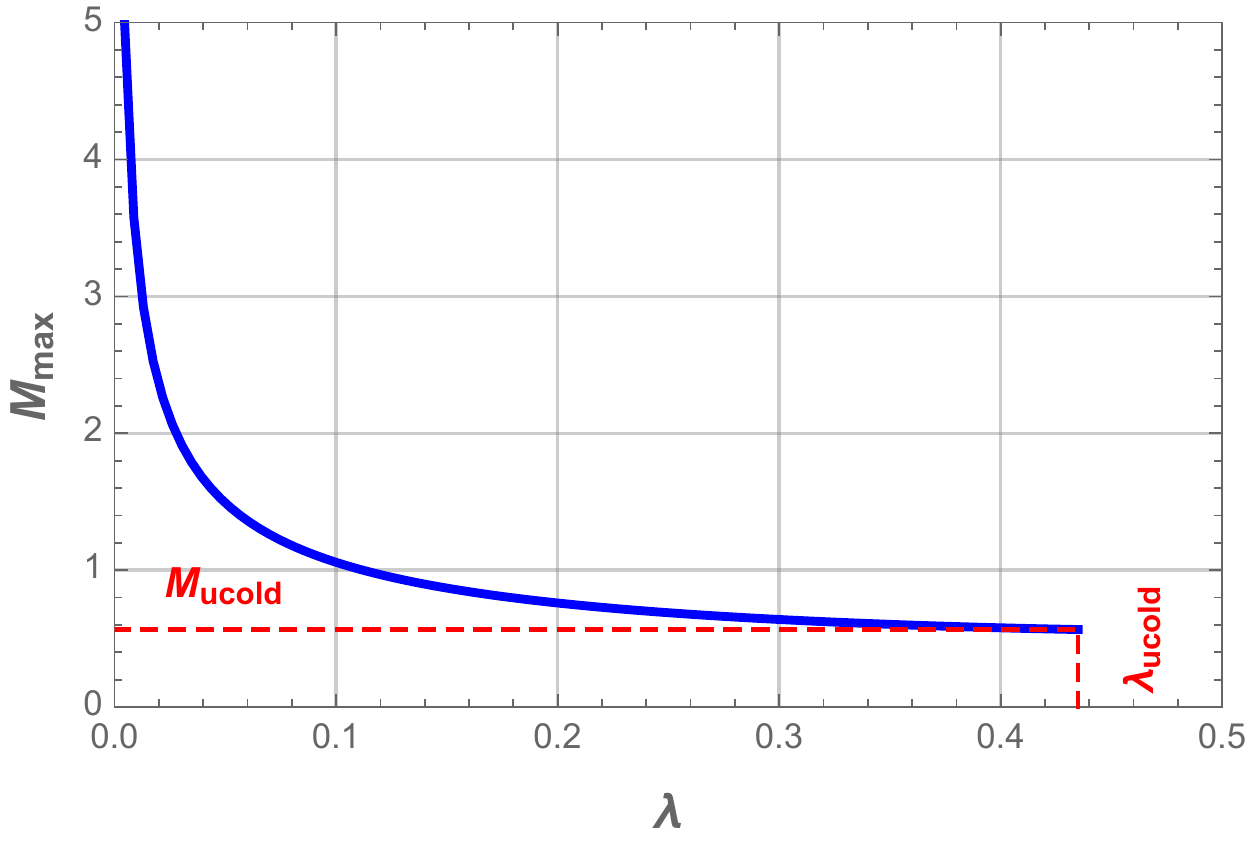}
		\caption{\footnotesize Nariai black hole horizon radius $M_{N}$ as a function of cosmological constant $\lambda$.}
		\label{f4_2}
	\end{subfigure}
	\caption{\footnotesize Nariai black hole horizon radius $\rho_{N}$ (red) and mass $M_{N}$ (blue) in terms of cosmological constant $\lambda$.}
	\label{f4}
\end{figure}

For the case $\lambda< \lambda_{ucold}$, the situation is quite different, on can note that the black hole  phase exists only for a range of  mass values, $M \in [M_{cold}, M_{N}]$, where $M_{cold}$ and $M_{N}$ are functions  of the charges and cosmological constant.  When $M=M_{N}$, the event horizon $\rho_{+}$ and the cosmological one $\rho_{c}$ coincide, this give rise to a new phase called Nariai black hole with an event horizon radius noted $\rho_{N}$. The latter and its associated mass are depicted in Fig.~\ref{f4} as a function of  the charge and cosmological constant.

From Fig.~\ref{f4} we notice that both the event horizon radius and mass of the Nariai black hole should decrease as the cosmological constant increases till the value $\lambda = \lambda_{ucold}$.

\begin{figure}[h!]
	\centering
	\begin{subfigure}[h]{0.45\textwidth}
		\centering \includegraphics[scale=0.5]{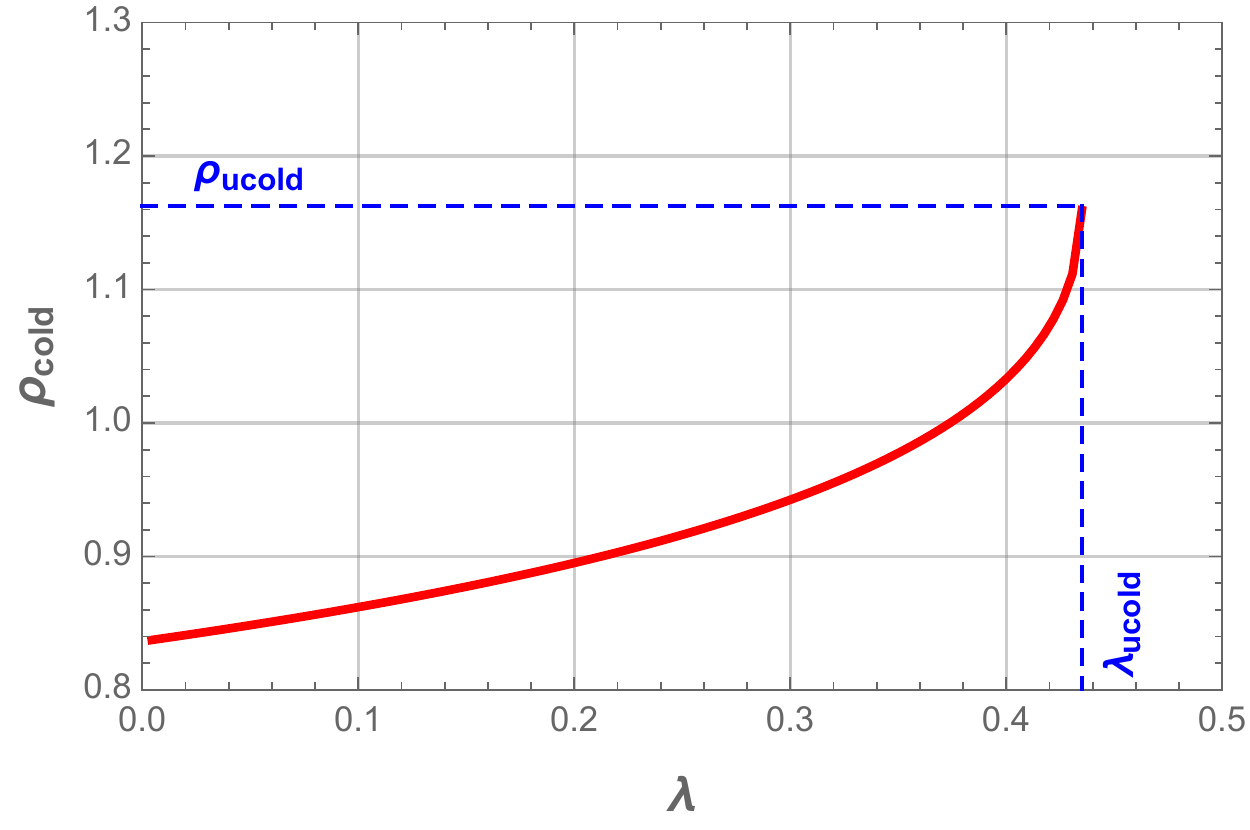}
		\caption{Cold black hole horizon radius $\rho_{cold}$ as a function of cosmological constant $\lambda$.}
		\label{f5_1}
	\end{subfigure}
	\hspace{1pt}	
	\begin{subfigure}[h]{0.45\textwidth}
		\centering \includegraphics[scale=0.5]{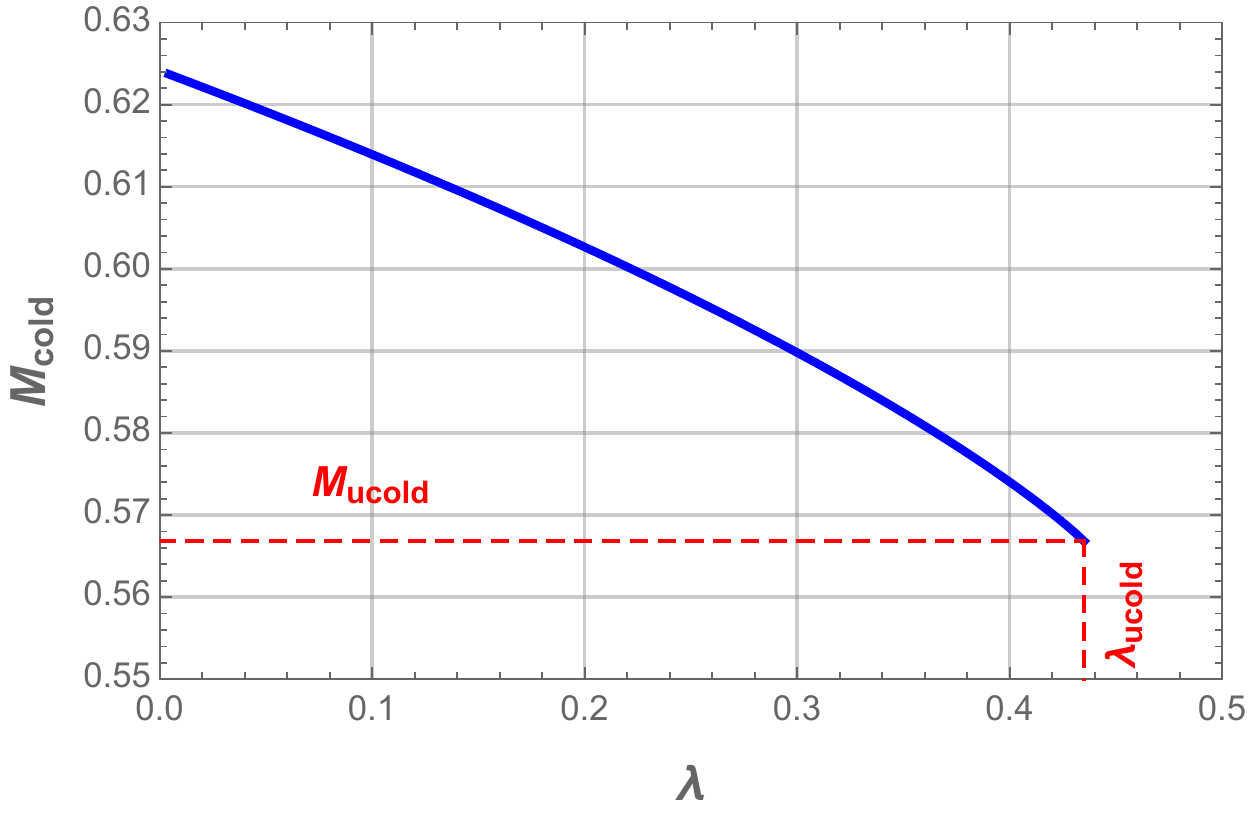}
		\caption{Cold black hole horizon radius $M_{cold}$ as a function of cosmological constant $\lambda$.}
		\label{f5_2}
		
	\end{subfigure}

	\caption{Cold black hole horizon radius $\rho_{cold}$ (red) and mass $M_{cold}$ (blue) as a function of cosmological constant $\lambda$.}
	\label{f5}
\end{figure}

When $M = M_{cold} $, the inner horizon $\rho_{-}$ and the event horizon $\rho_{+}$ coincide,  this phase is dubbed cold black hole. Fig.~\ref{f5} illustrates the plots of its event horizon radius $\rho_{cold}$ and corresponding mass as a function of the charge and  cosmological constant.

We clearly see that the event horizon radius of the cold black hole increases, but its mass decreases as the cosmological constant get larger  till the value  $\lambda = \lambda_{ucold}$.

After discussing various black hole possible solutions and their extremal case, the next section will be dedicated to the investigation of the thermodynamics from the local point of view. 

\section{Local black hole thermodynamics view }

In the extended phase space, the entropy $S$, the electric charge $Q$, Yukawa charge $\alpha$, and the thermodynamic pressure $P$ are interpreted as a complete set of the extensive thermodynamics variables. Their associated conjugate quantities are the temperature $T$, the pseudo-electrical potential $\Phi_{q}$, the pseudo-Yukawa potential $\Phi_{\alpha}$, and the thermodynamic volume $V$ respectively. Under these considerations, one can establish the first law of thermodynamics for each horizon: the event horizon, the cosmological and the inner one as follow, 
\begin{equation}\label{25}
\begin{split}
& dM = T_+ dS_+ +  V_+ dP + \Phi_{q_+}dQ+ \Phi_{\alpha_+}d\alpha, \\
& dM = -T_c dS_c+ V_c dP + \Phi_{q_c}dQ+\Phi_{\alpha_c}d\alpha,\\
&dM = -T_- dS_-+ V_- dP + \Phi_{q_-}dQ+\Phi_{\alpha_-}d\alpha.
\end{split}
\end{equation}
Note that the minus signs  in front of $T_c$ and $T_i$ are introduced to ensure an increasing entropy: As the cosmological and  inner horizon radius $r_{c,-}$ increase, the mass $M$ should decrease. Then, on can write the black hole mass, Eq.~\eqref{24}, in terms of 
 the event, inner and cosmological horizons, respectively as \begin{equation}\label{26}
\begin{split}
& M(\rho_{\pm,c},P,Q) = \dfrac{\rho_{\pm,c}}{2} - \dfrac{Q^2 \left(\rho_{\pm,c}^3 - \rho_{\pm,c}^2+ 2\rho_{\pm,c} - 6 \right) e^{-\rho_{\pm,c}} } {12 \rho_{\pm,c}} - \left( \dfrac{\lambda}{6} - \dfrac{Q^2}{12} \mathcal{E}_1 (\rho_{\pm,c})\right) \rho_{\pm,c}^3,\\
\end{split}
\end{equation}

Note that the Yukawa charge $\alpha$ is implicitly included in all variables by the re-scaling performed previously, thus we should take it into account when necessary. In what follows, to perform the thermodynamic analysis and phase transition for each horizon independently, we assume that the horizons are located far away from each other to avoid their overlapping. 

\begin{figure}[h!]
	\centering
	\begin{subfigure}[h]{0.45\textwidth}
		\centering \includegraphics[scale=0.5]{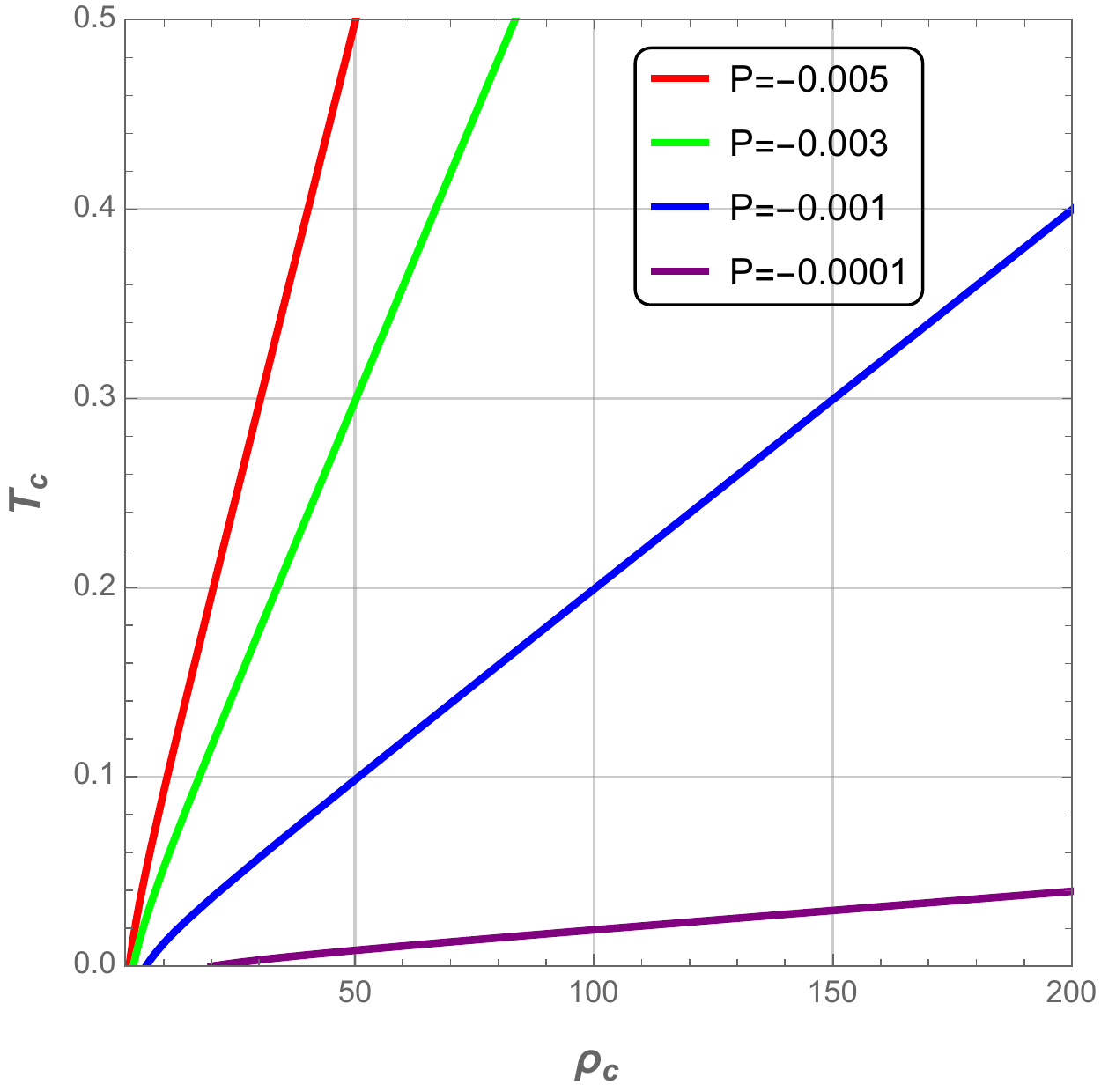}
		\caption{ \footnotesize  $T_c$ vs $\rho_{c}$.}
		\label{f6_1}
	\end{subfigure}
	\hspace{1pt}	
	\begin{subfigure}[h]{0.45\textwidth}
		\centering \includegraphics[scale=0.5]{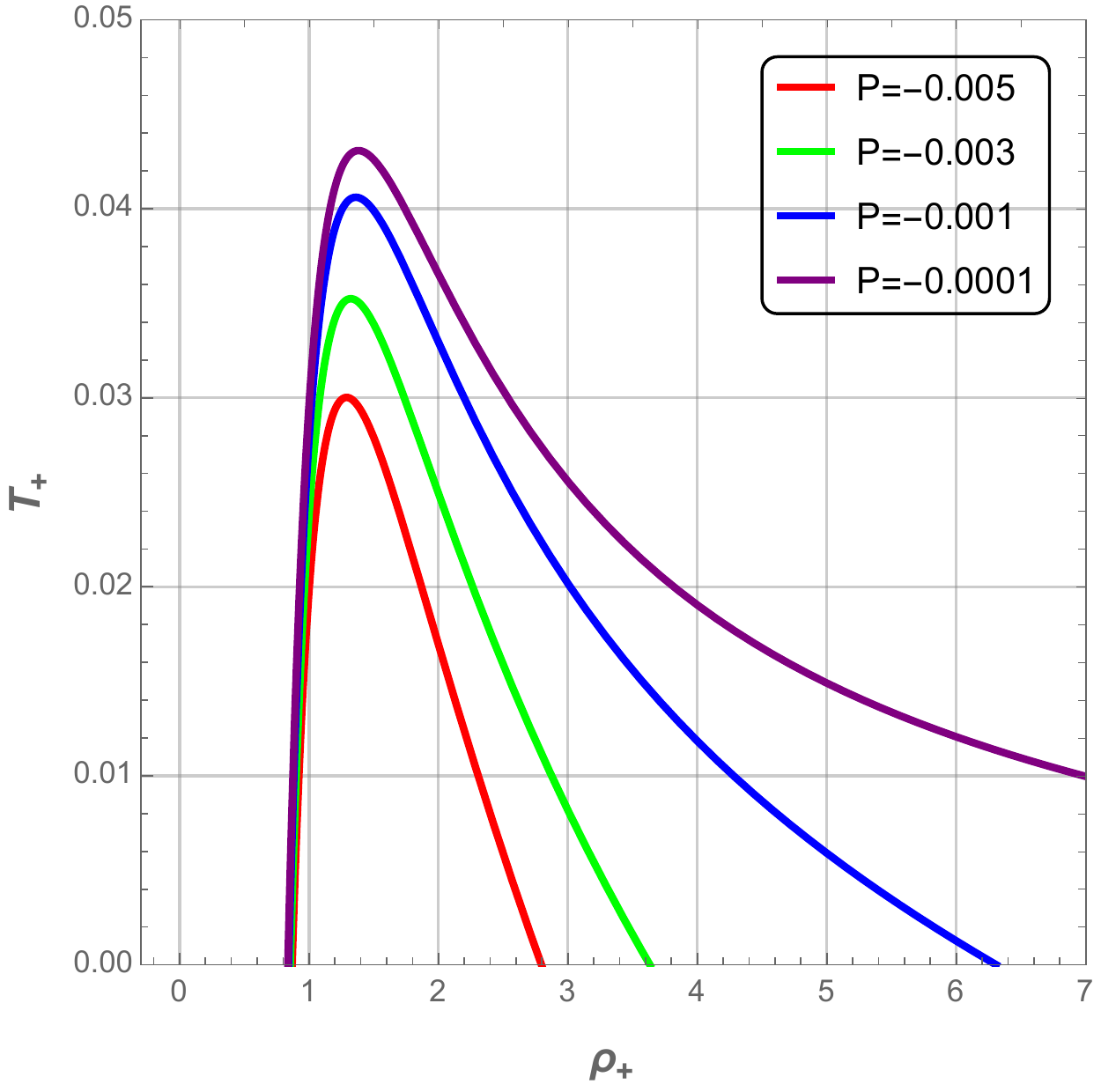}
		\caption{\footnotesize  $T_+$ vs $\rho_{+}$.}
		\label{f6_2}
		
	\end{subfigure}

	\caption{Cosmological and event horizon temperatures as a function of corresponding radius with $Q=1$.}
	\label{f6}
\end{figure}

The scaled Hawking temperature corresponding to each horizon is given by \cite{nam82,nam83}, 

\begin{equation}\label{27}
\begin{split}
 T_{\pm,c} &=\dfrac{1}{4\pi} \left. \dfrac{\partial f(\rho)}{\partial \rho}\right| _{\rho = \rho_{+}}\\ =& \dfrac{1}{8\pi \rho_{+}^3}\left[ \rho_{+}^2\left( 2 + 16 P \rho_{+}^2 + Q^2 \rho_+^2 \mathcal{E}(\rho_{+})\right) \right. 
  \left. - Q^2\left( 2+ \rho_{+}\left( 2+\rho_{+}\left( \rho_{+}-1\right) \right) \right)e^{-\rho_{+}} \right]  \\
 T_{-,c} =&-\dfrac{1}{4\pi} \left. \dfrac{\partial f(\rho)}{\partial \rho}\right| _{\rho = \rho_{-,c}}\\=& - \dfrac{1}{8\pi \rho_{-,c}^3}\left[ \rho_{-,c}^2\left( 2 + 16 P \rho_{-,c}^2 + Q^2 \rho_{-,c}^2 \mathcal{E}(\rho_{-,c})\right) \right. 
 \left. - Q^2\left( 2+ \rho_{-,c}\left( 2+\rho_{-,c}\left( \rho_{-,c}-1\right) \right) \right)e^{-\rho_{-,c}} \right]  \\
\end{split}
\end{equation}
where, we have defined the pressure to be $P = - \lambda/8 \pi < 0$. 
In Fig.~\ref{f6}, we plot the variation of the scaled temperatures as a function of  each reduced horizons. One can notice from Fig.~\ref{f6}
 that the temperatures $T_{-,c}$ are monotonous in terms of the inner and cosmological radius $\rho_{-,c}$,
whereas the event horizon temperature $T_+$ presents a maximum $T_{max}$ value listed in Table \ref{t1} with its corresponding radius, then decreases as the event horizon radius $\rho_{+}$ increases.  

\begin{table}
\centering	
\begin{tabular}{ |c|c|c|c|c| }
	\hline
	\bf Pressure & $P = -0.005 $ & $P = -0.003 $ & $P = -0.001  $& $P = -0.0001 $\\
	\hline
 $\bf T_{max}$   & $0.0300189$    & $0.0352443$ &   $0.0406112$ &   $0.0430807$\\
	\hline
 $\bf{\rho_{max}}$ &   $1.29066$  & $1.32294$   & $1.36185$ &   $1.38236$\\
	\hline
\end{tabular}
\caption{\footnotesize Maximal temperature $T_{max}$ and its corresponding radius $\rho_{max}$ for different values of pressure $P$ at event horizon with $Q = 1$.}
\label{t1}
\end{table}

The maximum temperature $T_{max}$ and its corresponding event horizon radius $\rho_{max}$ are illustrated in Fig.~\ref{7} where we see that both quantities are increasingly monotonous functions of pressure $P$.
 \begin{figure}[th!]
	\centering
	\begin{subfigure}[h]{0.45\textwidth}
		\centering \includegraphics[scale=0.5]{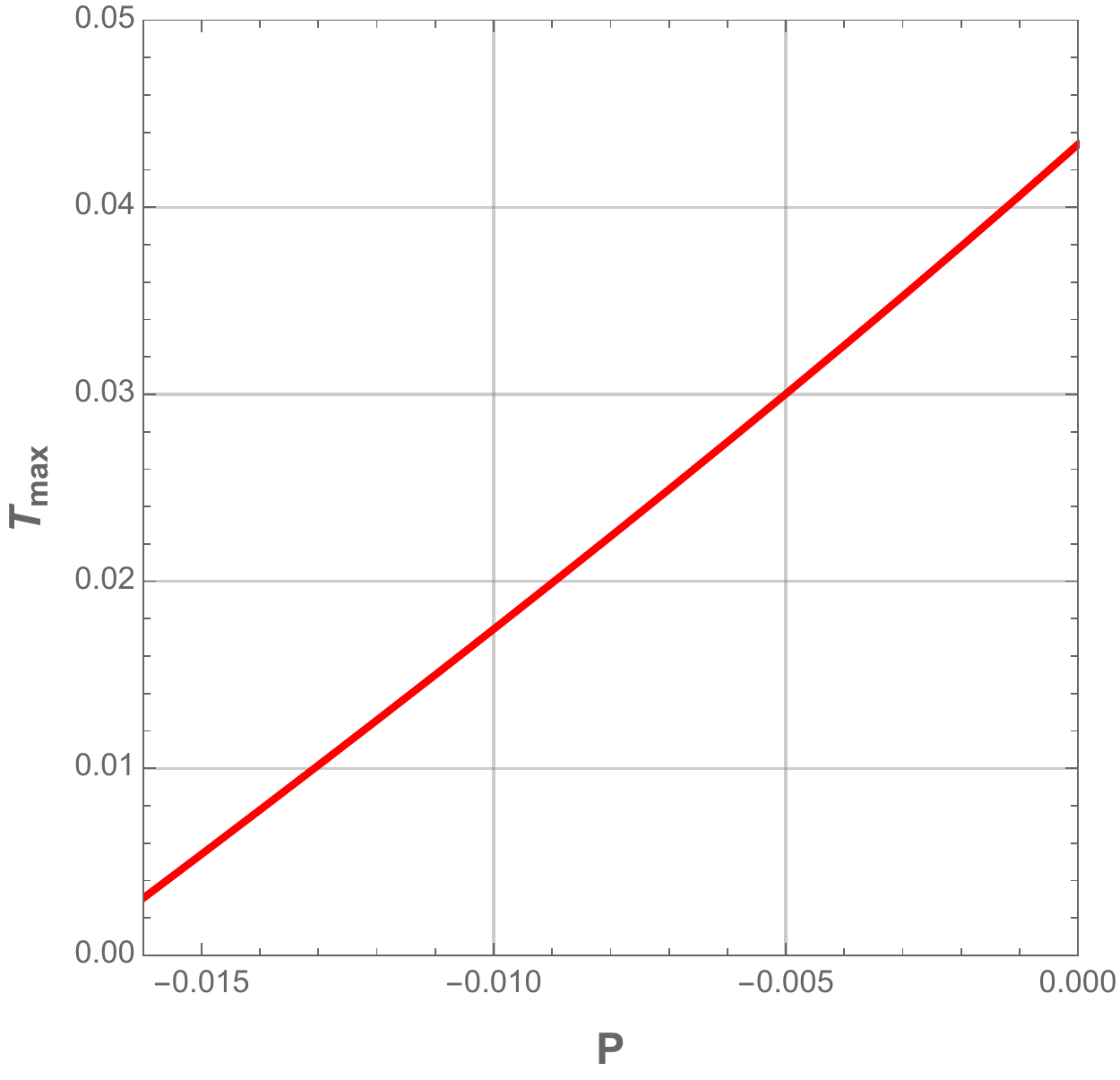}
		\caption{\footnotesize Maximum Temperature $T_{max}$ as function of pressure $P$. }
		\label{f7_1}
	\end{subfigure}
	\hspace{1pt}	
	\begin{subfigure}[h]{0.45\textwidth}
		\centering \includegraphics[scale=0.5]{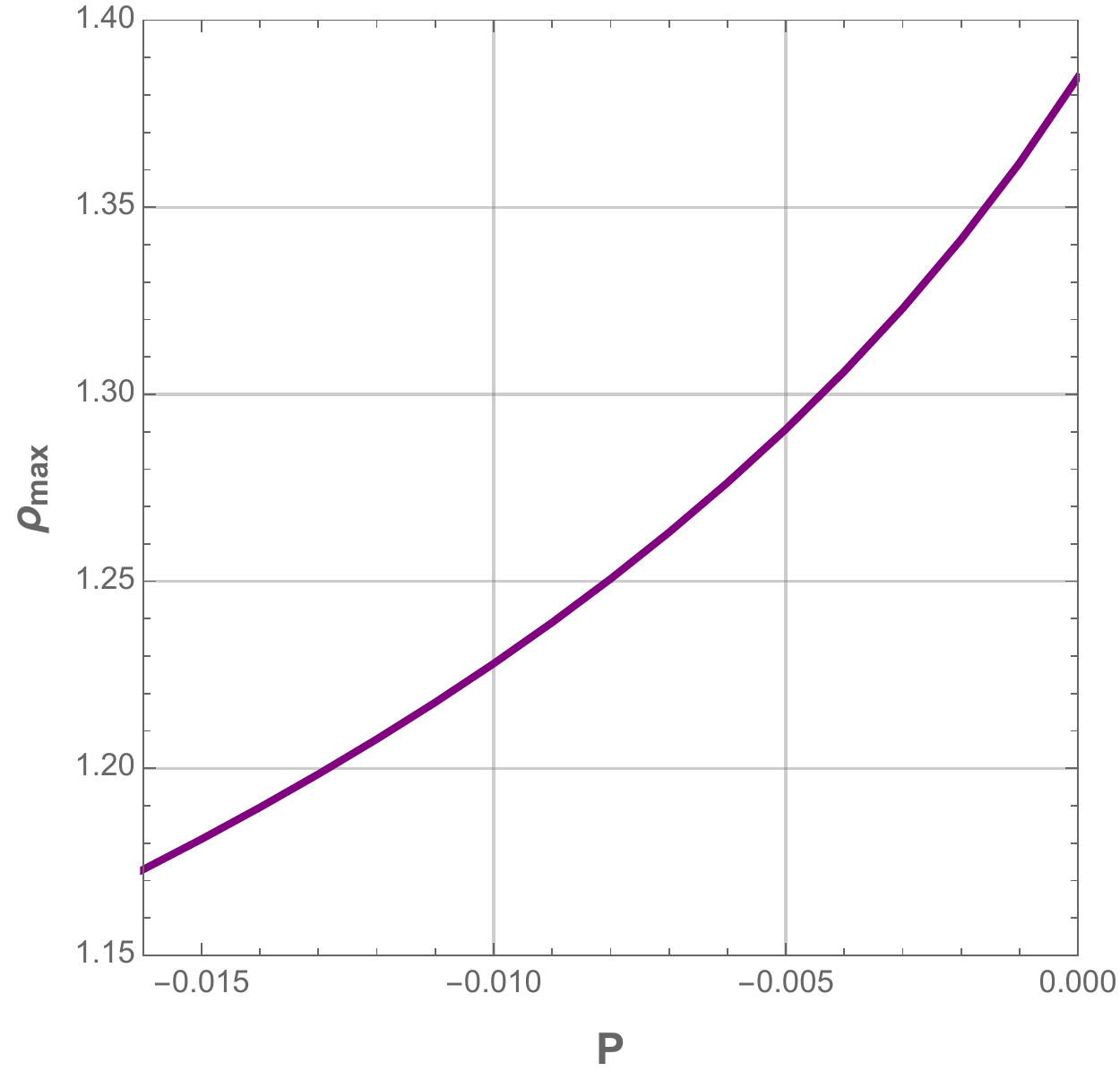}
		\caption{\footnotesize Event horizon radius $\rho_{max}$ as function of pressure $P$. }
		\label{f7_2}
	\end{subfigure}
	\caption{Maximum Temperature $T_{max}$ and its corresponding radius $\rho_{max}$ as function of pressure $P$ with $Q=1$.}
	\label{f7}
\end{figure}

\begin{figure}[h!]
	\centering \includegraphics[scale=0.7]{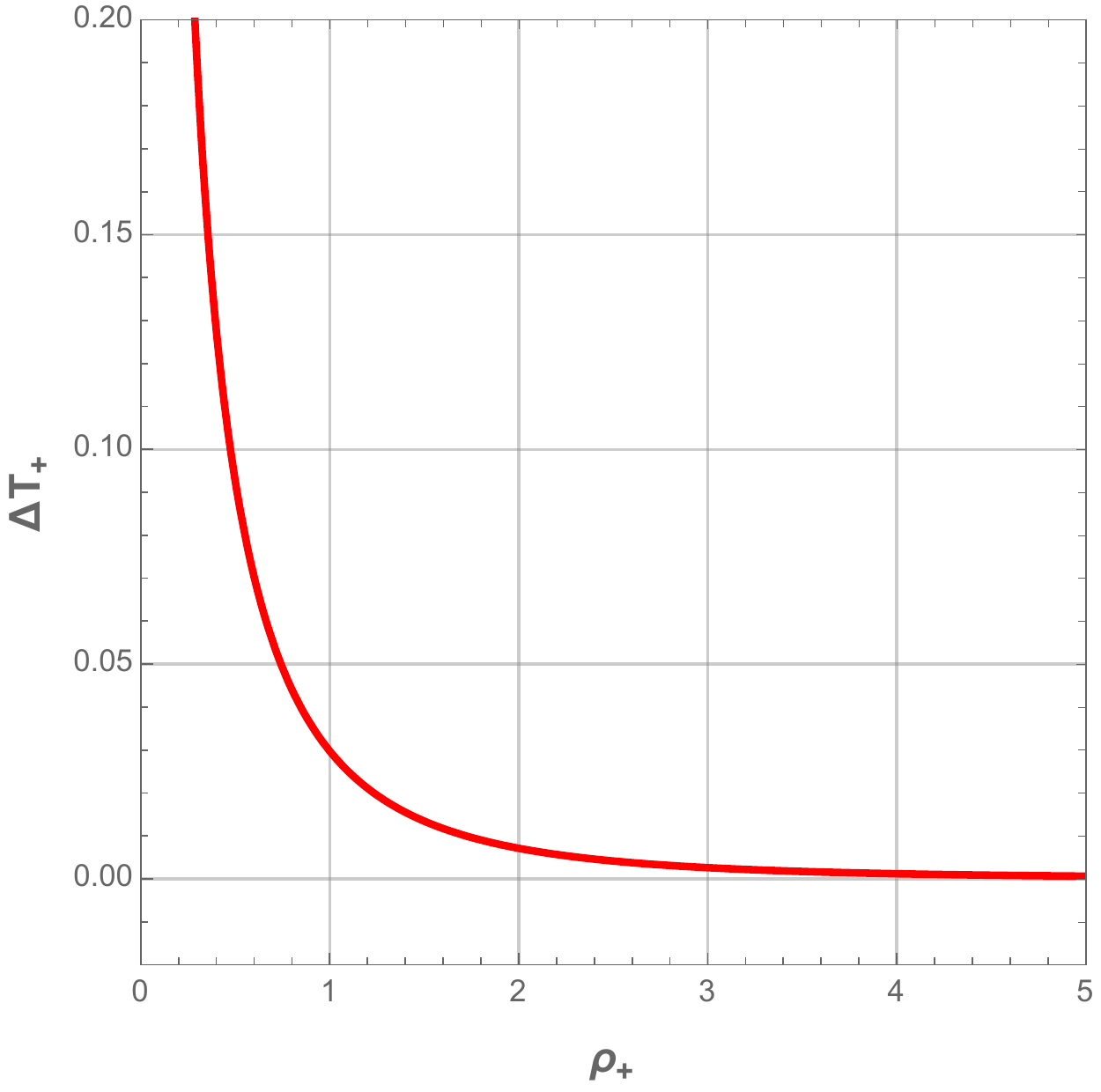}
	\caption{\footnotesize Difference of the temperature $\Delta T_+ $ between the ENLMY and usual charged dS black holes  as function of event horizon $\rho_{+}$ with $P = -0.001$ and $Q=1$. }
	\label{f8}
	
\end{figure}

Now, a comparison between the ENLMY and usual charged dS black holes with the same size, charge and the pressure is needed. For that, we depict  in Fig.~\ref{f8} the difference of  temperature as a function of horizon radius $\rho_+$. One can easily see that the ENLMY charged dS black hole is exponentially hotter than the usual charged black hole when $\rho \to 0$.

From the expression of the first law Eq.~\eqref{25}, the formulas of mass Eq.~\eqref{26} and temperature Eq.~\eqref{27}, we can evaluate the entropy corresponding to each horizon,

\begin{equation}\label{28}
S_+ =  \int \dfrac{1}{T_+} \dfrac{\partial M}{\partial \rho_+} d\rho_{+} = \pi \rho^2_+ ,
\end{equation}
\begin{equation}\label{29}
S_{-,c} = - \int \dfrac{1}{T_{-,c}} \dfrac{\partial M}{\partial \rho_{-,c}}d\rho_{-,c} = \pi \rho^2_{-,c}.
\end{equation}

Having obtained the black hole temperature and entropy associated with each horizon, we proceed further and evaluate other relevant thermodynamics variables composing the first law. The electric potential associated with the charge $q$ is : 

\begin{equation}\label{31}
\Phi_{q_{\pm,c}}  = \left. \dfrac{\partial M}{\partial Q}\right) _{S_{\pm,c},P,\alpha} = \dfrac{Q}{6\rho_{\pm,c}}\left[ \rho_{\pm,c}^4 \mathcal{E}(\rho_{\pm,c})  + \left( 6+ \rho_{\pm,c}\left( \rho_{\pm,c}\left( \rho_{\pm,c}-1\right) -2 \right) \right)e^{-\rho_{\pm,c}} \right]  
\end{equation}

while the potentials corresponding to Yukawa charge are given by,
\begin{equation}\label{32}
\Phi_{\alpha_{\pm,c}}  = \left. \dfrac{\partial M}{\partial \alpha}\right) _{S_{\pm,c},P,Q} = \dfrac{Q}{3}\left[ \rho_{\pm,c}^3 \mathcal{E}(\rho_{\pm,c})  + \left( \rho_{\pm,c}\left( \rho_{\pm,c}-1\right) -2 \right) e^{-\rho_{\pm,c}} \right], 
\end{equation}

and the thermodynamic volumes read as,
\begin{equation}\label{37}
V_{\pm,c}  = \left. \dfrac{\partial M}{\partial P}\right) _{S_{\pm,c},Q,\alpha} = \dfrac{4}{3}\pi  r_{\pm,c}^3
\end{equation}
All  above  thermodynamics variables  on the event and inner/cosmological horizons satisfy the Smarr and Smarr-like formulas which, by a scaling argument, can be expressed as,
 \begin{equation}\label{40}
 \begin{split}
 M & = 2 \left( T_+ S_+ - V_+ P\right) + Q \Phi_{q_+} -\alpha \Phi_{\alpha_+}, \\
  M & = -2 \left( T_{-,c} S_{-,c} + V_{-,c} P\right) + Q \Phi_{q_+} -\alpha \Phi_{\alpha_{-,c}}, \\
 \end{split}
 \end{equation}
 Note that the Yakawa potential energy is simply given by the formula:
  \begin{equation}\label{41}
E_Y = q \Phi_{q} -\dfrac{\alpha}{2} \Phi_{\alpha} = \dfrac{q^2}{r} e^{-\alpha r} = q \phi(r).
 \end{equation}

Hence, we can express the heat capacity at constant pressure on the event, inner and cosmological horizons respectively as,
 \begin{equation}\label{42}
 C_{P+}  = \left. \dfrac{\partial M}{\partial T_+}\right)_{P} , \hspace{5mm}  C_{P-,c}  = - \left. \dfrac{\partial M}{\partial T_-,c}\right)_{P}, \hspace{5mm}  
 \end{equation}
 
\begin{figure}[ht!]
	\centering
	\begin{subfigure}[h]{0.45\textwidth}
		\centering \includegraphics[scale=0.5]{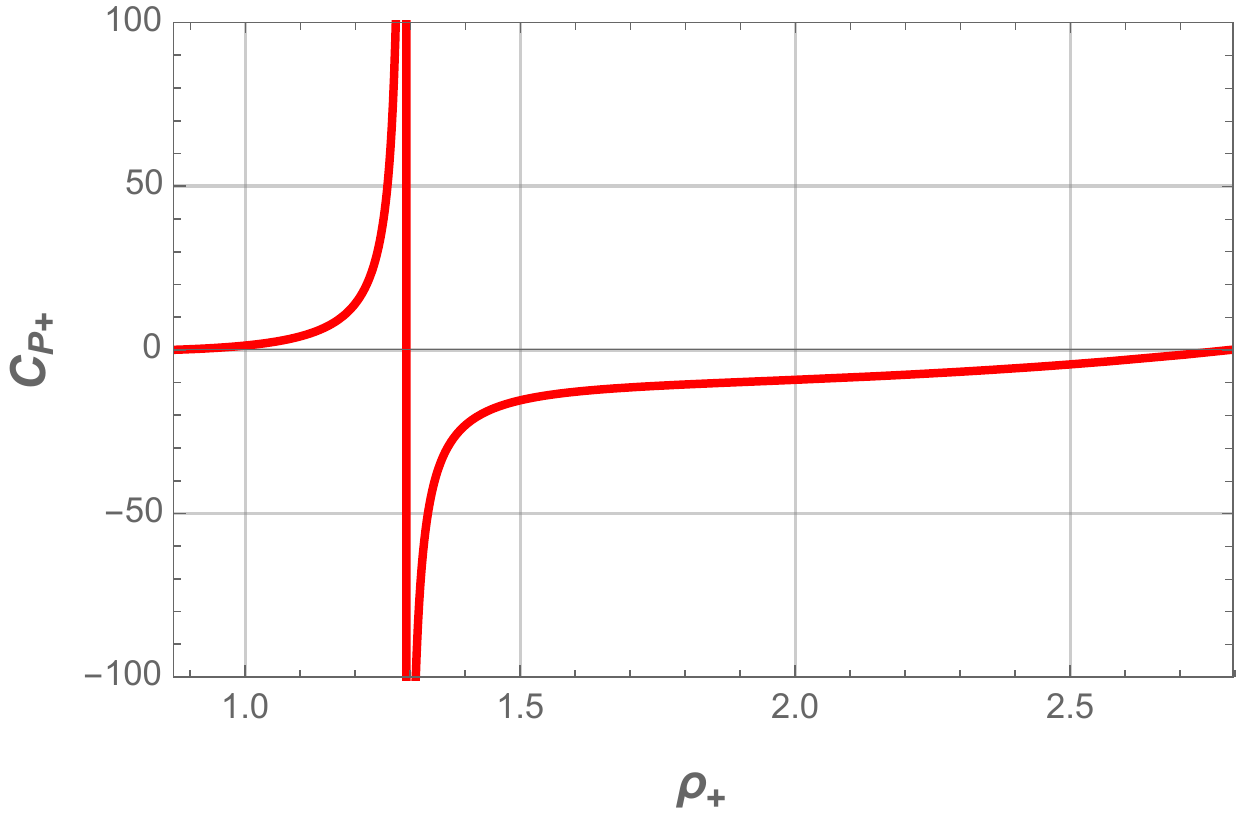}
		\caption{\footnotesize $P = -0.005$.}
		\label{f9_1}
	\end{subfigure}
	\hspace{1pt}	
	\begin{subfigure}[h]{0.45\textwidth}
		\centering \includegraphics[scale=0.5]{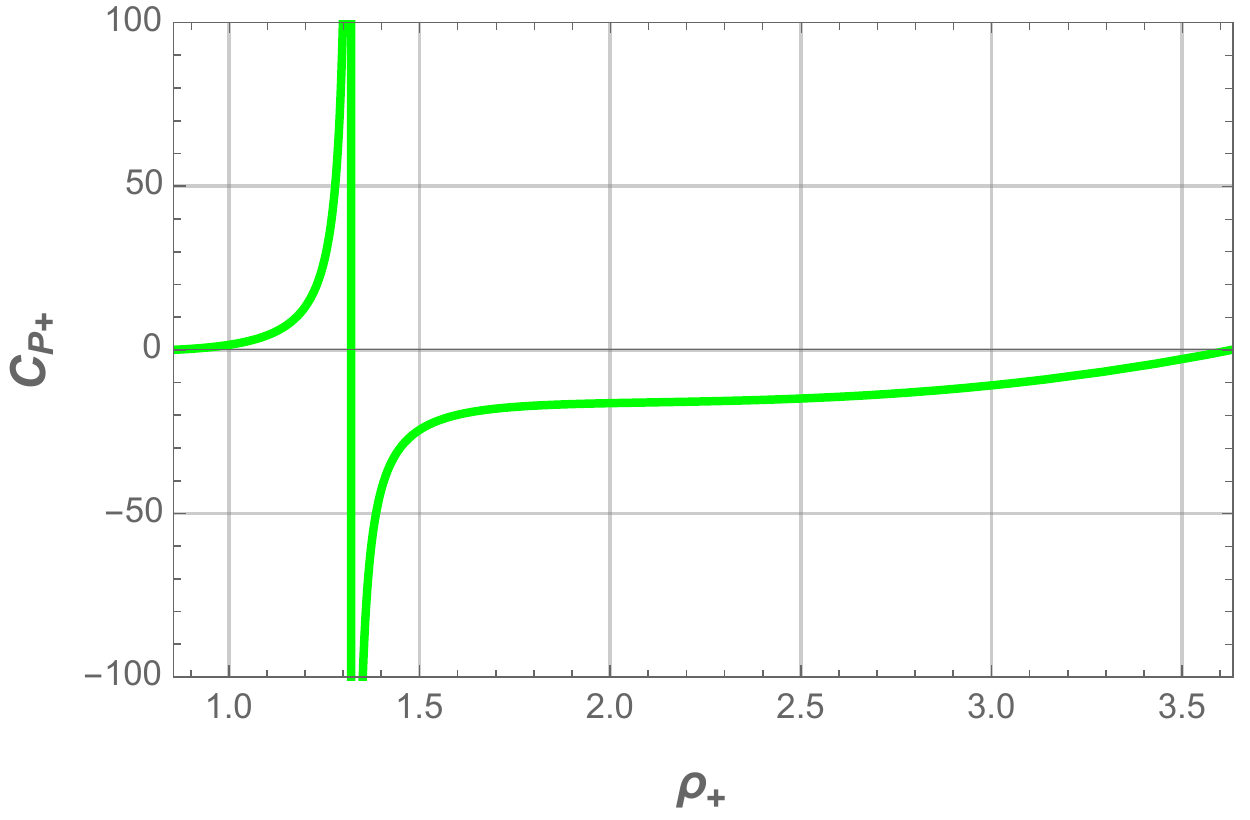}
		\caption{\footnotesize $P = -0.003$.}
		\label{f9_2}
		
	\end{subfigure}
	\begin{subfigure}[h]{0.45\textwidth}
		\centering \includegraphics[scale=0.5]{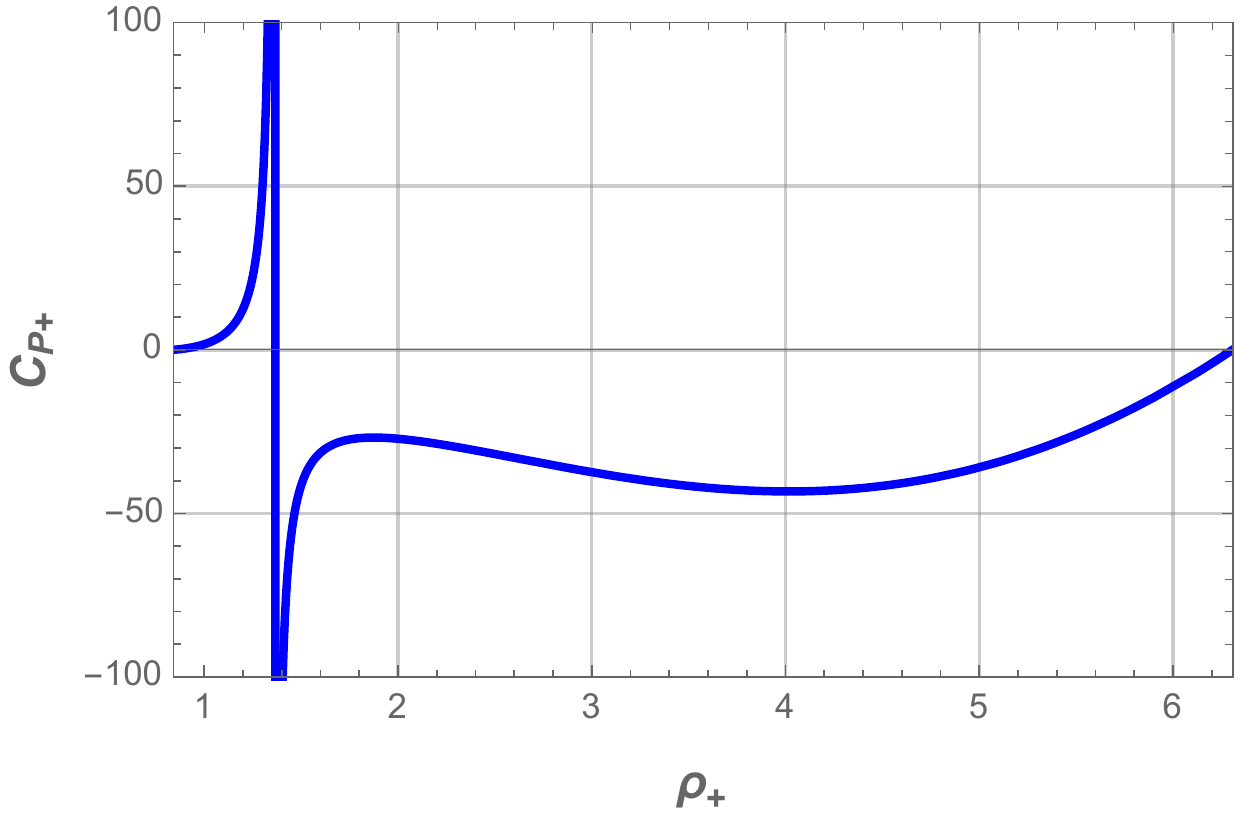}
		\caption{\footnotesize $P = -0.001$.}
		\label{f9_3}
	\end{subfigure}
	\hspace{1pt}	
	\begin{subfigure}[h]{0.45\textwidth}
		\centering \includegraphics[scale=0.5]{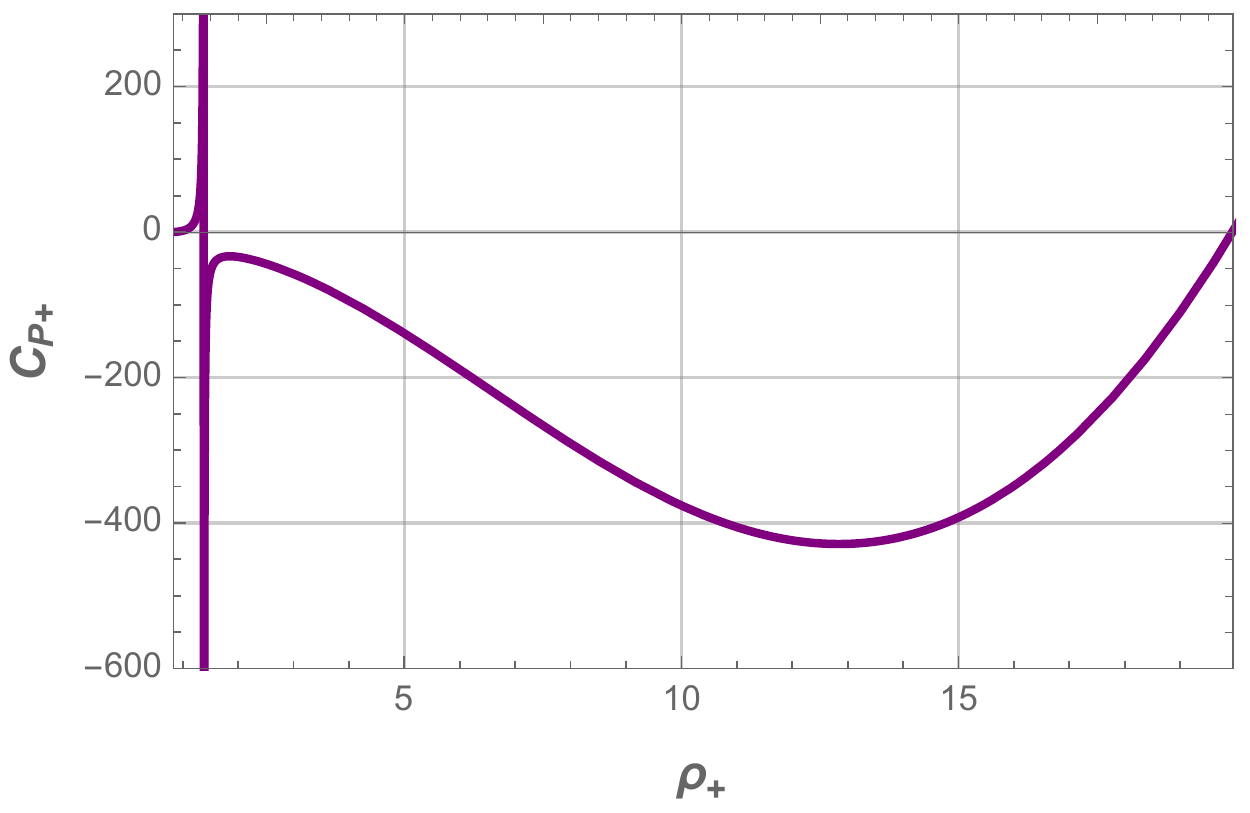}
		\caption{\footnotesize $P = -0.0001$.}
		\label{f9_4}
	\end{subfigure}
	\caption{\footnotesize Event horizon heat capacity $C_{P+}$ as a function of event horizon radius $\rho_+$ with the electric charge $Q=1$ .}
	\label{f9}
\end{figure}

\begin{figure}[h!]
	\centering
	\begin{subfigure}[h]{0.45\textwidth}
		\centering \includegraphics[scale=0.5]{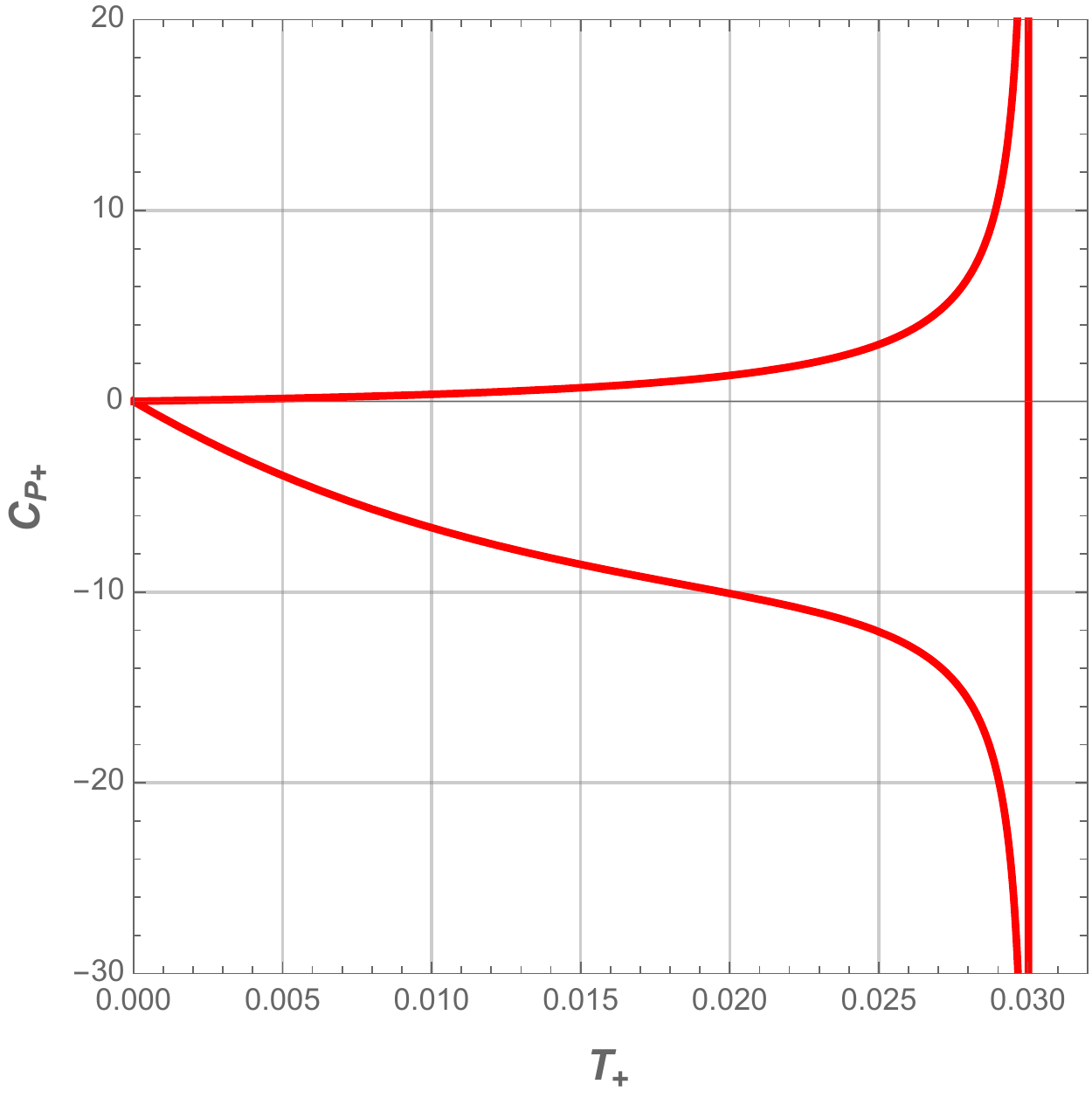}
		\caption{\footnotesize $P = -0.005$.}
		\label{f91_1}
	\end{subfigure}
	\hspace{1pt}	
	\begin{subfigure}[h]{0.45\textwidth}
		\centering \includegraphics[scale=0.5]{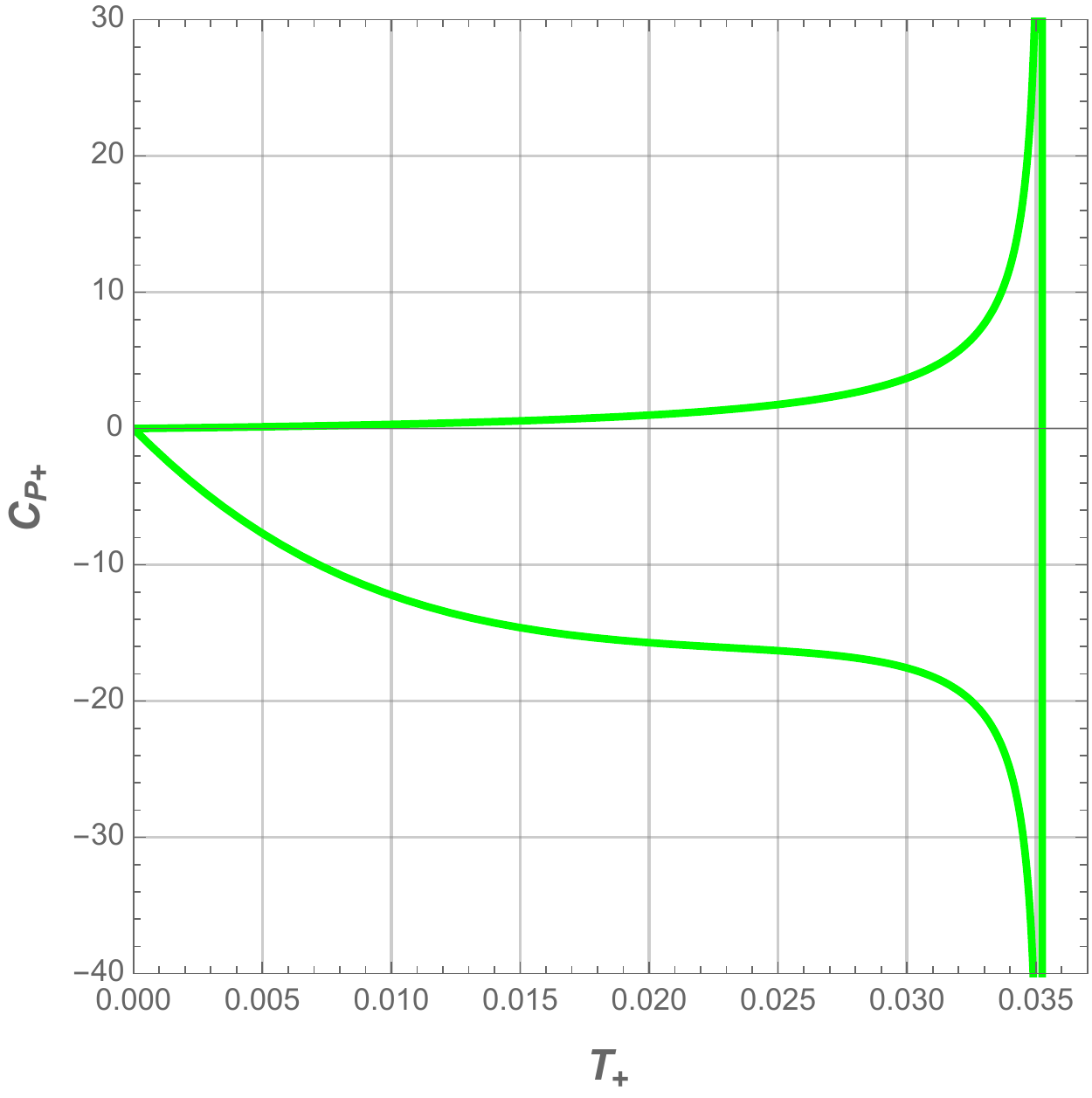}
		\caption{\footnotesize $P = -0.003$.}
		\label{f91_2}
		
	\end{subfigure}
	\begin{subfigure}[h]{0.45\textwidth}
		\centering \includegraphics[scale=0.5]{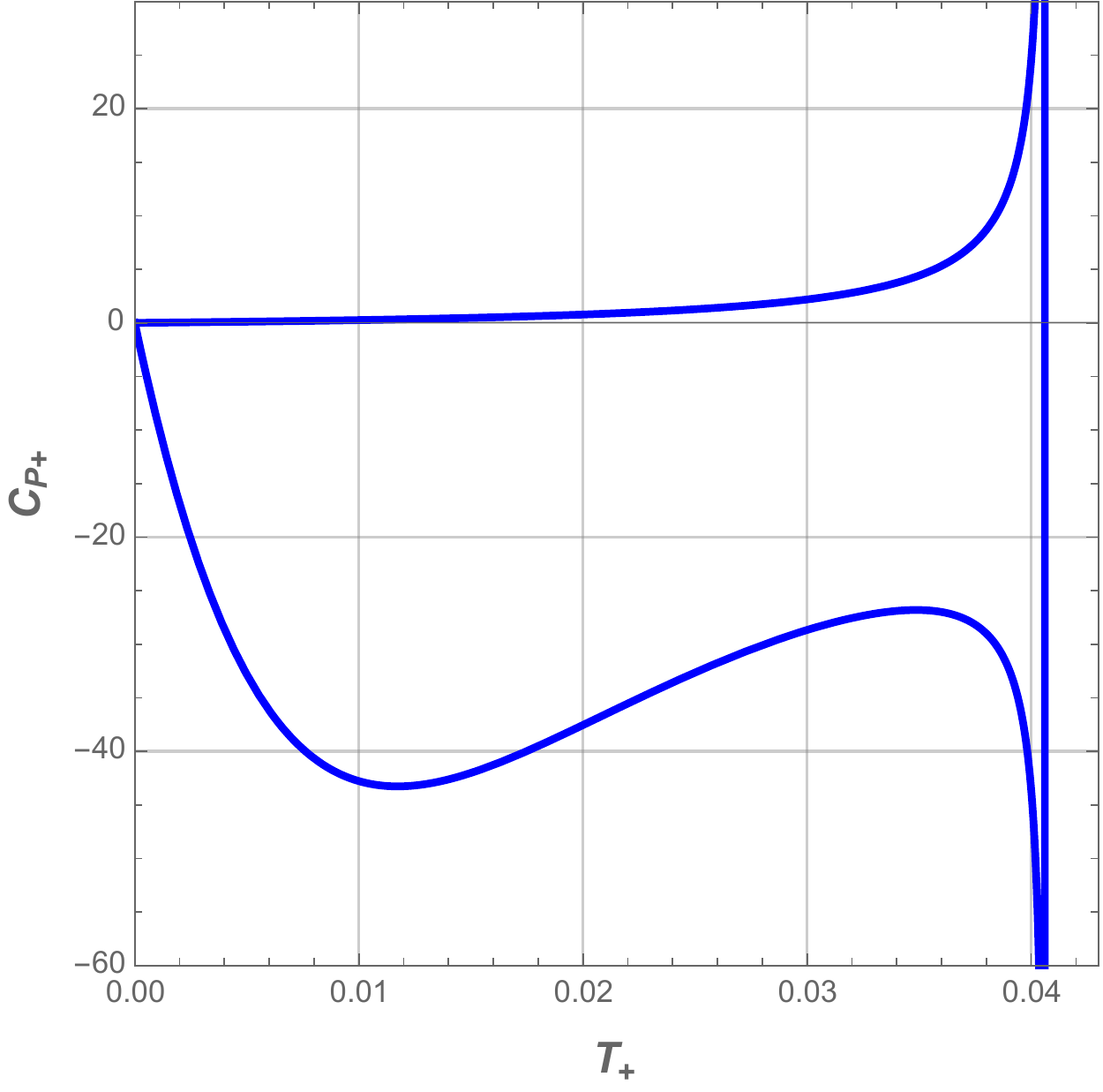}
		\caption{\footnotesize $P = -0.001$.}
		\label{f91_3}
	\end{subfigure}
	\hspace{1pt}	
	\begin{subfigure}[h]{0.45\textwidth}
		\centering \includegraphics[scale=0.5]{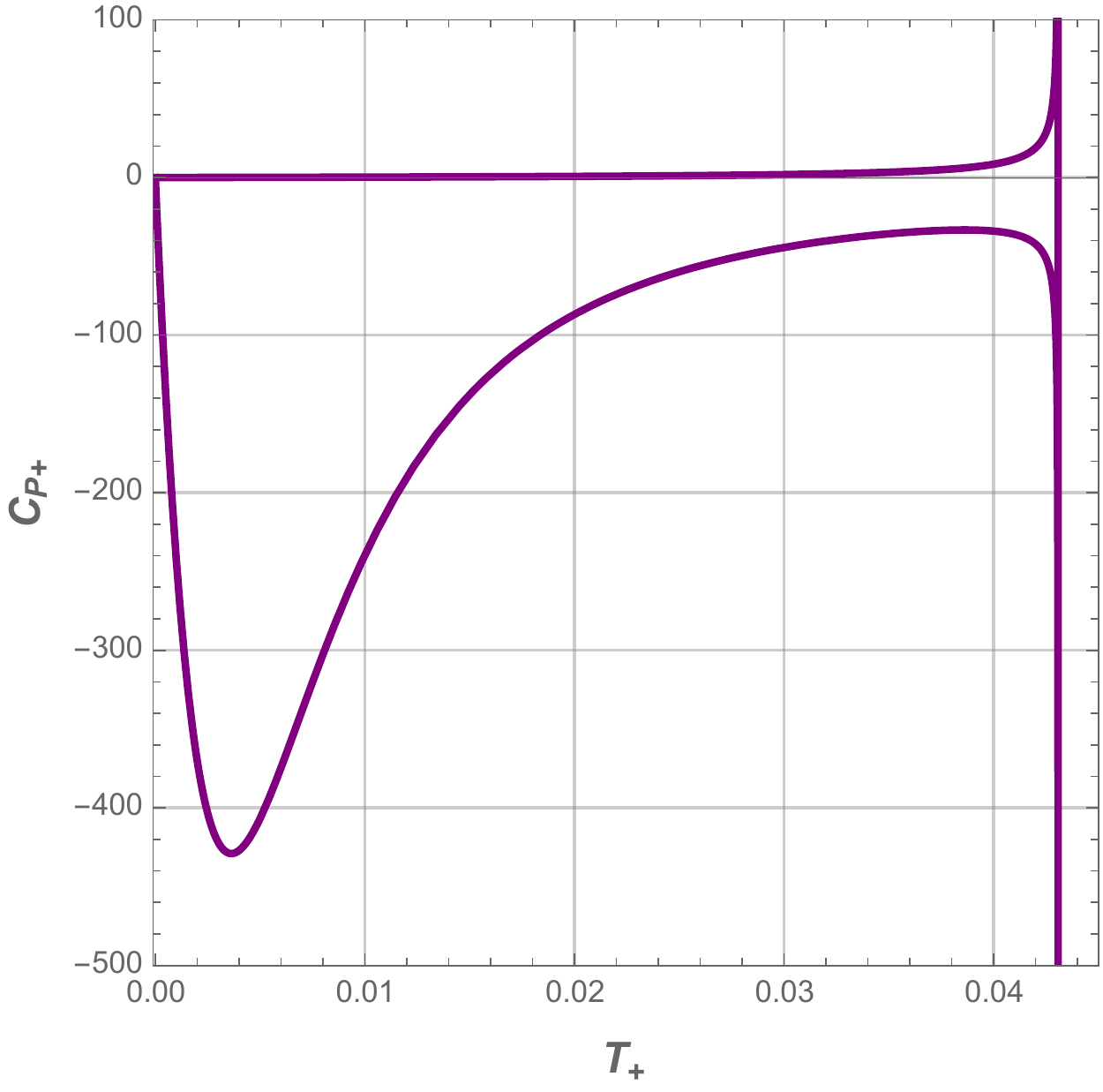}
		\caption{\footnotesize $P = -0.0001$.}
		\label{f91_4}
		
	\end{subfigure}

	\caption{\footnotesize Event horizon heat capacity $C_{P+}$ as a function of event temperature $T_+$ with the electric charge $Q=1$ .}
	\label{f91}
\end{figure}

 Next we display in Fig.~\ref{f9} and Fig.~\ref{f91} the heat capacity $C_{P+}$ as a function of the reduced horizon $\rho$ and temperature respectively, for different values of the pressure $P$. 

In all panels of the above figures, we easily deduce that the heat capacity $C_{P+}$ presents a discontinuity at the local maximum $T_{max}$. This singularity involves a thermal phase transition between a stable small black hole and unstable large black one at this critical point. We also note that the system is dominated by the unstable phase when the pressure $P$ decreases.

\begin{figure}[ht!]
	\centering
	\includegraphics[scale=0.7]{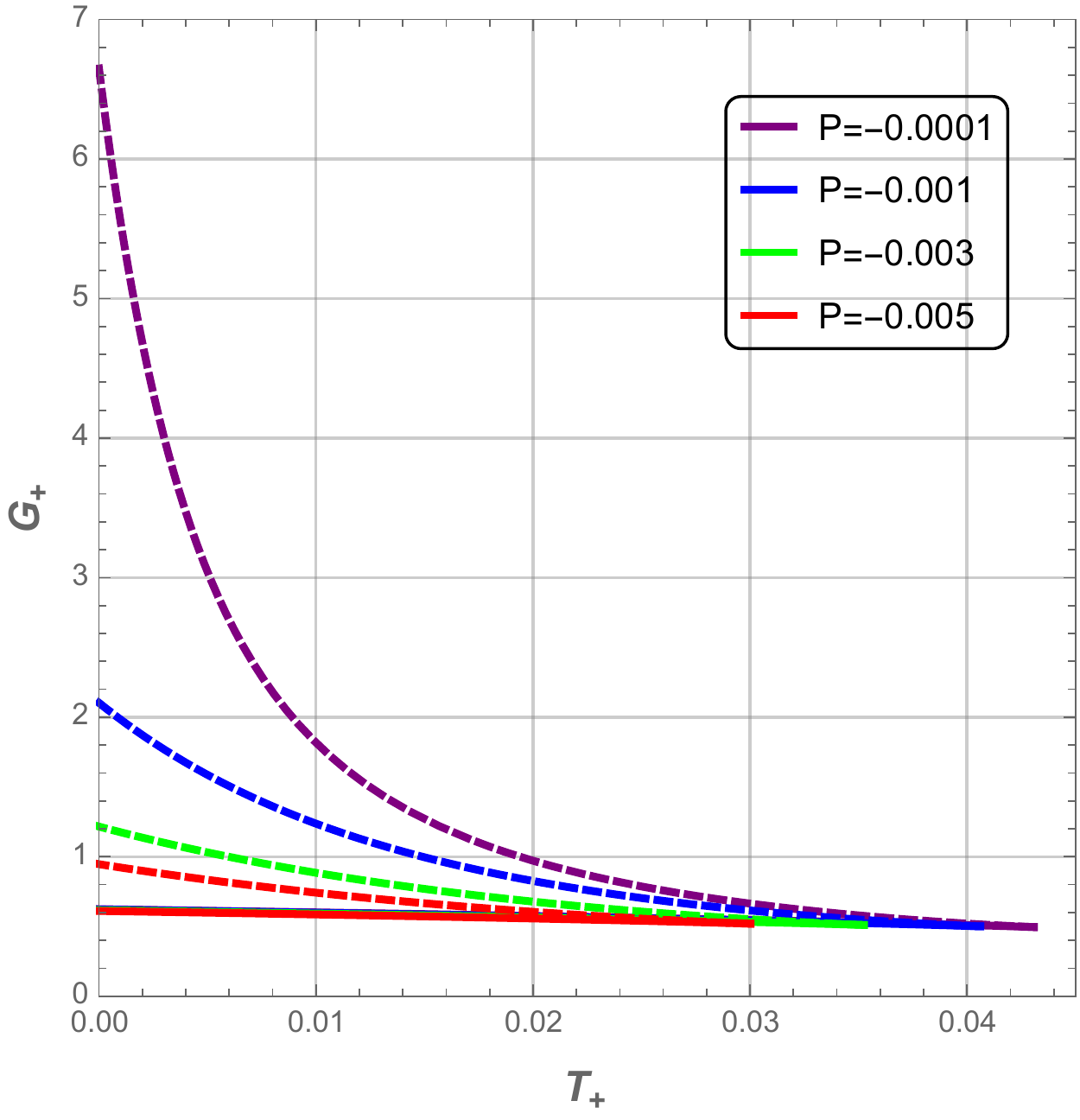}
	\caption{\footnotesize Free Gibbs energy $G_+$ as function of the black hole temperature $T_+$, with different values of the thermodynamic pressure $P$, and the electric charge $Q=1$.}
	\label{f10}
	
\end{figure}

\begin{figure}[ht!]
	\centering
	\begin{subfigure}[h]{0.45\textwidth}
		\centering \includegraphics[scale=0.5]{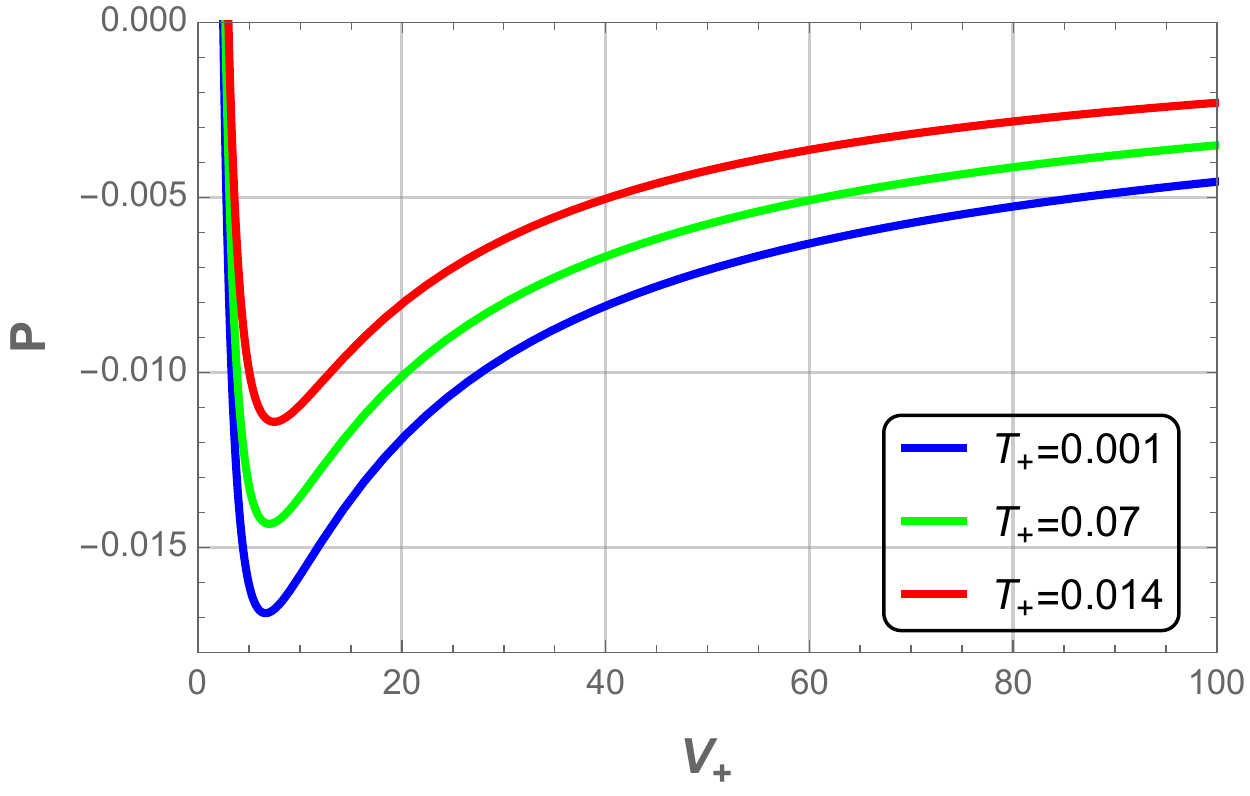}
		\caption{\footnotesize Event horizon.}
		\label{f11_1}
	\end{subfigure}
	\hspace{1pt}	
	\begin{subfigure}[h]{0.45\textwidth}
		\centering \includegraphics[scale=0.5]{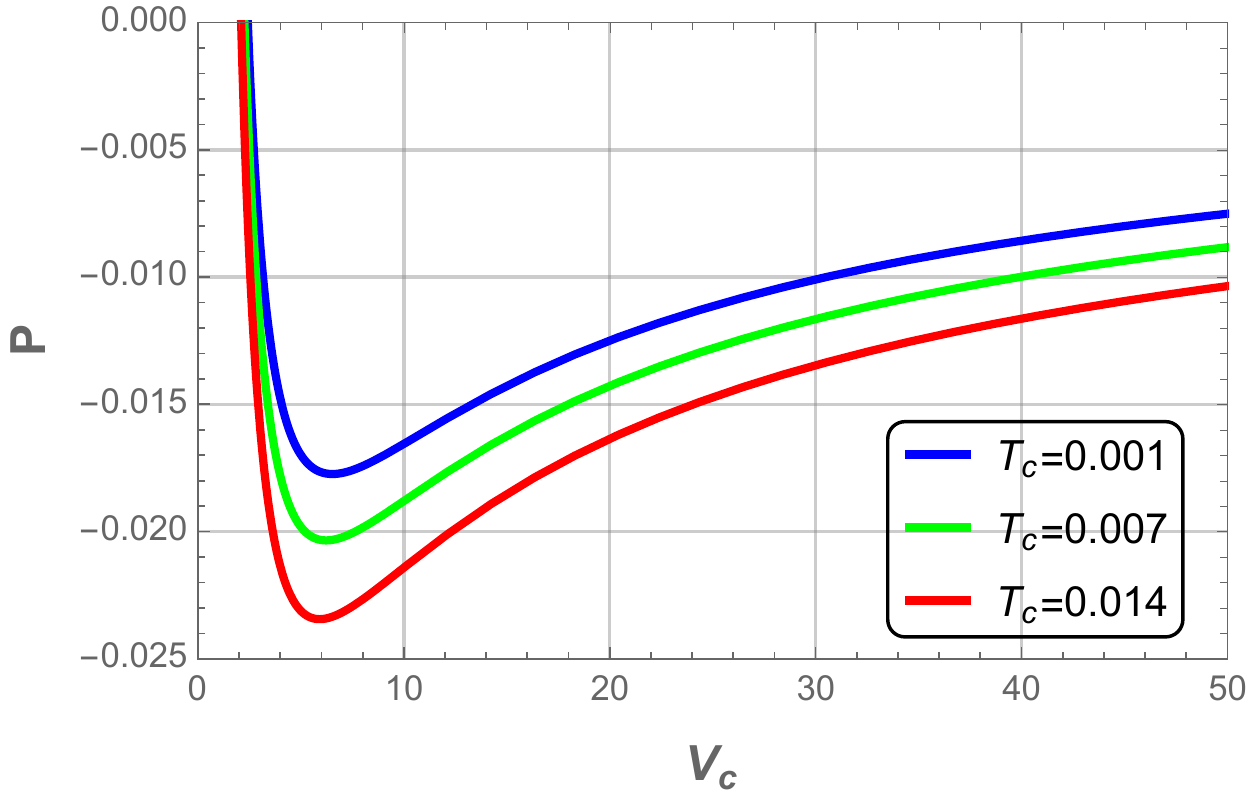}
		\caption{\footnotesize Cosmological horizon.}
		\label{f11_2}
	\end{subfigure}	
	
	\caption{\footnotesize Thermodynamic pressure $P$ as a function of the thermodynamics volume $V$, with different values of the temperature, and the electric charge $Q=1$.}
	\label{f11}
\end{figure}

In Fig.~\ref{f10}, the characteristic swallowtail behavior of the Gibbs free energy $G_+ = M - T_+ S_+$ shows that first-order phase transition occurs between large and small black holes. It is worth to notice that both branches meet at the cusp: The lower branch corresponds to small black hole stable phase, while the upper one  to the large black hole unstable phase.
With respect to the cosmological horizon, the heat capacity $C_{Pc}$ is always negative and thus the cosmological horizon becomes larger.  Next step, we focus on the study of the equation of state. To this end, we illustrate in Fig.~\ref{f11} the variation of the pressure $P$ as a function of the horizon radius $\rho$ for fixed charge $Q$.  
From the $P-\rho$ diagram we see that the event and  cosmological horizons have the same shape. However, by comparison to the event horizon, the larger temperatures cosmological horizon correspond to smaller pressures.

After presenting the local thermodynamics context, where each horizon is treated separately,  we will focus in the next section on the global thermodynamics view where the horizons are simultaneously analyzed.
\newpage
\section{Global black hole thermodynamics view }

To study all the horizons simultaneously, we note that
\begin{align} \label{43}
G_c\left( \rho_c, P, Q\right) &= - G\left( \rho_+ \to \rho_c, P, Q\right) ,\\
G_-\left( \rho_-, P, Q\right) &= - G\left( \rho_+ \to \rho_-, P, Q\right) ,
\end{align}
This means that the information about the thermodynamics of all horizons is in fact encoded in $G_+$, provided we extend its validity to “all admissible radii” $\rho_+$, and hence to negative temperatures $T_+$. The latter correspond to positive temperature of the cosmological and inner horizons. Therefore, in these regions, the thermodynamic equilibrium occurs for the global maximum (rather than minimum) of $G_+$.

\begin{figure}[h!]
	\centering
	\includegraphics[scale=0.7]{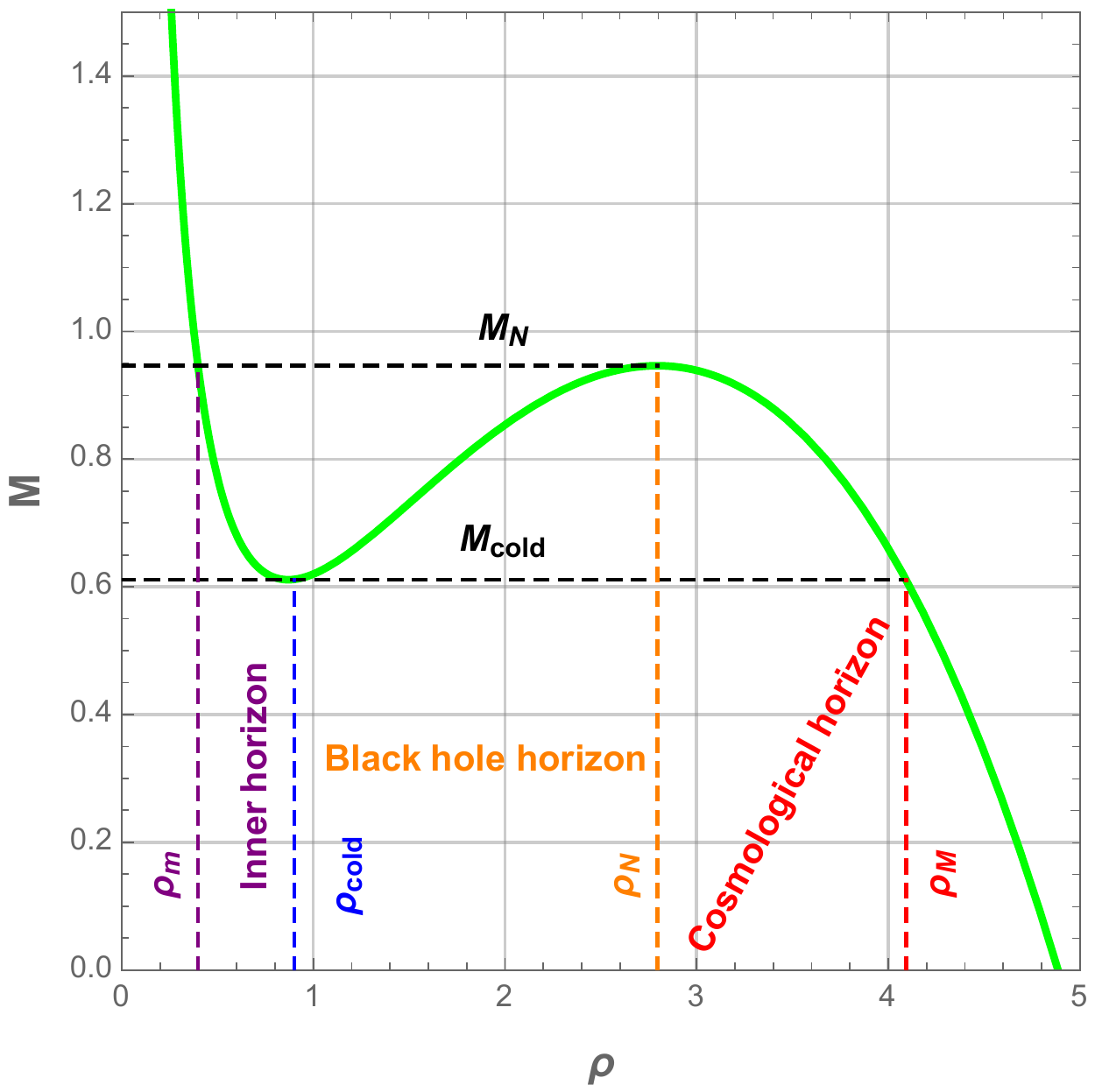}
	\caption{ \footnotesize Black hole mass $M$ as a function of $\rho$ with the electric charge $Q=1$ and the pressure $P = -0.005$.}
	\label{f12}
	
\end{figure}

Now, for a fixed pressure $P$ and charge $Q$, we introduce the radius of  cold black hole $\rho_{cold}$  and the radius associated with  Nariai black hole $\rho_N$.  We also denote by $\rho_m$ the minimal radius of the inner horizon, which occurs for the Nariai limit, and  $\rho_M$ the maximal radius of the cosmological horizon, happening at the extremal black hole . This is illustrated by Fig.~\ref{f12} where the black hole mass variation is plotted as a function of the horizon radius.

\begin{figure}[h!]
	\centering
	\includegraphics[scale=0.6]{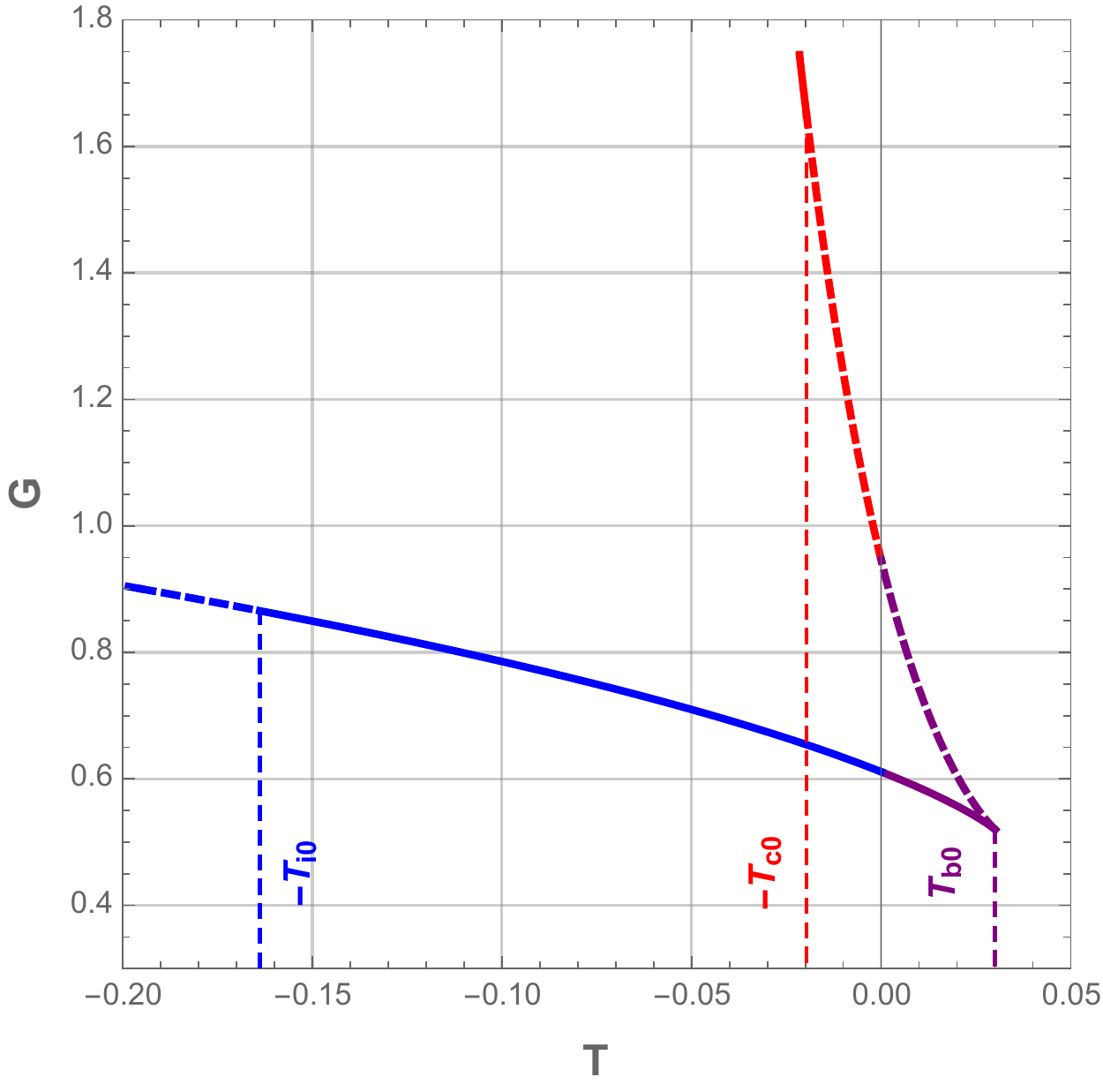}
	\caption{\footnotesize Black hole Gibbs free energy as a function of temperature of all horizons with the electric charge $Q=1$ and the pressure $P = -0.005$ : Inner horizon (blue), black hole horizon (purple) and cosmological horizon (Red). Dashed curves correspond to unstable large black hole phase.}
	\label{f13}
	
\end{figure}

In Fig.~\ref{f13}, we plot the Gibbs free energy function $G = G(T)$ for the range $ \rho \in [\rho_m, \rho_M]$, with  an interpretation of each thermodynamic quantities 
  for different ranges of $\rho$. 
 \begin{itemize} 
\item  In the range of radius $ \rho \in [r_m, \rho_{cold}]$, the $ G = - G_-$ and $T = - T_-$ are the inner horizon quantities and the global maximum of $G$ corresponds to a thermodynamic preferred state. 
  
\item  The domain $ \rho \in [\rho_{cold}, \rho_N]$, the $ G = G_+$ and $T = T_+ $ denote black hole horizon quantities while $G$ is minimized by the thermodynamic preferred state. 
  
\item  Finally, the interval  $ \rho \in [\rho_N, \rho_M]$,the $G = - G_c$ and $T = - T_c$ are seen  as cosmological horizon quantities and the preferred state again corresponds to the maximum of $G$. As is illustrated in Fig.~\ref{f13}.
 \end{itemize}

\begin{figure}[h!]
	\centering
	\includegraphics[scale=0.6]{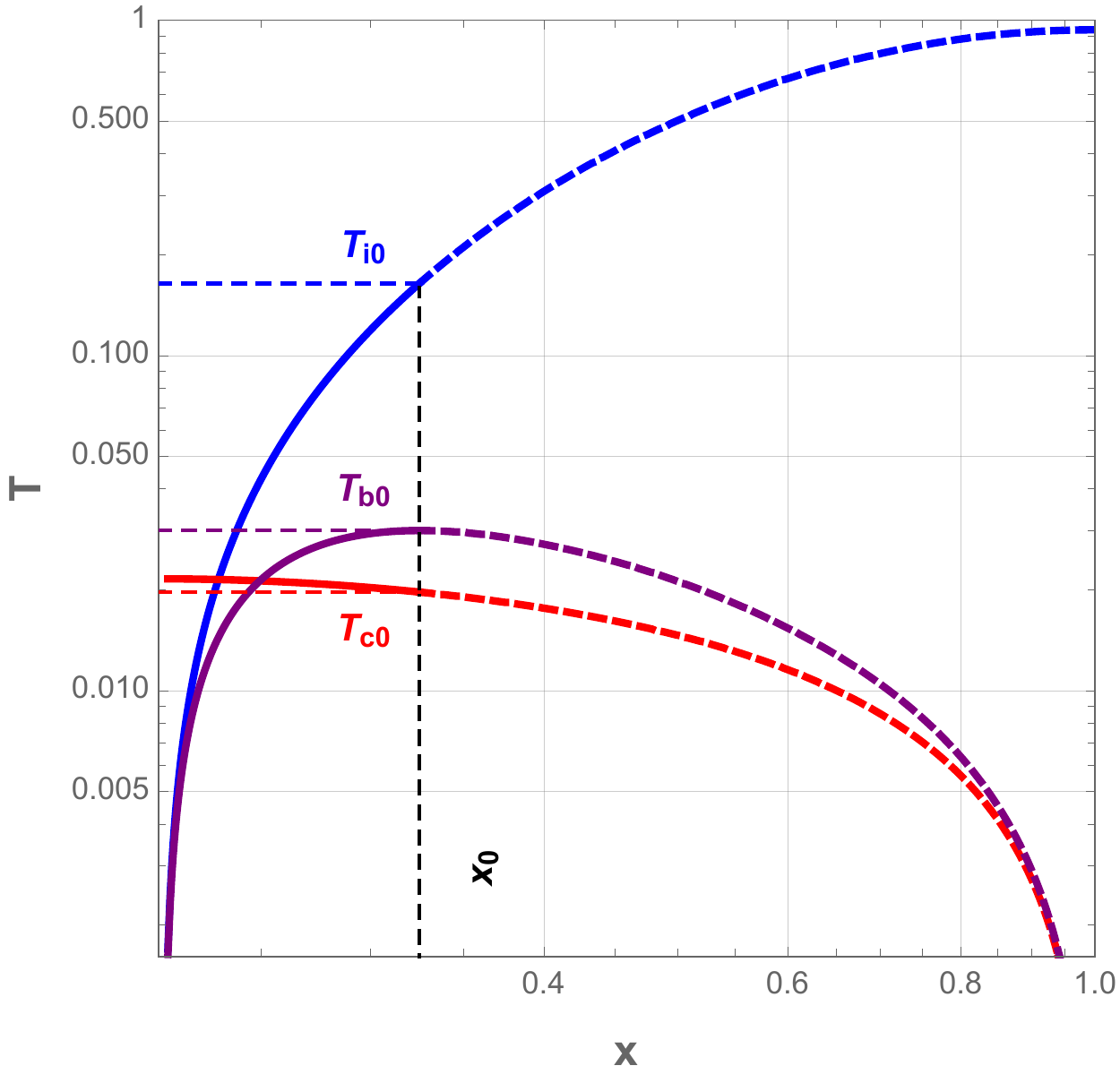}
	\caption{\footnotesize Black hole Temperatures as a function of temperature of all horizons with the electric charge $Q=1$ and the pressure $P = -0.005$ : Inner horizon (blue), black hole horizon (purple) and cosmological horizon (Red). Dashed curves correspond to unstable large black hole phase.}
	\label{f14}	
\end{figure}

Having defined the Gibbs free energy $G$ that describes the thermodynamic behavior of each involved horizons, we will focus on the stability and the phase structure of such a black hole. 
In Fig.~\ref{f14}, we illustrate the temperatures of all horizons  in terms of the ratio $x = \rho_+/\rho_c$ taking into account that 
 only admissible values of $x$ which range between extremal and Nariai' bounds, $ x \in [x_m=\frac{\rho_{cold}}{\rho_N}, 1]$.

From the previous plots, we directly observe how the temperatures behave when $\rho_+$ $(x)$ increases. We  also see from Fig.~\ref{f13} that for small variations of $T$ (till the cusp), the black hole keeps moving along a lower branch with low Gibbs free energy, and hence is in a  preferred thermal phase. However, by crossing the cusp, the black hole  migrates to  the upper black hole branch which is unstable. Therefore, we conclude that the instability regions exist not only for the black hole horizon but also for the cosmological and inner horizon branches.

This explains how the thermodynamics of each horizon interplay towards the understanding of the thermodynamic behavior of the whole system.
\section{Effective black hole thermodynamics of two horizons}\label{twohorizon}

The effective black hole thermodynamics of dS black hole has attracted increasing interest in literature \cite{Urano:2009xn,Ma:2013aqa,Zhao:2014raa,Zhang:2014jfa,Ma:2014hna,Guo:2015waa,Guo:2016eie,Bhattacharya:2015mja,Kubiznak:2016qmn}, thus, it is justified to investigate such a black hole where the main ingredient for such a formalism is to focus on an observer located  between the black hole horizon and the cosmological one,  and attribute to the system an “effective temperature”. 

If the event and cosmological horizons are not located far away, one cannot analyze the thermodynamics and thermal phase transition in an independent way. Since the temperatures on the event and cosmological horizons are different, the black hole generally cannot be in thermodynamic equilibrium, except for the degenerate case $ \rho_+ =  \rho_c$

In the effective thermodynamics framework, the thermodynamic first law and  Smarr-like formula  \cite{Kubiznak:2016qmn} read as:
\begin{equation}\label{44}
dM = T_{eff} dS_{eff}+V_{eff} dP + \Phi_{q_{eff}} dQ + \Phi_{\alpha_{eff}} d\alpha,
\end{equation}
\begin{equation}\label{45}
M = -2 (T_{eff} S_{eff}+V_{eff} P) + \Phi_{q_{eff}} Q - \Phi_{\alpha_{eff}} \alpha.
\end{equation}

then, by using Eq.~\eqref{25} and Eq.~\eqref{40}, they can be written as,
\begin{equation}\label{47}
\begin{split}
dM =& -\dfrac{T_+T_c}{T_++T_c} d(S_c-S_+)+ \dfrac{V_+T_c + V_c T_+}{T_++T_c}dP
+\dfrac{\Phi_{q_+}T_c + \Phi_{q_c} T_+}{T_++T_c}dQ +\dfrac{\Phi_{\alpha_+}T_c + \Phi_{\alpha_c} T_+}{T_++T_c}d\alpha  ,
\end{split}
\end{equation}

\begin{equation}\label{48}
\begin{split}
M =& -2\left(  \dfrac{T_+T_c}{T_++T_c} (S_c-S_+)+ \dfrac{V_+T_c + V_c T_+}{T_++T_c}P\right) 
+\dfrac{\Phi_{q_+}T_c + \Phi_{q_c} T_+}{T_++T_c}Q -\dfrac{\Phi_{\alpha_+}T_c + \Phi_{\alpha_c} T_+}{T_++T_c}\alpha  .
\end{split}
\end{equation}

The effective thermodynamic variables are given by,
\begin{equation}\label{49}
\begin{split}
&T_{eff} = \dfrac{T_+T_c}{T_++T_c}, \hspace{5mm} V_{eff} = \dfrac{V_+T_c + V_c T_+}{T_++T_c},\\
 &\Phi_{q_{eff}} = \dfrac{\Phi_{q_+}T_c + \Phi_{q_c} T_+}{T_++T_c}, \hspace{5mm}  \Phi_{\alpha_{eff}} = \dfrac{\Phi_{\alpha_+}T_c + \Phi_{\alpha_c} T_+}{T_++T_c}
\end{split}
\end{equation}

while the effective entropy is defined as in \cite{Bhattacharya:2015mja,Kubiznak:2016qmn},
\begin{equation}\label{46}
	S_{eff} = S_c - S_+
\end{equation}

From Fig.~\eqref{40} it is worth to notice that the pressure can be expressed in terms of the horizon radii, especially if one considers black hole thermodynamics with a fixed cosmological horizon. This scenario  will be examined thereafter.

\begin{figure}[h!]
	\centering
	\includegraphics[scale=0.6]{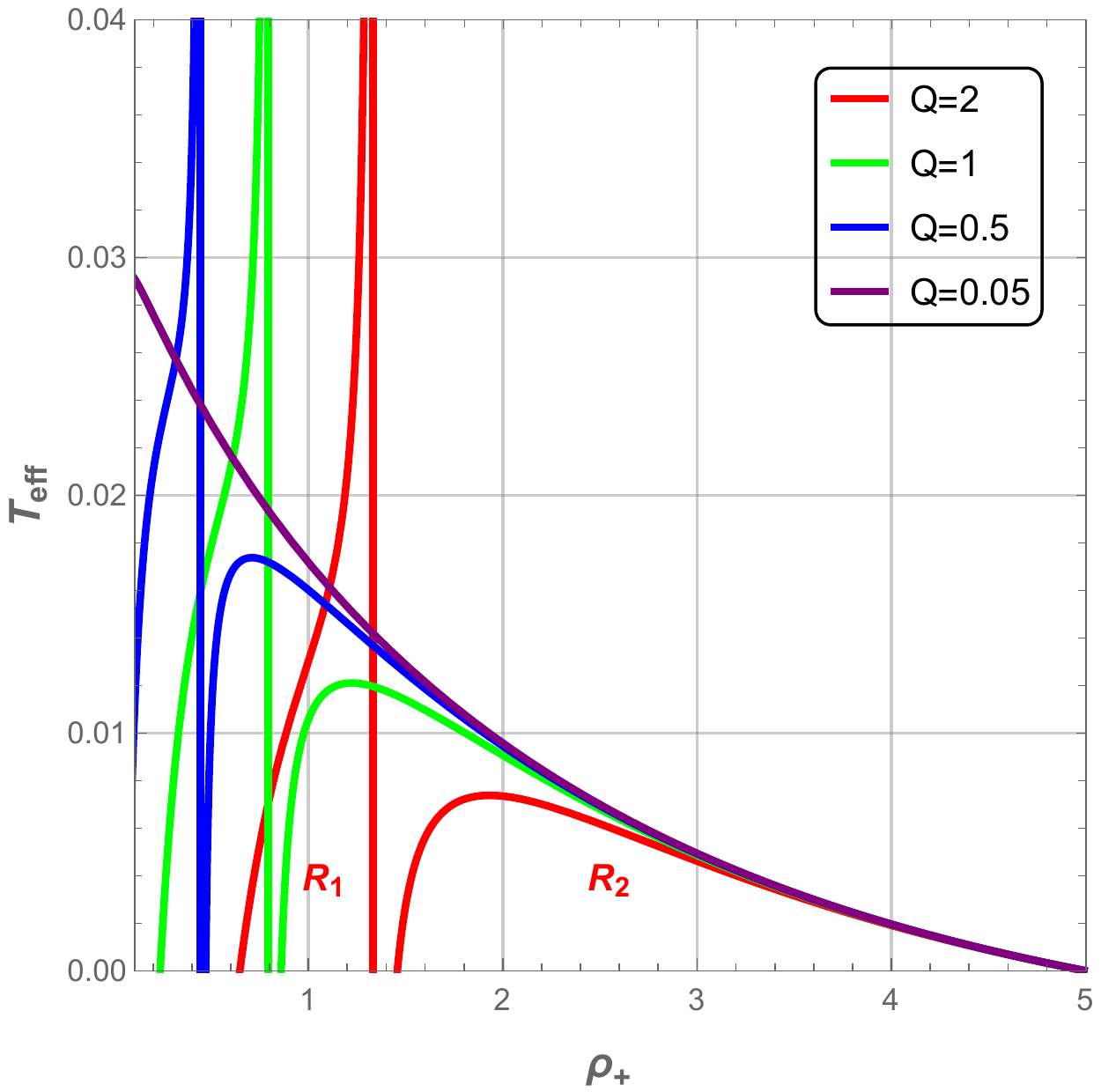}
	\caption{\footnotesize Effective temperature as a function of the event horizon radius $\rho_+$ with different values of the electric charge $Q$ and a fixed cosmological horizon radius $\rho_c=5$.}
	\label{f16}
\end{figure}

In Fig.~\ref{f16}, we plot the behavior of the effective temperature $T_{eff}$ as a function of the horizon radius $\rho_+$ for different values of electric charge $Q$.
For large $Q$, there exist two regions, $R_1$ and $R_2$ which are separated by negative temperatures forbidden region $\rho_+ \in \left[ \rho_{div}, \rho_0\right] $ , where $\rho_{div}$ and $\rho_0$ denote a  divergent point and a zero-temperature point of $T_{eff}$ respectively. In the region $R_1$, $T_{eff}$ is an increasingly monotonous function of $\rho_+$ with an inflection point, while in the second region $R_2$,  $T_{eff}$  first increases until a maximum value, then decreases when $\rho_+$ becomes large. Due to the existence of forbidden region, the black hole freezes in one of these two regions without possible transition to the other one.

In the next step in our analysis, we evaluate  the effective heat capacity at constant pressure,
$C_{P_{eff}}$, and reveal its singularities with the aim to probe its thermal phase structure. The effective heat capacity $C_{P_{eff}}$ can be defined as:
\begin{equation}\label{50}
C_{P_{eff}} = \left. \dfrac{\partial M}{\partial T_{eff} }\right) _P = \left( \dfrac{\partial M}{\partial \rho_{+} } - \dfrac{\partial M}{\partial \rho_c } \dfrac{\frac{\partial P}{\partial \rho_+}}{\frac{\partial P}{\partial \rho_c}}\right)  \left( \dfrac{\partial T_{eff}}{\partial \rho_{+} } - \dfrac{\partial T_{eff}}{\partial \rho_c } \dfrac{\frac{\partial P}{\partial\rho_+}}{\frac{\partial P}{\partial \rho_c}}\right)^{-1} .
\end{equation}

\begin{figure}[h!]
	\centering
	\begin{subfigure}[h]{0.45\textwidth}
		\centering \includegraphics[scale=0.5]{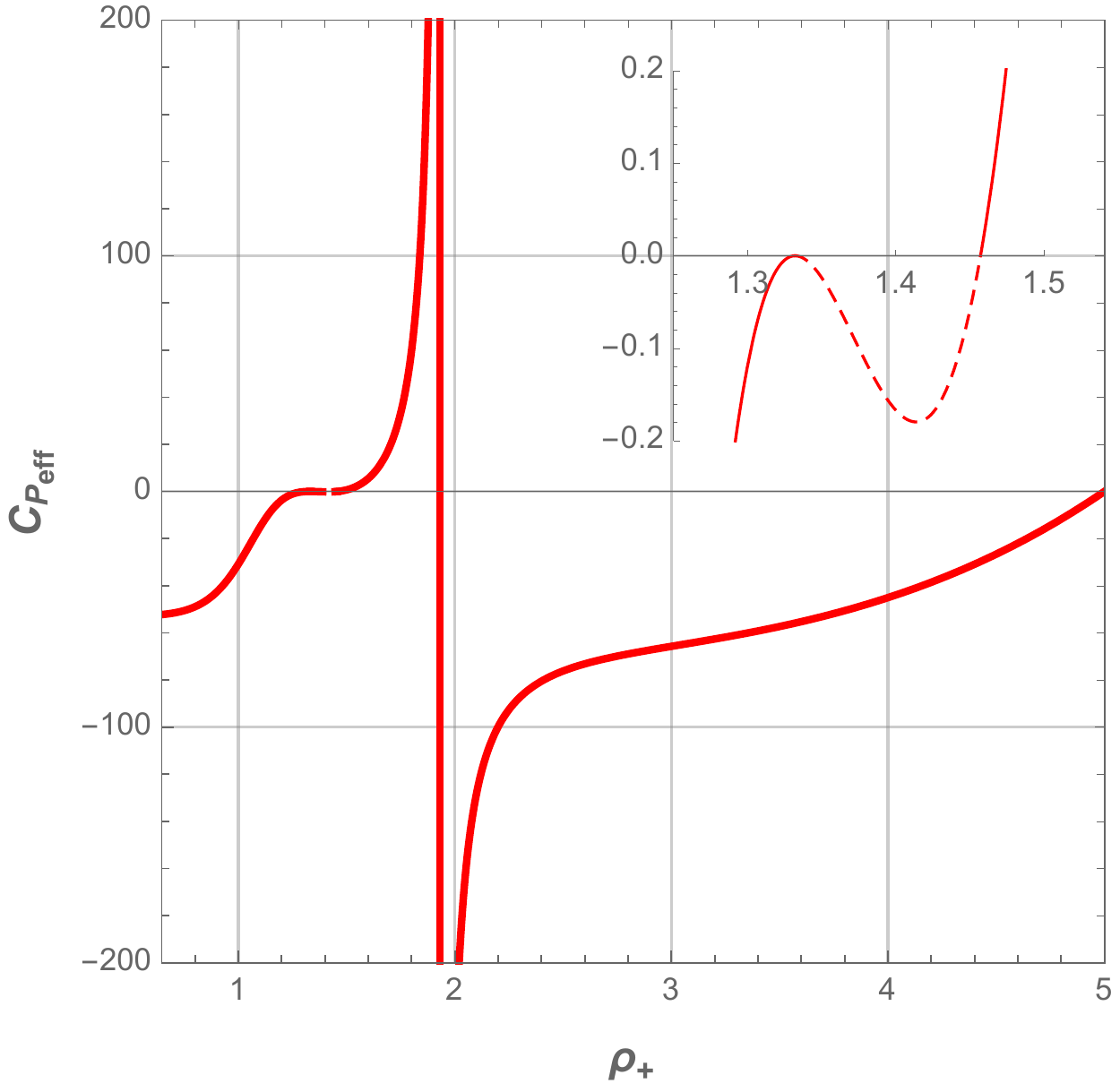}
		\caption{\footnotesize$Q=2$ }
		\label{f17_1}
	\end{subfigure}
	\hspace{1pt}
	\begin{subfigure}[h]{0.45\textwidth}
		\centering \includegraphics[scale=0.5]{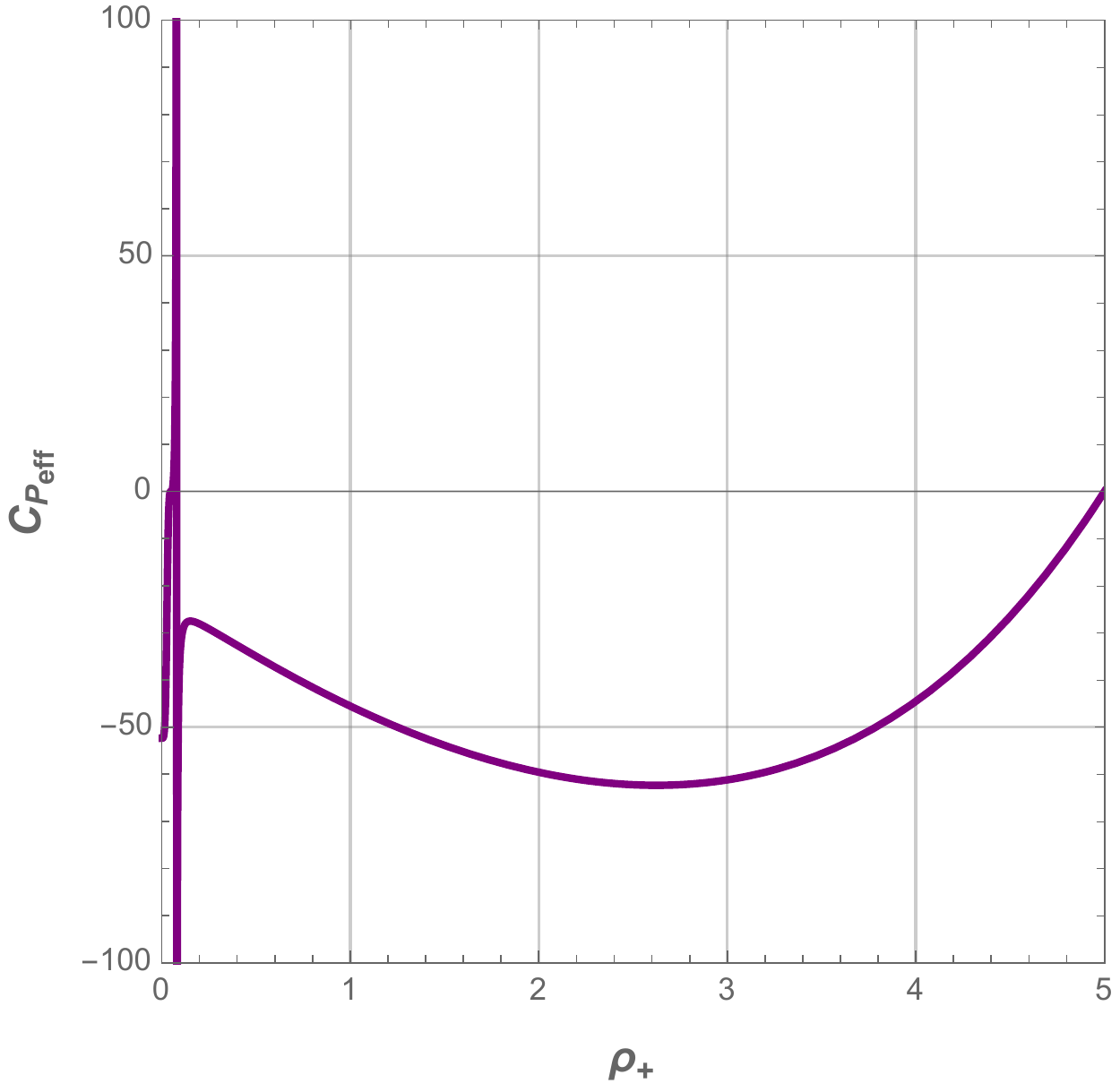}
		\caption{ \footnotesize$Q=0.05$}
		\label{f17_2}
	\end{subfigure}
	\vspace{1pt}
	\caption{\footnotesize Effective heat capacity $C_{P_{eff}}$ as a function of event  horizon radius $\rho_+$, with different values of the electric charge, and  $\rho_c = 5$.}
	\label{f17}
\end{figure}

In  Fig.~\ref{f17}, we plot the heat capacity as a function of the horizon radius $\rho_+$ with different values of the electric charge $Q$. 

 From this figure, one can easily deduce the thermodynamic stability of such a system and derive its thermal phase transition. For large electric charge $Q$, if the black hole resides in the first region $R_1$, the heat capacity $C_{P_{eff}}$ is negative, hence the black hole is  unstable and no thermal phase transition occurs. Alternatively, if the black hole subsists in the region $R_2$, the heat capacity $C_{P_{eff}}$ presents a singularity that can be interpreted as a thermal phase transition between an unstable large black hole ($C_{P_{eff}} < 0$ ) and a  stable small black hole ($C_{P_{eff}} > 0$ ).

However, for the small electric charge scenario, the first region $R_1$ should disappear,  thus the black hole remains in the second region $R_2$. This means that the black hole undergoes a thermal phase transition between an unstable and stable phases. In this regime, when $Q$ decreases below a certain small value, the black hole is again unstable with no possible thermal phase transition, a situation similar to that where the black hole resides in  $R_1$.

\begin{figure}[h!]
	\centering
	\begin{subfigure}[h]{0.45\textwidth}
		\centering \includegraphics[scale=0.5]{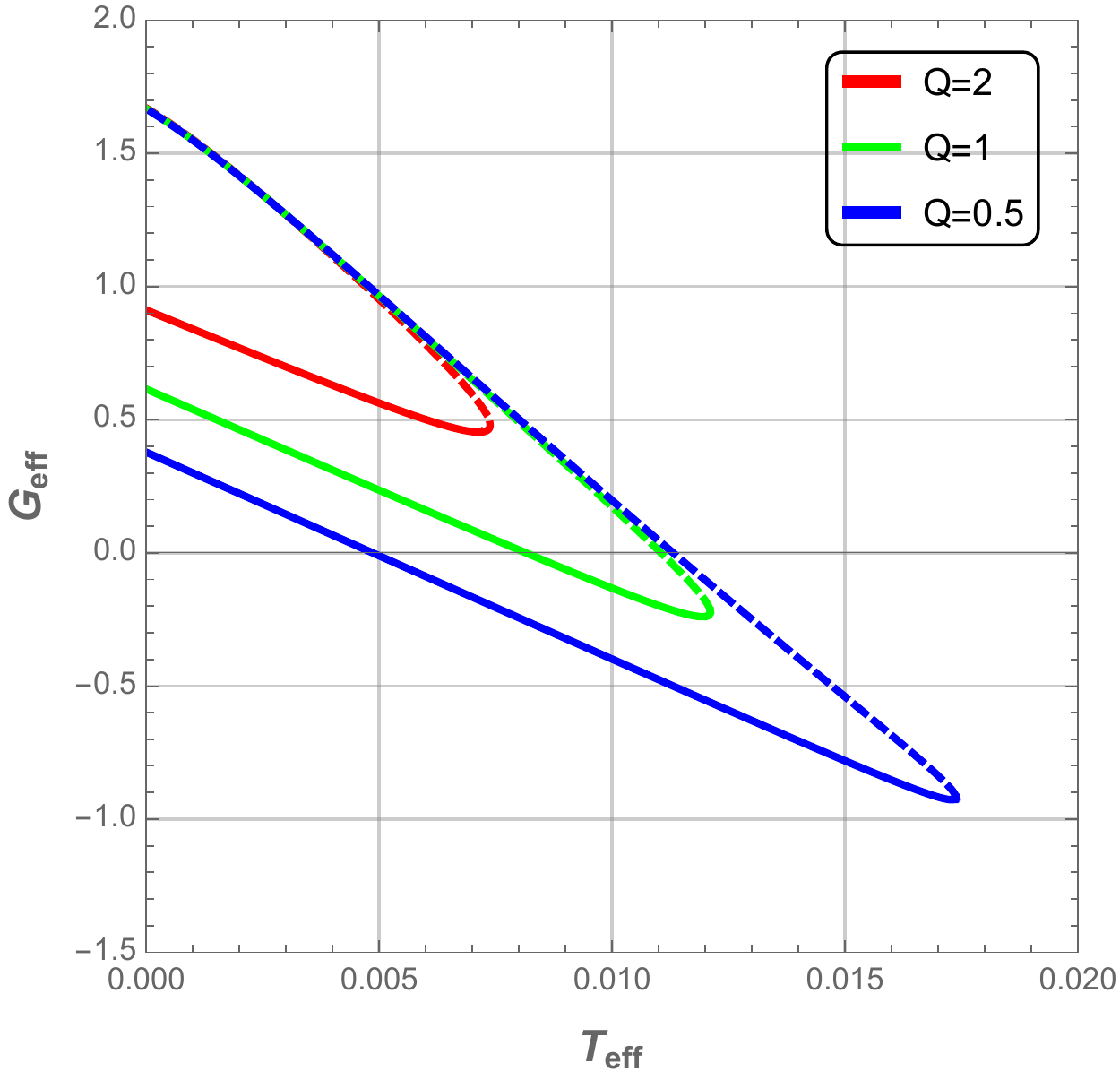}
		\caption{\footnotesize Region $R_2$  }
		\label{f18_1}
	\end{subfigure}
	\hspace{1pt}
	\begin{subfigure}[h]{0.45\textwidth}
		\centering \includegraphics[scale=0.5]{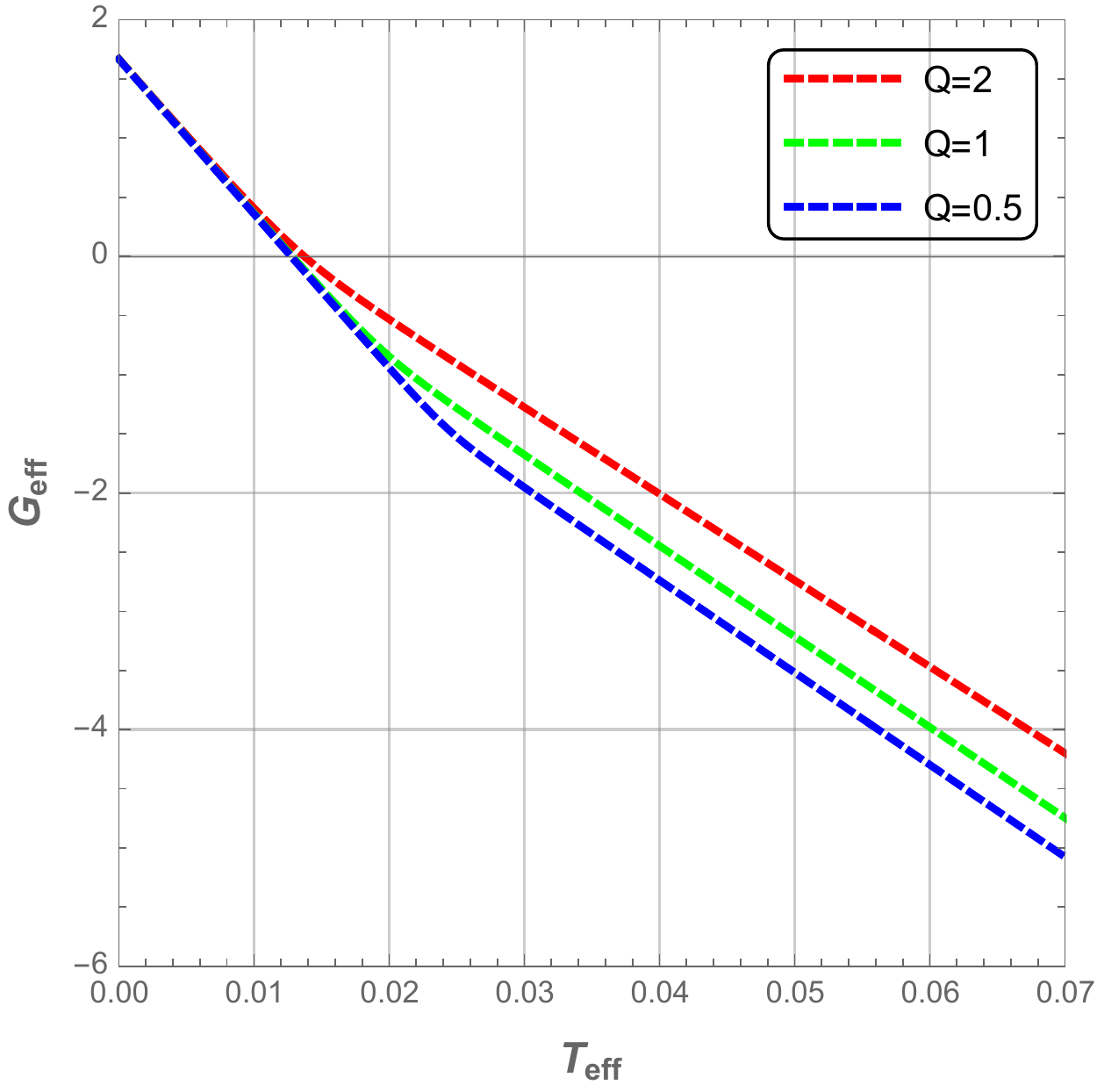}
		\caption{ \footnotesize Region $R_1$}
		\label{f18_2}
	\end{subfigure}
	\caption{\footnotesize Effective free Gibbs energy $G_{eff}$ as a function of the effective temperature $T_{eff}$, with different values of the electric charge $Q$, with $\rho_c = 5$.}
	\label{f18}
\end{figure}

The previous suggested phase transitions can also be recovered through analysis of the Gibbs free energy  $G_{eff}$  given by, 
\begin{equation}\label{51}
G_{eff} = M - T_{eff}S_{eff},
\end{equation}
and plotted in Fig.~\ref{f18}.

From Fig.~\ref{f18}, one can clearly see that for the black hole residing in the region $R_1$, the effective free Gibbs energy $G_{eff}$  is a decreasingly monotonous linear function of the effective temperature $T_{eff}$ with the presence of a  point where $G_{eff}$ changes the slope. This change of slope can suggest a thermal phase transition at this point (See next section). Otherwise, for the black hole staying in the region $R_2$, $G_{eff}$  is a multi-valued function as indicated by the presence of the swallowtail figure. Thus a thermal phase transition between an unstable phase (large black hole \textit{i.e} dashed line) and stable phase (small black hole \textit{i.e} solid line) should occur. Here, the black hole with a larger electric charge has a smaller critical effective temperature.

\section{Strong-Weak electric interaction transition}

\begin{figure}[h!]
	\centering
	\begin{subfigure}[h]{0.45\textwidth}
		\centering \includegraphics[scale=0.5]{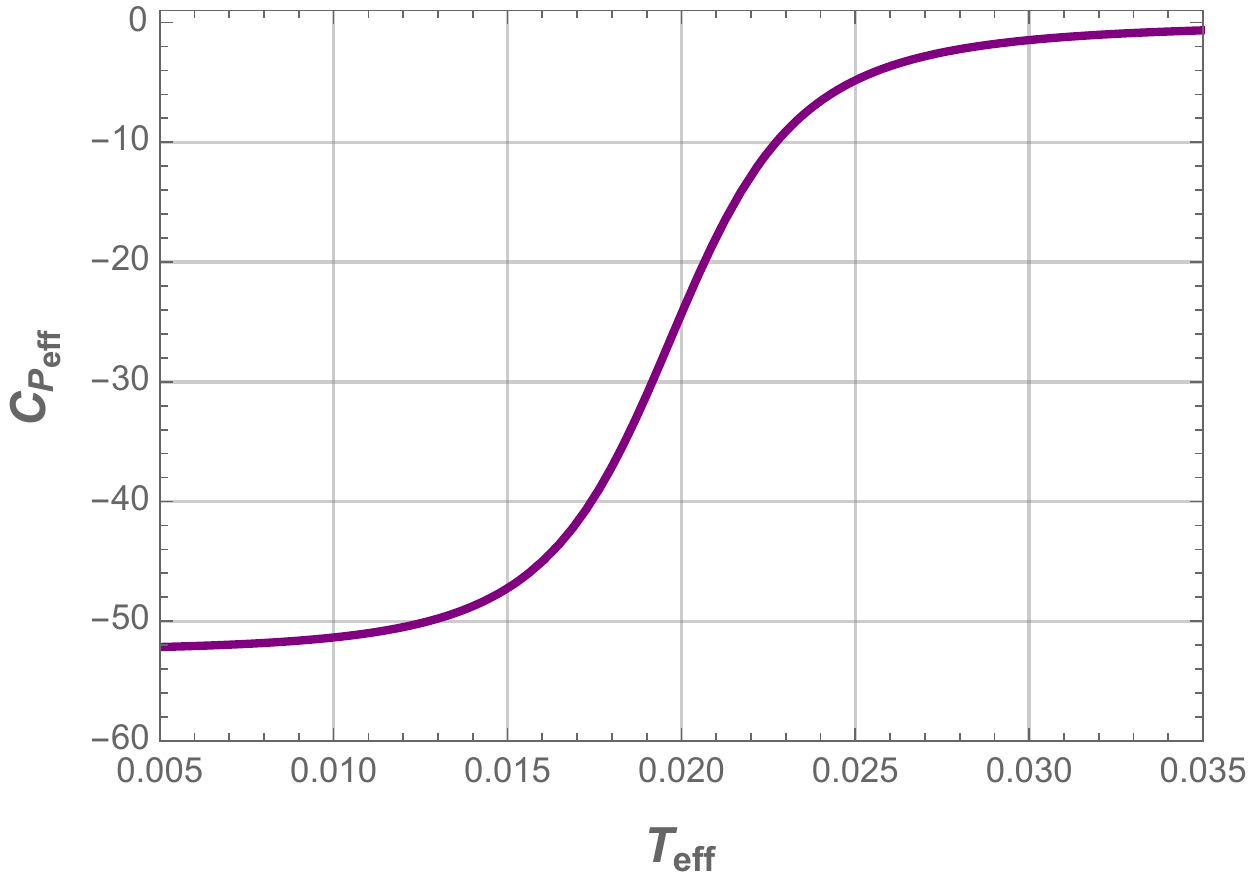}
		\caption{}
		\label{f20_1}
	\end{subfigure}
	\hspace{1pt}
	\begin{subfigure}[h]{0.45\textwidth}
		\centering \includegraphics[scale=0.54]{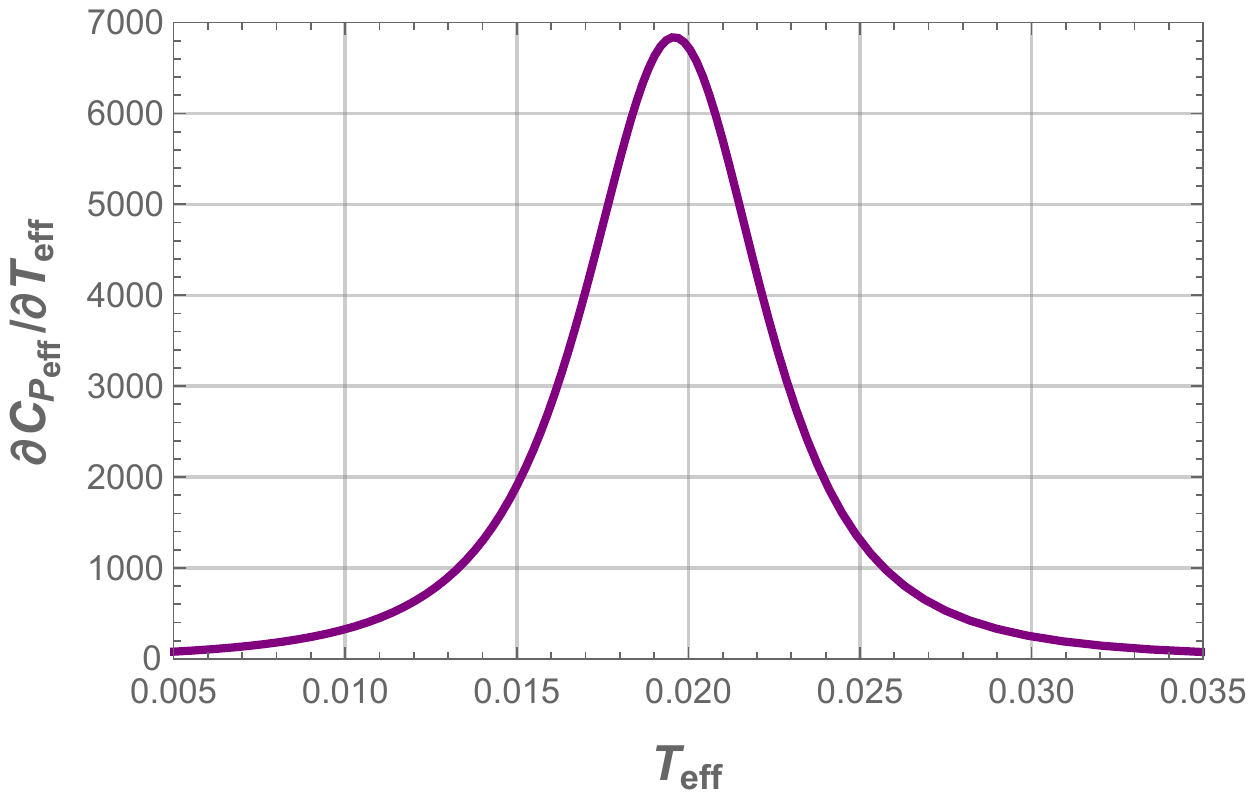}
		\caption{}
		\label{f20_2}
	\end{subfigure}
	\caption{\footnotesize  Effective heat capacity $C_{P_{eff}}$ (a) and its first partial derivative $\frac{\partial C_{P_{eff}}}{\partial T_{eff}}$ (b), in first region $R_1$, as a function of effective temperature $T_{eff}$  with  $Q=1$ and  $\rho_c = 5$.}
		\label{f20}
	\end{figure}

\begin{figure}[h!]
	\centering
	\begin{subfigure}[h]{0.45\textwidth}
		\centering \includegraphics[scale=0.5]{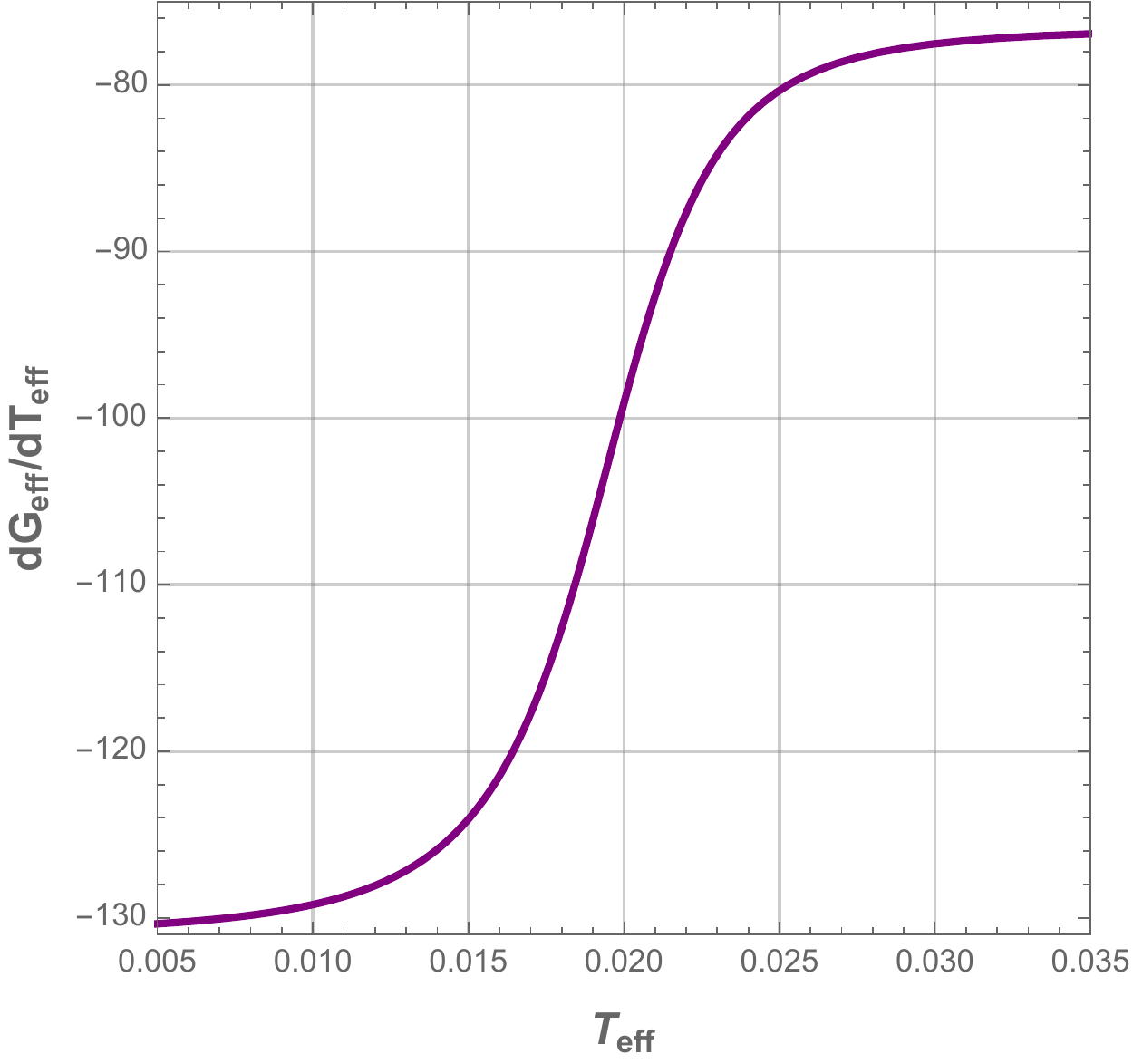}
		\caption{ }
		\label{f21_1}
	\end{subfigure}
	\hspace{1pt}
	\begin{subfigure}[h]{0.45\textwidth}
		\centering \includegraphics[scale=0.5]{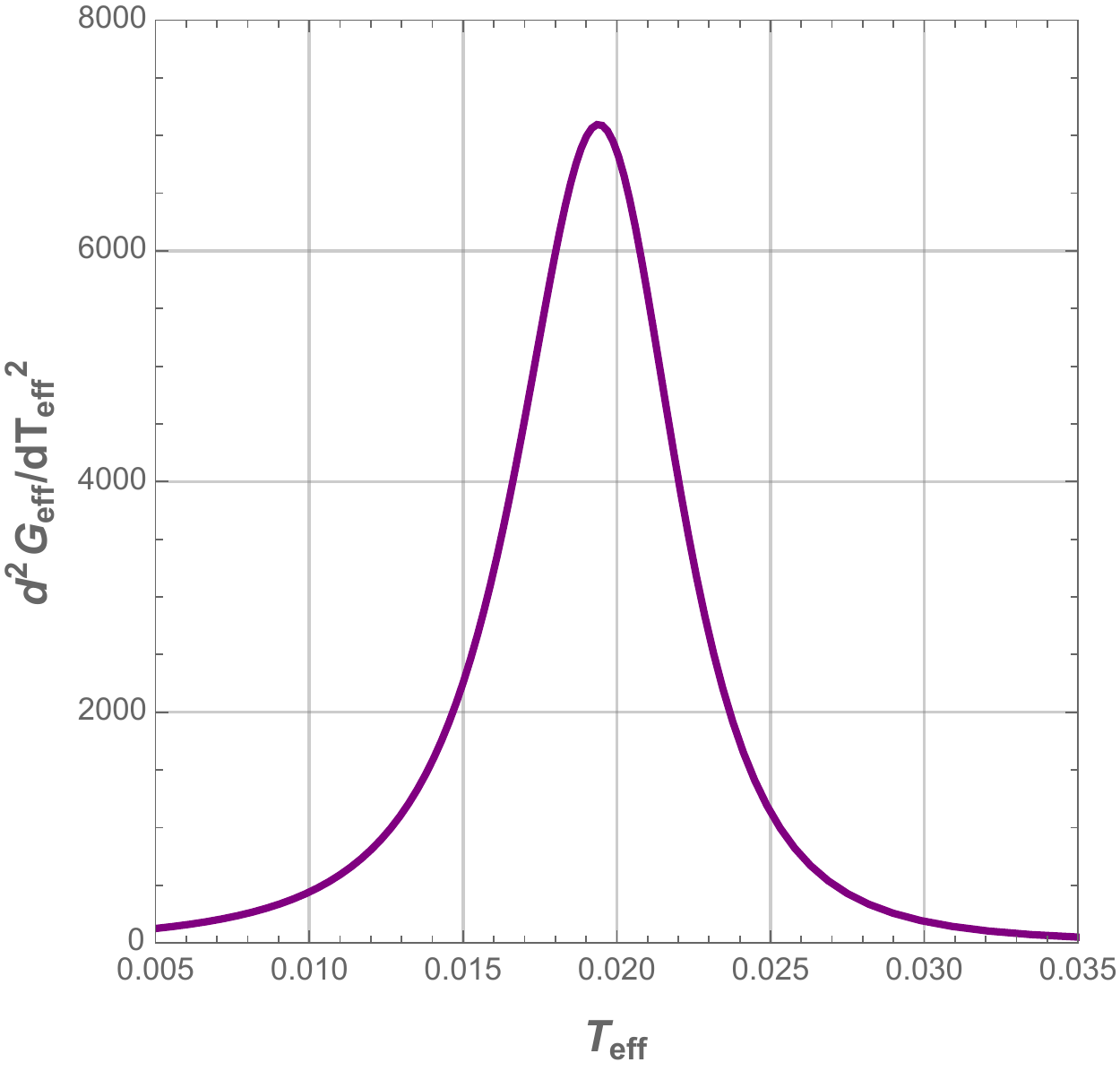}
		\caption{}
		\label{f21_2}
	\end{subfigure}
	
	\caption{\footnotesize First (a) and second (b) total derivative of effective free Gibbs energy,  $\frac{dG_{eff}}{d T_{eff}}$  and  $\frac{d^2G_{eff}}{d T_{eff}^2}$, in first region $R_1$, as a function of effective temperature $T_{eff}$, with  $Q=1$ and  $\rho_c = 5$.}
	\label{f21}
\end{figure}

In the previous section, we have seen that the effective free Gibbs energy in the first region $R_1$ changes its slope at a certain temperature which means that the derivative of $G_{eff}$ may probably show a discontinuity at this point. In order to look at this, we study first the effective heat capacity $C_{P_{eff}}$ and its first partial derivative which we plot in Fig.~\ref{f20}.
We see that the effective heat capacity suffers form a discontinuity which correspond to a divergent point in its partial derivative. 
If we look further at the first and second derivatives of the effective Gibbs free energy  $G_{eff}$,  we notice from Fig.~\ref{f21}, that $\frac{dG_{eff}}{d T_{eff}}$ (panel a) has a discontinuity, as the effective heat capacity, whereas $\frac{d^2G_{eff}}{d T_{eff}^2}$ (panel b) diverges at this point\footnote{Here we assume that very large values are assimilated to infinity.}.

\begin{figure}[h!]
	\begin{subfigure}[h]{0.45\textwidth}
	\centering \includegraphics[scale=0.5]{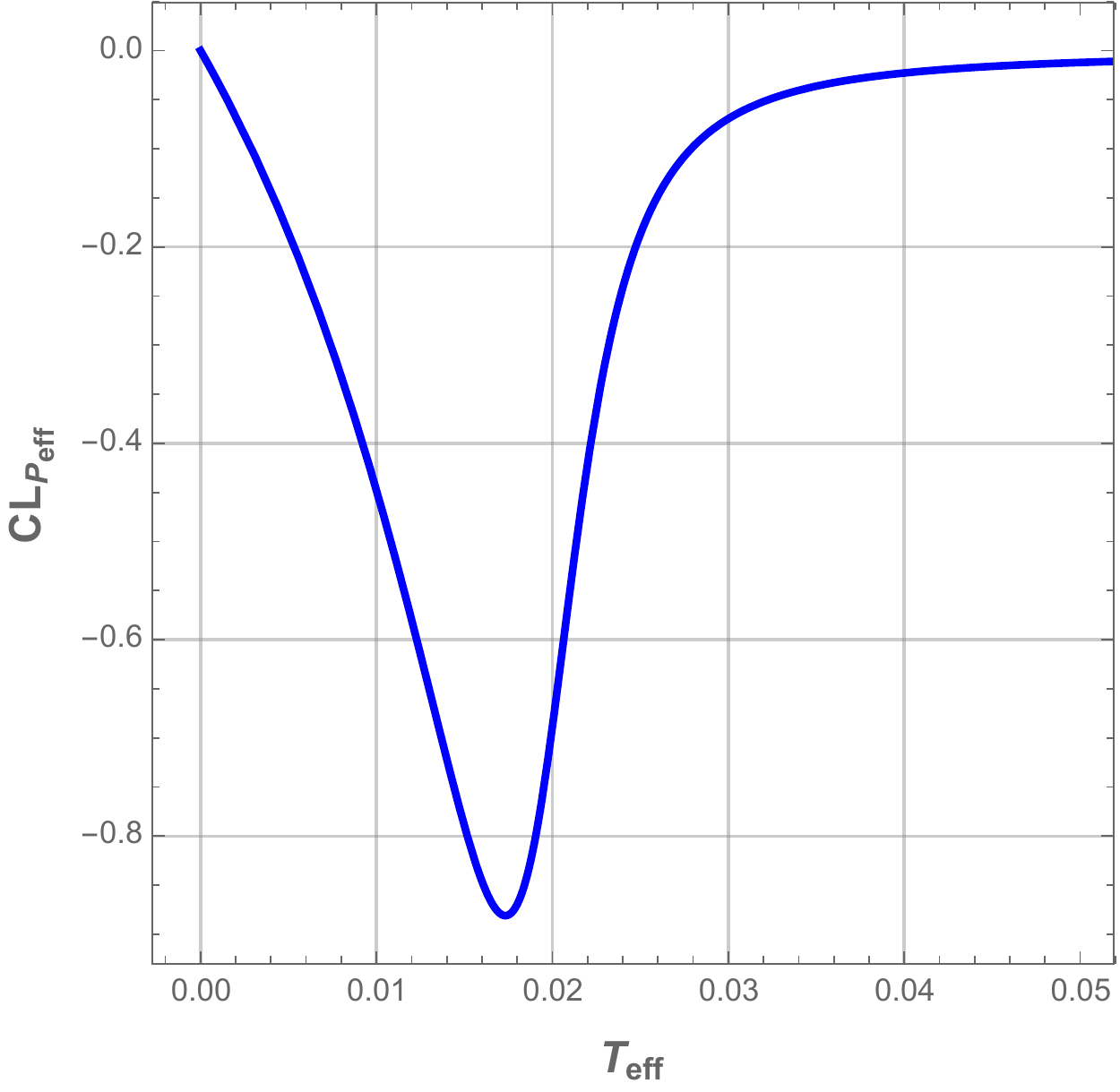}
	\caption{ }
	\label{f22_1}
\end{subfigure}
\hspace{1pt}
\begin{subfigure}[h]{0.45\textwidth}
	\centering \includegraphics[scale=0.5]{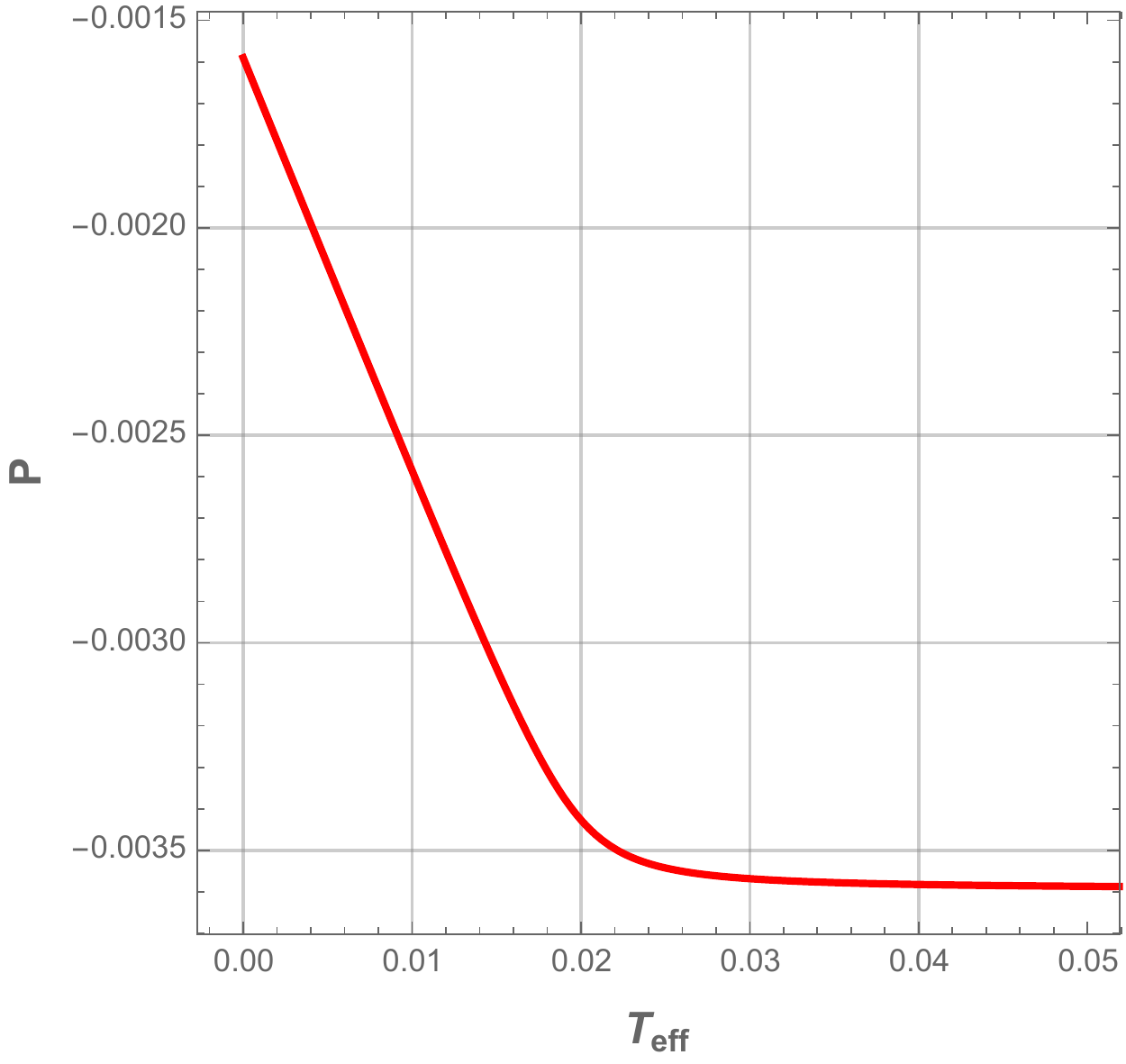}
	\caption{}
	\label{f22_2}
\end{subfigure}

\caption{\footnotesize Effective logarithmic heat capacity $CL_{P_{eff}}$ (a) and pressure (b), in the first region $R_1$, as a function of effective temperature $T_{eff}$, with $Q=1$ and  $\rho_c = 5$.}
\label{f22}
\end{figure}

Finally, probably for the first time in black hole thermodynamics, we introduce the logarithmic effective heat capacity $C_{P_{eff}}$ by analogy with the $\beta$-function introduced in \cite{Abrahams1979} to study the metal-insulator transition in condensed matter physics \footnote{$\beta$-function is defined by $\beta (G) = \frac{d Log (G)}{d Log(L)}$, where $G$ is the conductance and $L$ is the system size. $\beta (G)$ is positive in the metallic phase and negative in the insulator phase. This function changes its monotony due to the interactions as it was demonstrated in \cite{Cherroret2014}}. We accordingly define the logarithmic effective heat capacity $C_{P_{eff}}$ as,
\begin{equation}\label{52}
\begin{split}
CL_{P_{eff}}& = \left. \dfrac{\partial \log(M)}{\partial \log(T_{eff}) }\right) _P \\
& = \left( \dfrac{\partial \log(M)}{\partial \rho_{+} } - \dfrac{\partial \log(M)}{\partial \rho_c } \dfrac{\frac{\partial P}{\partial \rho_+}}{\frac{\partial P}{\partial \rho_c}}\right)  \left( \dfrac{\partial \log(T_{eff})}{\partial \rho_{+} } - \dfrac{\partial \log(T_{eff})}{\partial \rho_c } \dfrac{\frac{\partial P}{\partial\rho_+}}{\frac{\partial P}{\partial \rho_c}}\right)^{-1} ,
\end{split}
\end{equation}
that we illustrate in Fig.~\ref{f22} (panel a) as a function of the effective temperature. We also show in the panel b, the behavior of pressure as a function of the effective temperature. 
 
\begin{figure}[h!]
	\begin{subfigure}[h]{0.45\textwidth}
		\centering \includegraphics[scale=0.5]{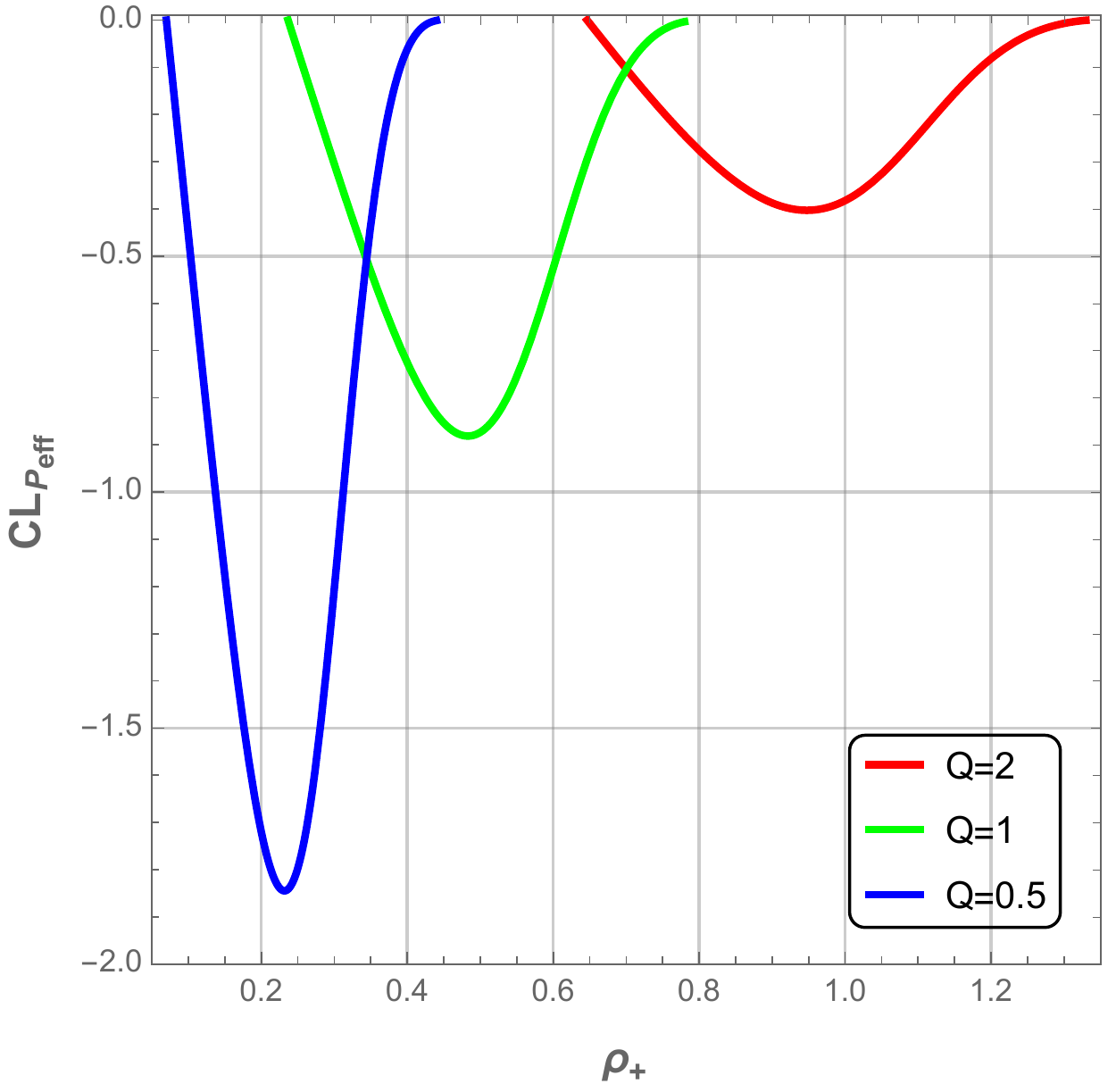}
		\caption{ }
		\label{f23_1}
	\end{subfigure}
	\hspace{1pt}
	\begin{subfigure}[h]{0.45\textwidth}
		\centering \includegraphics[scale=0.5]{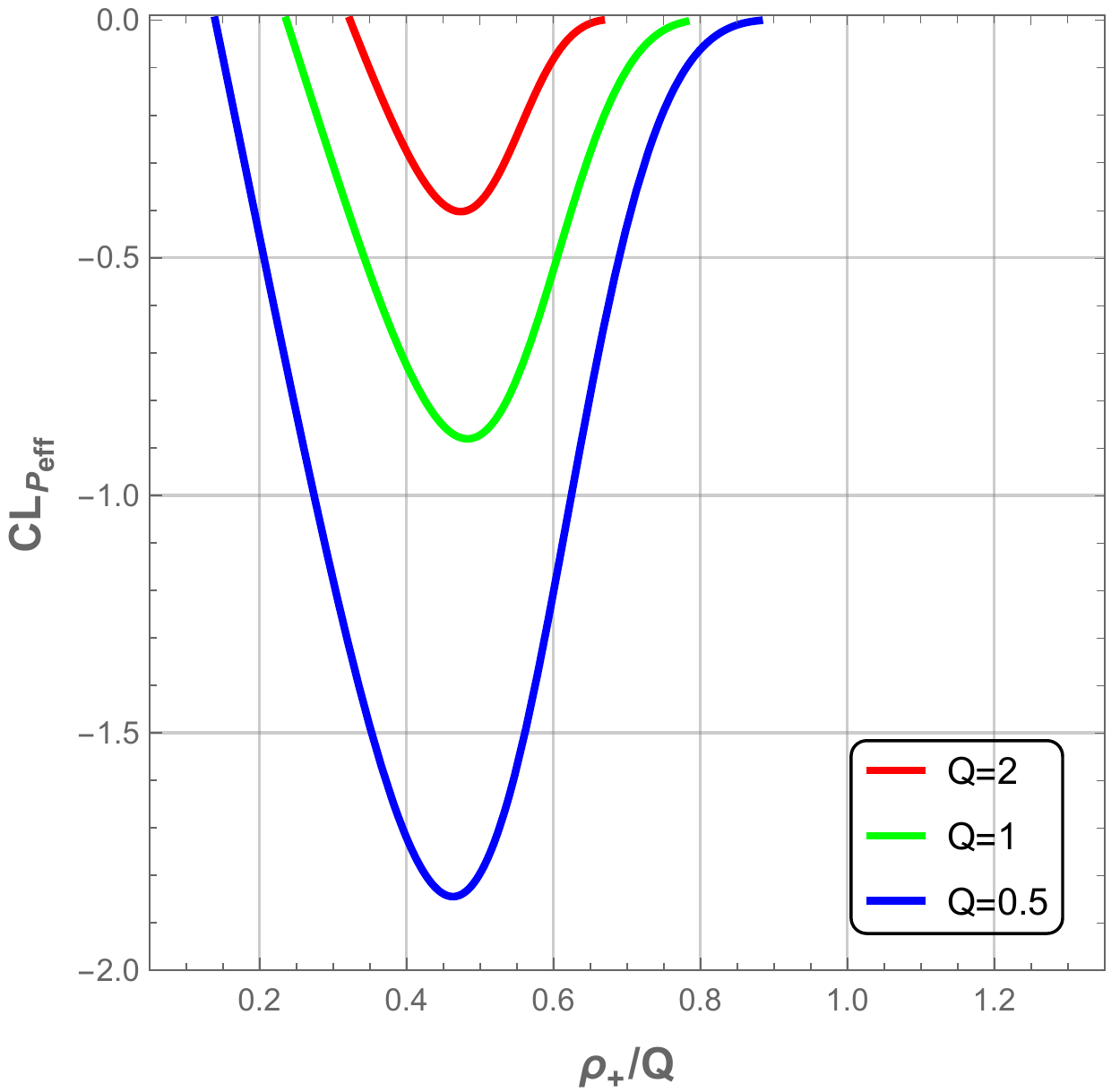}
		\caption{}
		\label{f23_2}
	\end{subfigure}
	
	\caption{\footnotesize Effective logarithmic heat capacity $CL_{P_{eff}}$ as a function of event horizon radius $\rho_{+}$ (a) and reduced event horizon radius $\frac{\rho_{+}}{Q}$ (b) for different values of electric charge $Q$ with $\rho_c = 5$.}
	\label{f23}
\end{figure}

 We clearly see that we have a thermal transition between the strong interacting phase and the weak interacting phase. These two phases are unstable. The critical radius $\rho_{+s}$ of this transition is proportional to electric charge as it is shown in Fig.~\ref{f23}, where we have plotted the logarithmic effective heat capacity $CL_{P_{eff}}$ as a function of the radius $\rho_+$ (left panel),  and as a function of the ratio $\rho_+/ Q$ in the right panel.

\begin{figure}[h!]
	\begin{subfigure}[h]{0.45\textwidth}
		\centering \includegraphics[scale=0.5]{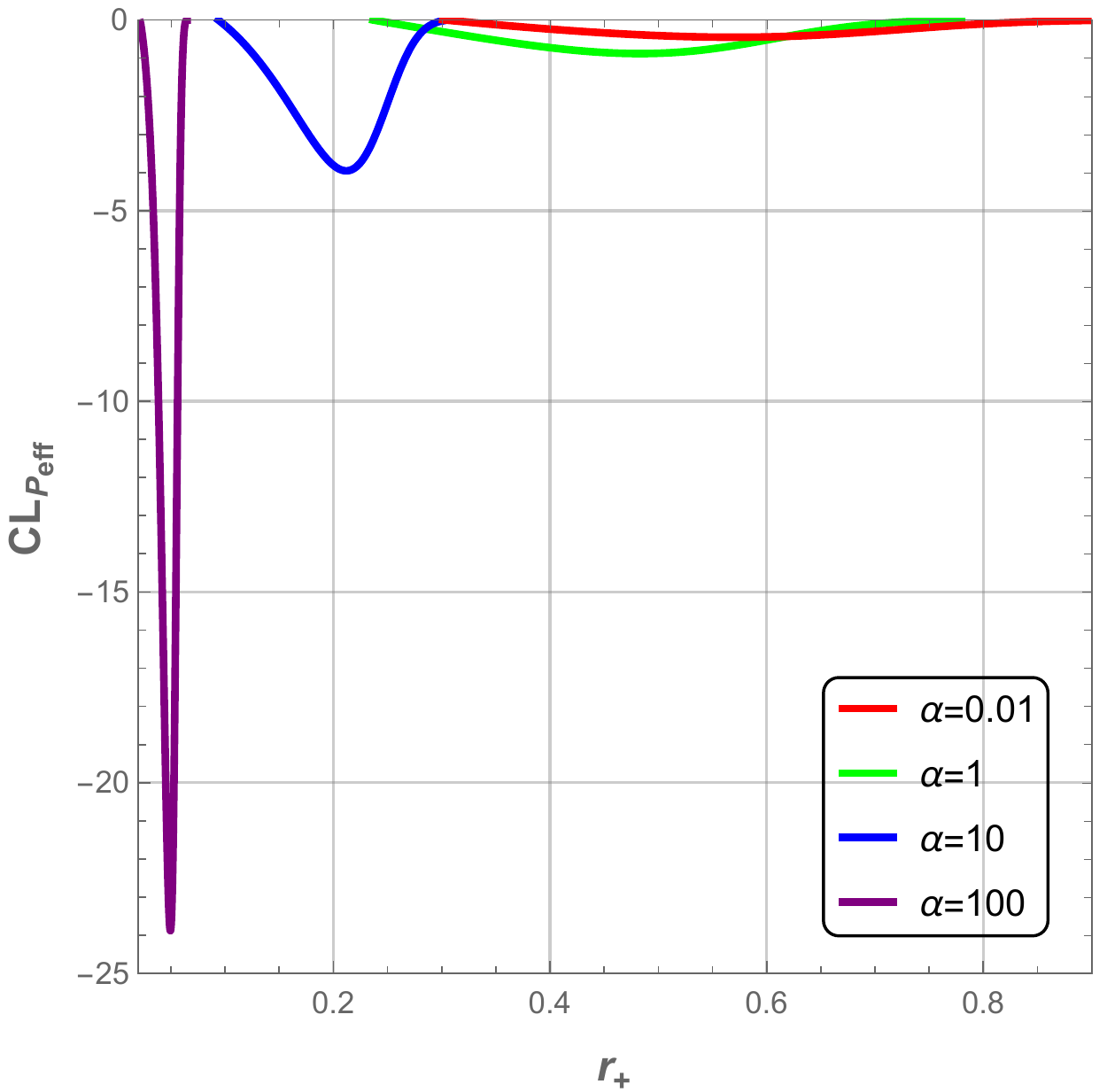}
		\caption{ }
		\label{f24_1}
	\end{subfigure}
	\hspace{1pt}
	\begin{subfigure}[h]{0.45\textwidth}
		\centering \includegraphics[scale=0.5]{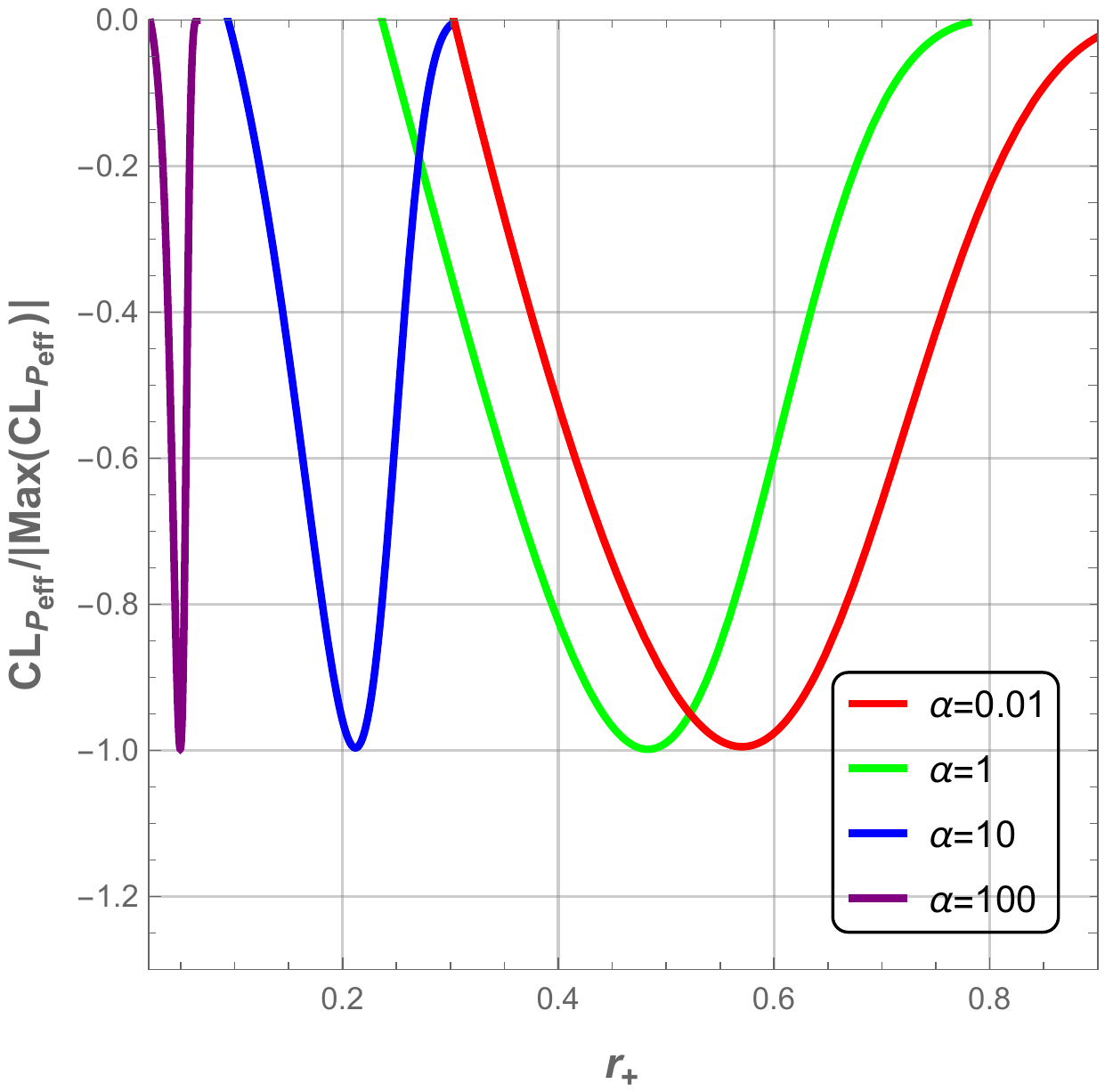}
		\caption{}
		\label{f24_2}
	\end{subfigure}
	
	\caption{\footnotesize Effective logarithmic heat capacity $CL_{P_{eff}}$ (a) and its normalized values (b) as a function of event horizon radius $r_{+}$  for different values of Yukawa charge $\alpha$ with electric charge $Q=1$ and  $r_c = 5$.}
	\label{f24}
\end{figure}

We also note that the electric interaction is weaker in ENLMY dS black hole than the usual charged black hole due to the exponential decreasing of Yukawa potential. Besides; when $\rho_+ \to 0$, the interactions become stronger and thus the black hole is strongly unstable. Hence, when the Yukawa charge is quite large the critical non-rescaled event horizon radius is almost vanishing as illustrated  by Fig.~\ref{f24}.

\begin{figure}[h!]
	\centering
	\includegraphics[scale=0.6]{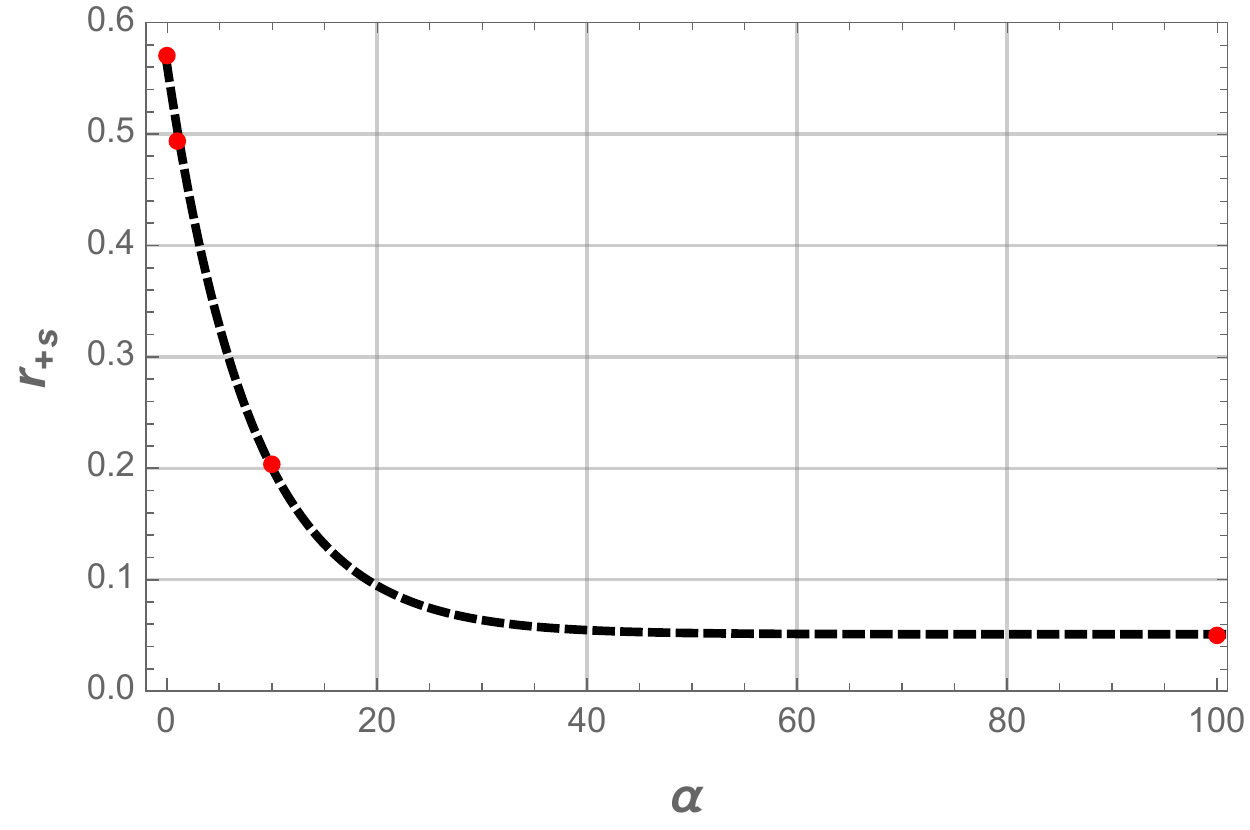}
	\caption{\footnotesize Critical radius $r_s$ of strong-weak electric transition as a function of Yukawa charge $\alpha$ (red points). The black dashed line represents the exponential fitting : $r_{+s} = 0.05 + 0.51 e^{-0.12 \alpha}$}
	\label{f24_3}
	
\end{figure}

Besides, the effect of the Yukawa charge $\alpha$ on the critical radius $r_{+s}$ is plotted in Fig.~\ref{f24_3}.
The critical radius $r_{+s}$ of strong-weak electric transition decreases exponentially when Yukawa charge $\alpha$ increases as it is displayed in Fig.~\ref{f24_3}. By fitting our result, we can describe the behavior of $r_{+s}$ quantity by a nice formula given by:
\begin{equation}
r_{+s} = 0.05 + 0.51 e^{-0.12 \alpha}.
\end{equation}

\section{Effective thermodynamics of three horizons}

In this last section, we extend the results obtained in section \ref{twohorizon} related to effective thermodynamics with two horizons to the three horizons. First we define the effective entropy of the three horizons as, 
\begin{equation}\label{54}
S_{eff} = S_c +S_- - S_+
\end{equation}
then, from Eq.~\eqref{25} and Eq.~\eqref{40}, we assume that  the effective first thermodynamics law and the effective Smarr-like formula of three horizons can read as:
\begin{equation}\label{55}
\begin{split}
dM =& -\dfrac{T_+T_cT_-}{T_+T_c+T_+T_-+T_-T_c} d(S_c+S_--S_+)
+ \dfrac{V_+T_cT_- + V_c T_+T_-+ V_- T_+T_c}{T_++T_c+T_-}dP\\
&+\dfrac{\Phi_{q_+}T_cT_- + \Phi_{q_c} T_+T_-+ \Phi_{q_-} T_+T_c}{T_++T_c+T_-}dQ+\dfrac{\Phi_{\alpha_+}T_cT_- + \Phi_{\alpha_c} T_+T_-+ \Phi_{\alpha_-} T_+T_c}{T_++T_c+T_-}d\alpha  .
\end{split}
\end{equation}
\begin{equation}\label{56}
\begin{split}
M =& -2 \dfrac{T_+T_cT_-}{T_+T_c+T_+T_i+T_-T_c} (S_c+S_--S_+)-2 \dfrac{V_+T_cT_- + V_c T_+T_i+ V_- T_+T_c}{T_++T_c+T_-}P \\
&+\dfrac{\Phi_{q_+}T_cT_- + \Phi_{q_c} T_+T_-+ \Phi_{q_-} T_+T_c}{T_++T_c+T_-}Q -\dfrac{\Phi_{\alpha_+}T_cT_- + \Phi_{\alpha_c} T_+T_-+ \Phi_{\alpha_-} T_+T_c}{T_++T_c+T_-}\alpha  .
\end{split}
\end{equation}
Hence, the effective thermodynamic variables are given by,
\begin{equation}\label{57}
\begin{split}
&T_{eff} = \dfrac{T_+T_cT_-}{T_+T_c+T_+T_-+T_-T_c}, \\
& V_{eff} = \dfrac{V_+T_cT_- + V_c T_+T_-+ V_- T_+T_c}{T_++T_c+T_-},\\
&\Phi_{q_{eff}} =\dfrac{\Phi_{q_+}T_cT_- + \Phi_{q_c} T_+T_-+ \Phi_{q_-} T_+T_c}{T_++T_c+T_-}, \\
 & \Phi_{\alpha_{eff}} = \dfrac{\Phi_{\alpha_+}T_cT_- + \Phi_{\alpha_c} T_+T_-+ \Phi_{\alpha_-} T_+T_c}{T_++T_c+T_-}.
\end{split}
\end{equation}

We note that from Eq.~\eqref{40} the pressure and electric charge could be expressed as a function of horizon radii, particularly if we consider black hole thermodynamics with fixed cosmological and inner horizons (treated below).

First we define the characteristic  ratio $a_\rho = \frac{\rho_c}{\rho_-}$, that varies from $1$ to $+\infty$, and plot the effective temperature as a function of the event horizon radius. 
\begin{figure}[h!]
	\begin{subfigure}[h]{0.45\textwidth}
		\centering \includegraphics[scale=0.5]{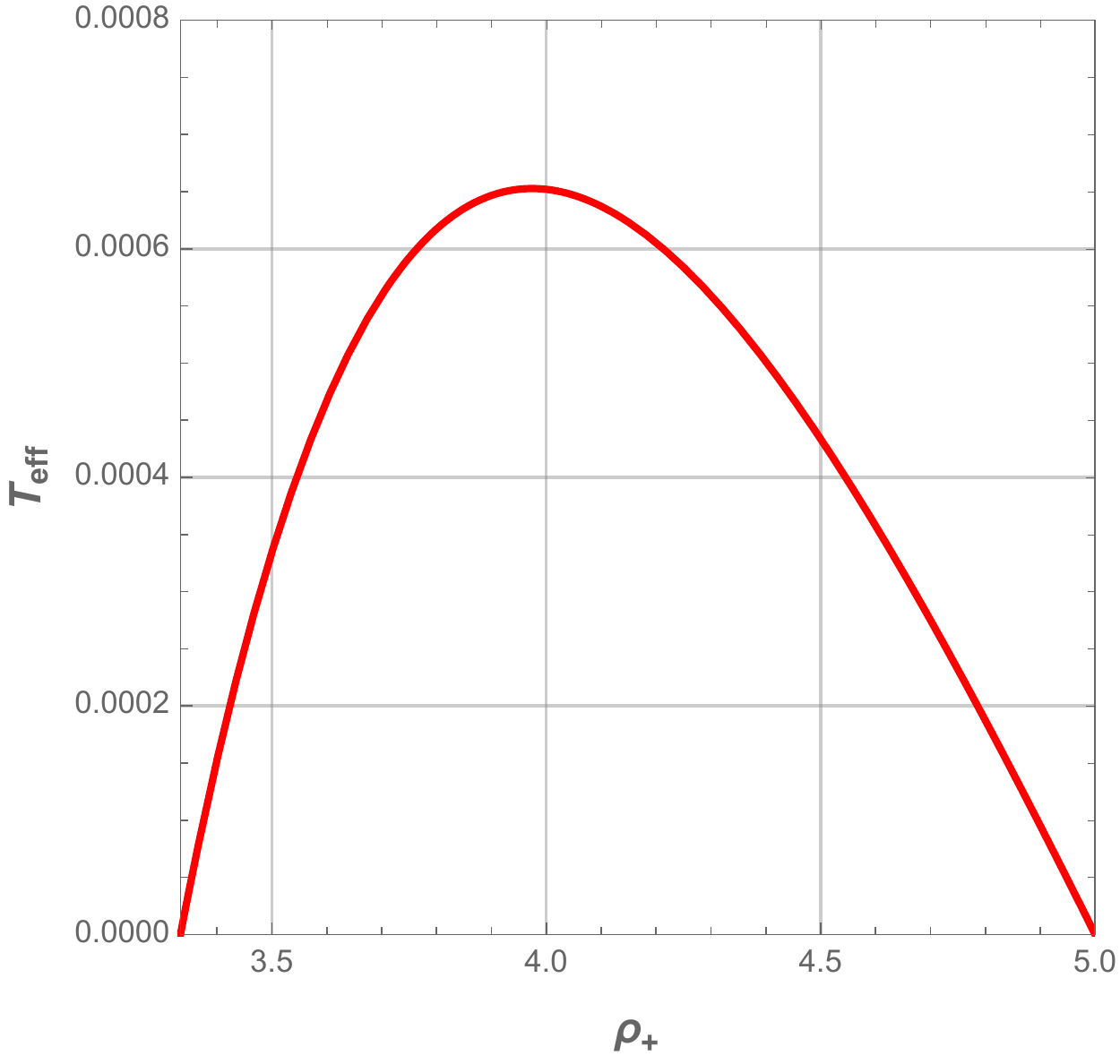}
		\caption{\footnotesize $a_\rho = 1.5$.}
		\label{f25_1}
	\end{subfigure}
	\begin{subfigure}[h]{0.45\textwidth}
		\centering \includegraphics[scale=0.5]{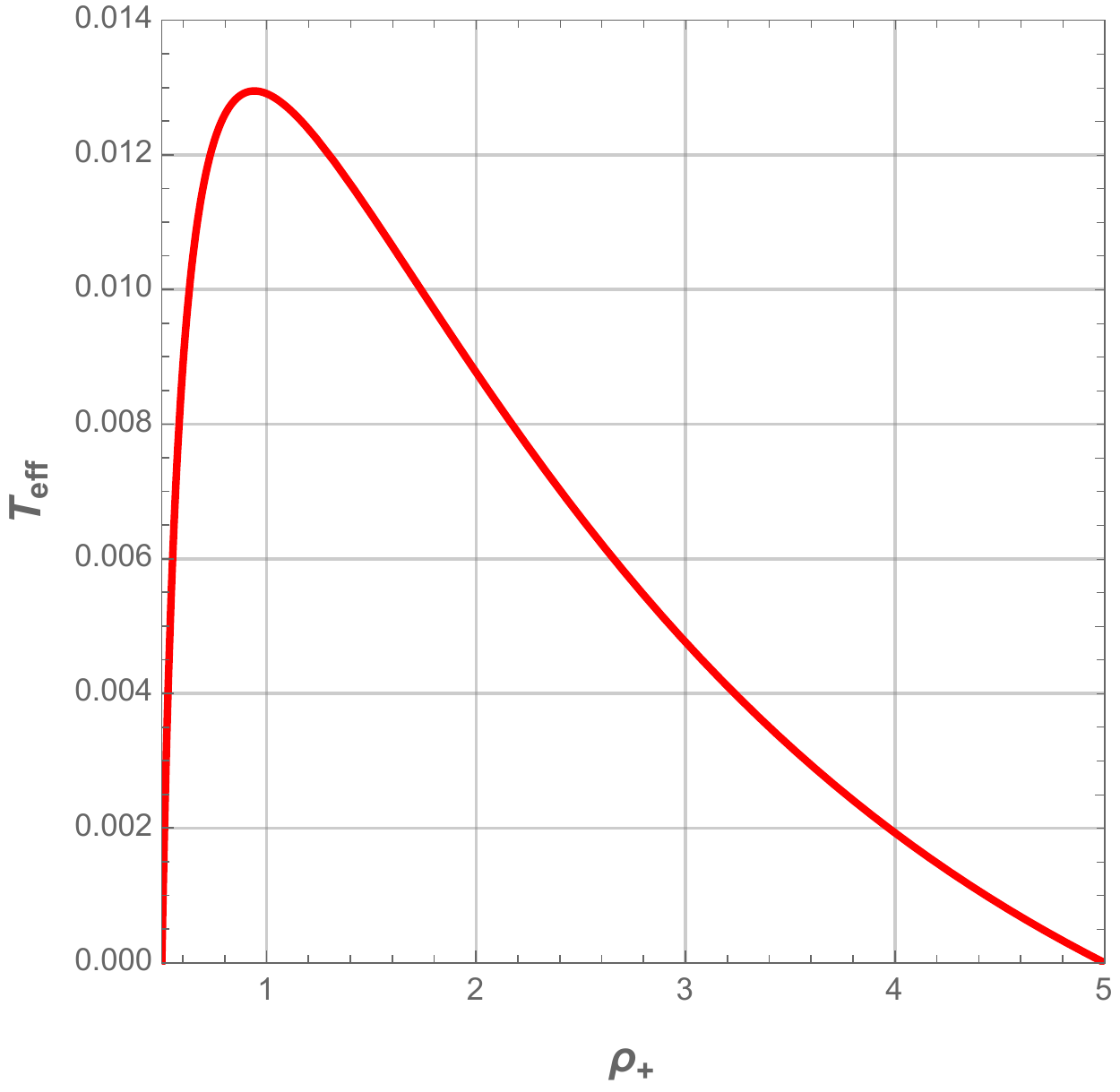}
		\caption{\footnotesize  $a_\rho = 10$.}
		\label{f2_2}
	\end{subfigure}
	\begin{subfigure}[h]{0.45\textwidth}
	\centering \includegraphics[scale=0.5]{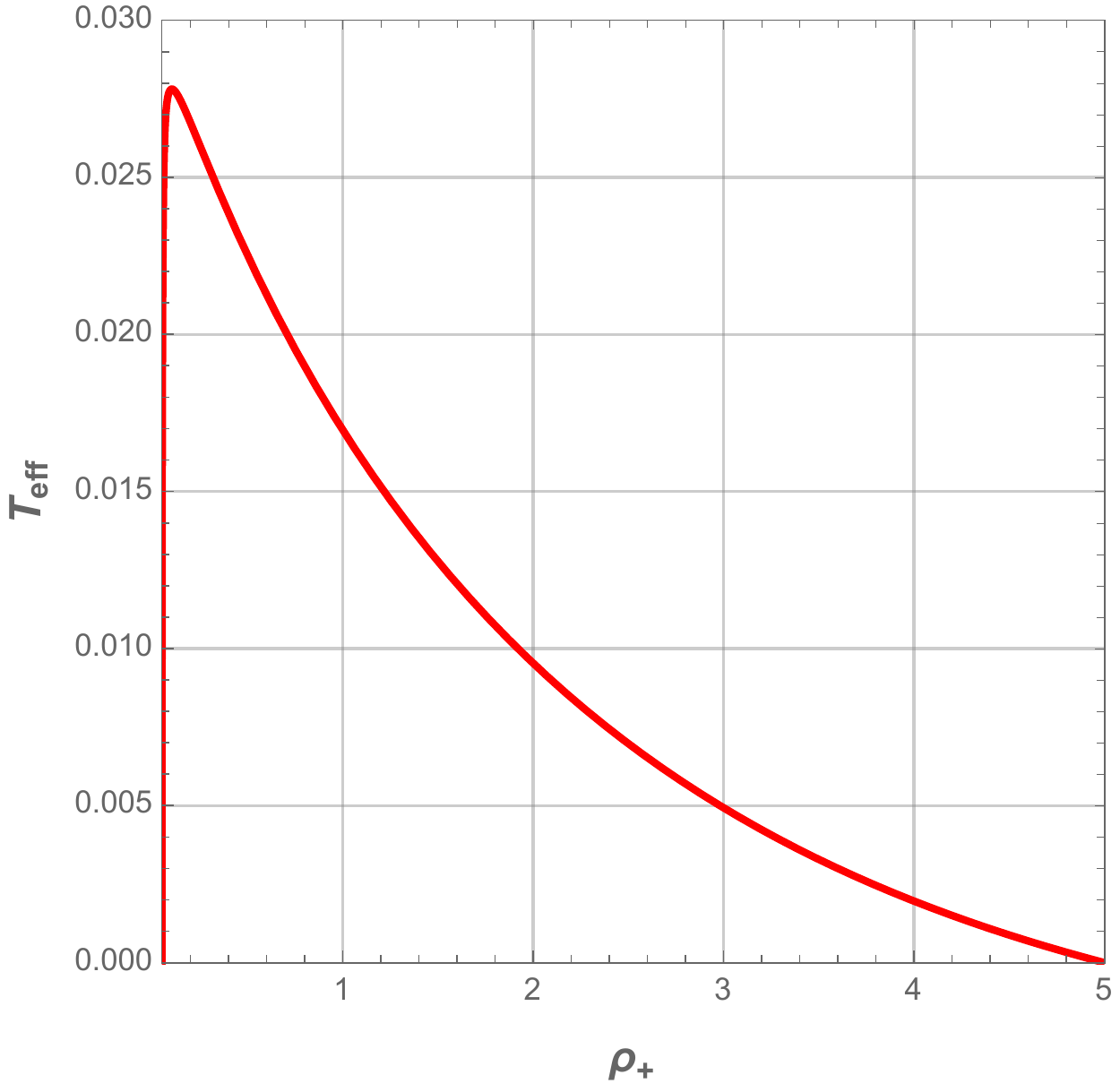}
	\caption{\footnotesize $a_\rho = 100$. }
	\label{f25_3}
\end{subfigure}
\hspace{12mm}
\begin{subfigure}[h]{0.45\textwidth}
	\centering \includegraphics[scale=0.5]{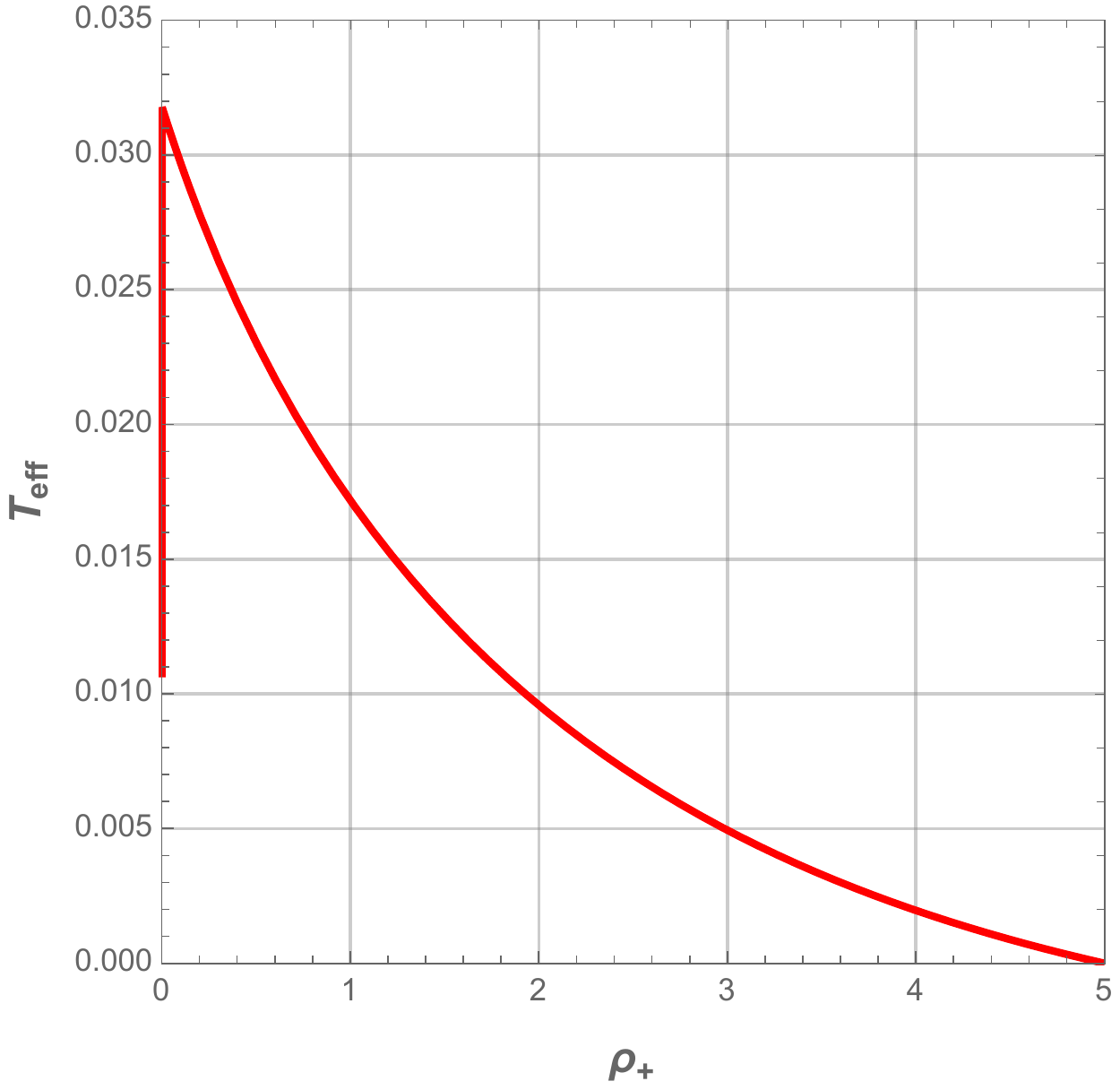}
	\caption{\footnotesize $a_\rho = 10000$.}
	\label{f25_4}
\end{subfigure}
	\caption{\footnotesize Effective Temperature of the three black hole horizons as a function of $\rho_+$ with $\rho_c=5$.}
	\label{f25}
\end{figure}

From Fig.~\ref{f25}, it can be seen that we have just the second region $R_2$ determined previously in Section 6, while the first region $R_1$ has vanished.  Since the behavior of the effective temperature $T_{eff}$ depends on the ratio $a_\rho$, for small $a_\rho$, $T_{eff}$ should first increase until a maximum, then decreases when $\rho_+$ gets large.
For large $a_\rho$, with the increasing part of $T_{eff}$ almost disappearing,   the effective temperature behaves as  decreasing monotonous function of $\rho_+$.

\begin{figure}[h!]	
	\begin{subfigure}[h]{0.45\textwidth}
		\centering \includegraphics[scale=0.5]{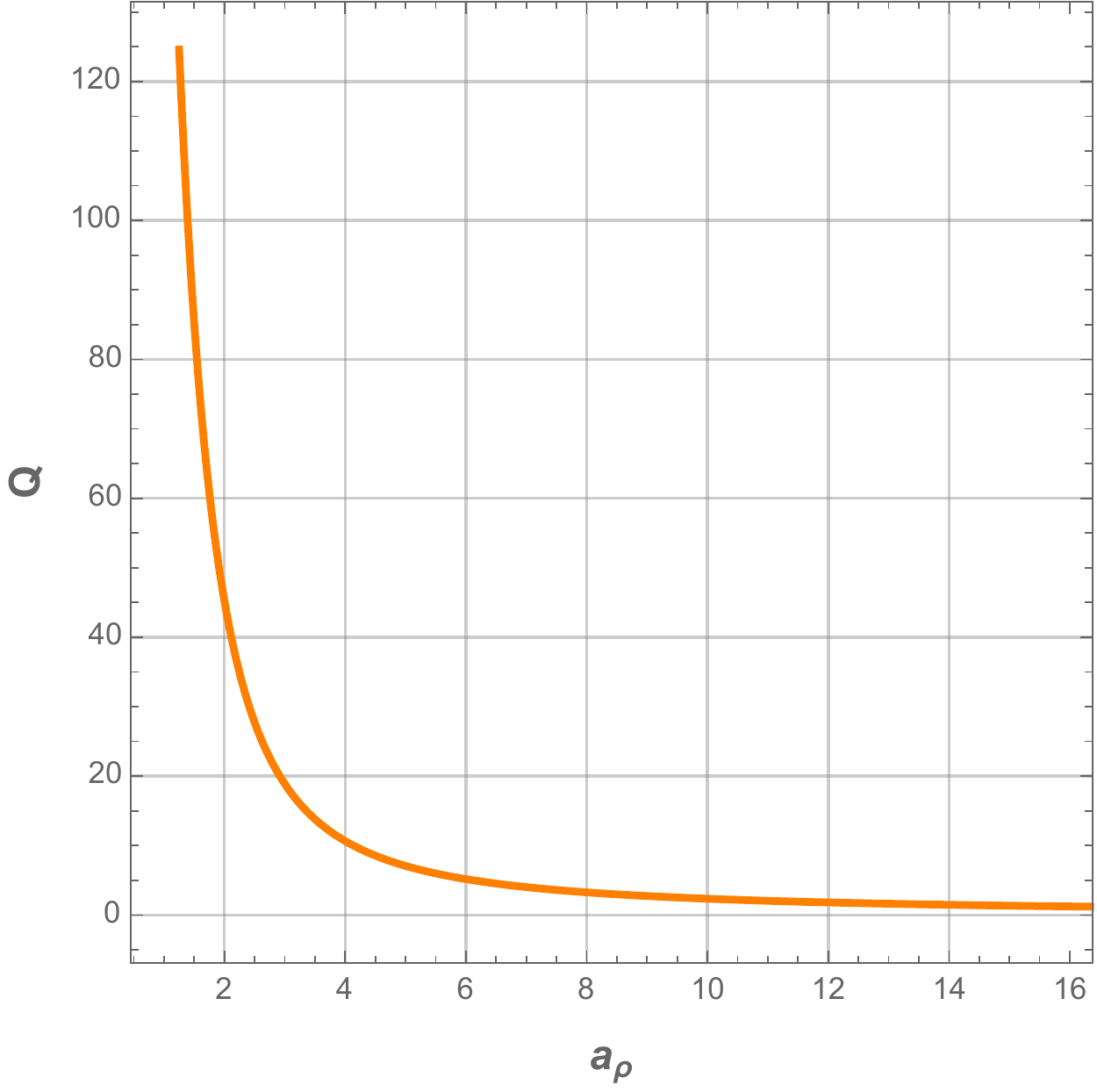}
		\caption{}
		\label{f26_1}
	\end{subfigure}
	\begin{subfigure}[h]{0.45\textwidth}
		\centering \includegraphics[scale=0.5]{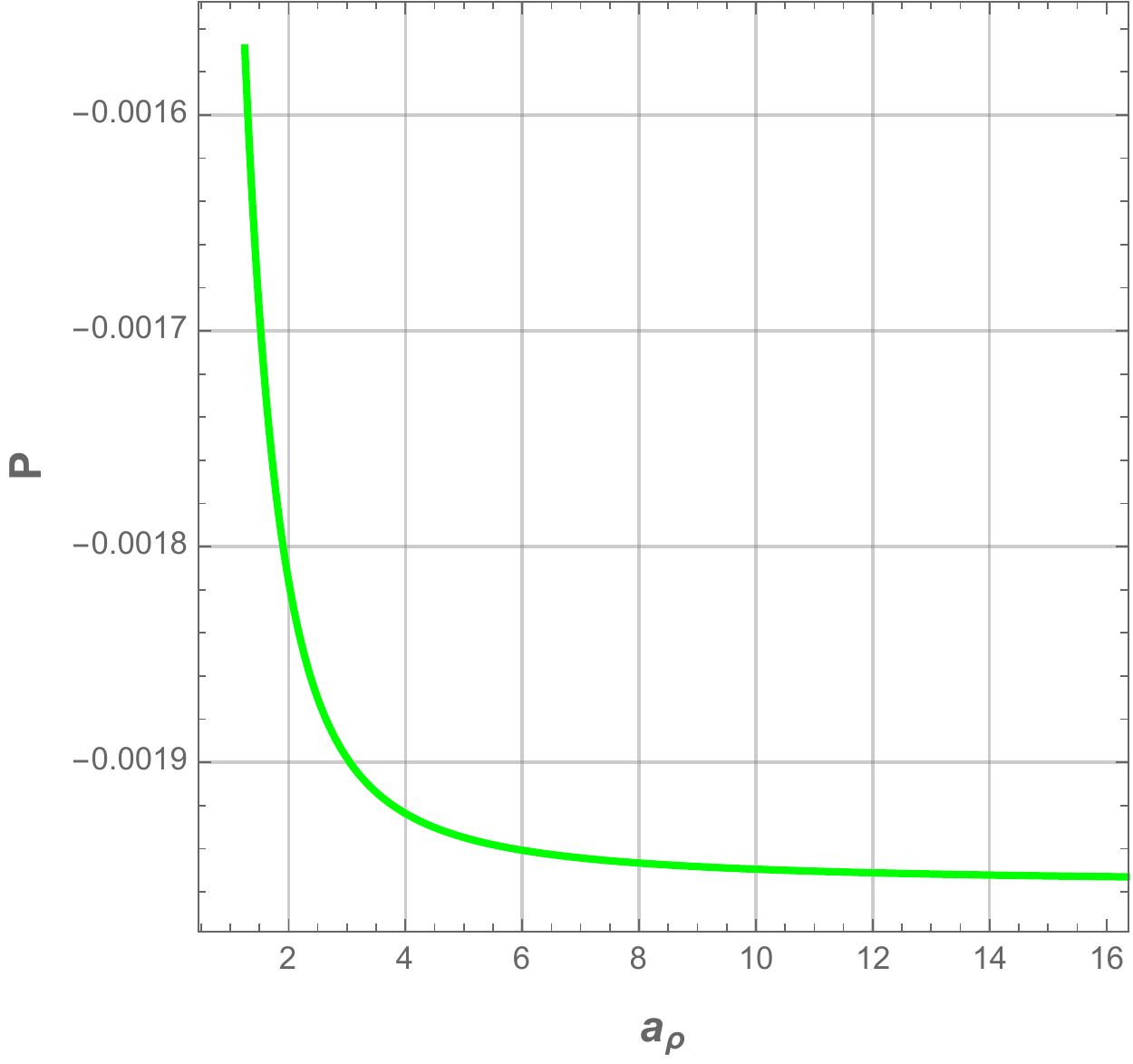}
		\caption{}
		\label{f26_2}
	\end{subfigure}

	\caption{\footnotesize Electric charge $Q$ (a) and thermodynamic pressure $P$ (b) as a function of  the characteristic ratio $r_\rho$ at $\rho_+ = 4$ with $\rho_c=5$.}
\label{f26}
\end{figure}

We plot in Fig.~\ref{f26} the electric charge $Q$ and thermodynamic pressure $P$ as a function of the characteristic ratio $a_\rho$.  Their behaviors  show that the black hole can undergo a thermodynamic phase transition only if the electric charge $Q$ and cosmological constant are quite large. When $Q\to 0$ and $P \to 0$ the black hole phase is in unstable phase.

\begin{figure}[h!]

	\begin{subfigure}[h]{0.45\textwidth}
		\centering \includegraphics[scale=0.5]{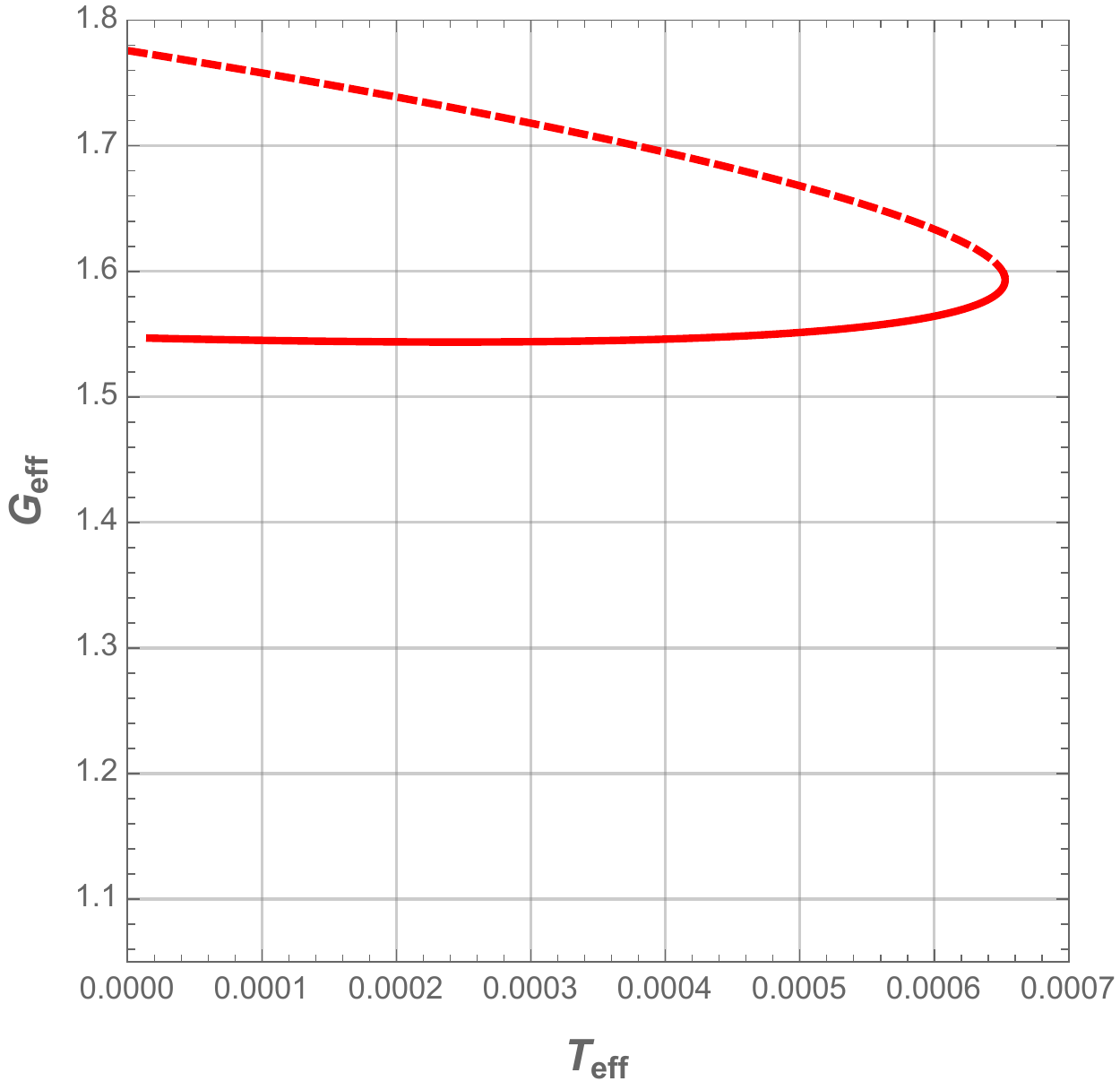}
		\caption{\footnotesize $a_\rho = 1.5$ .}
		\label{f27_1}
	\end{subfigure}
	\begin{subfigure}[h]{0.45\textwidth}
		\centering \includegraphics[scale=0.5]{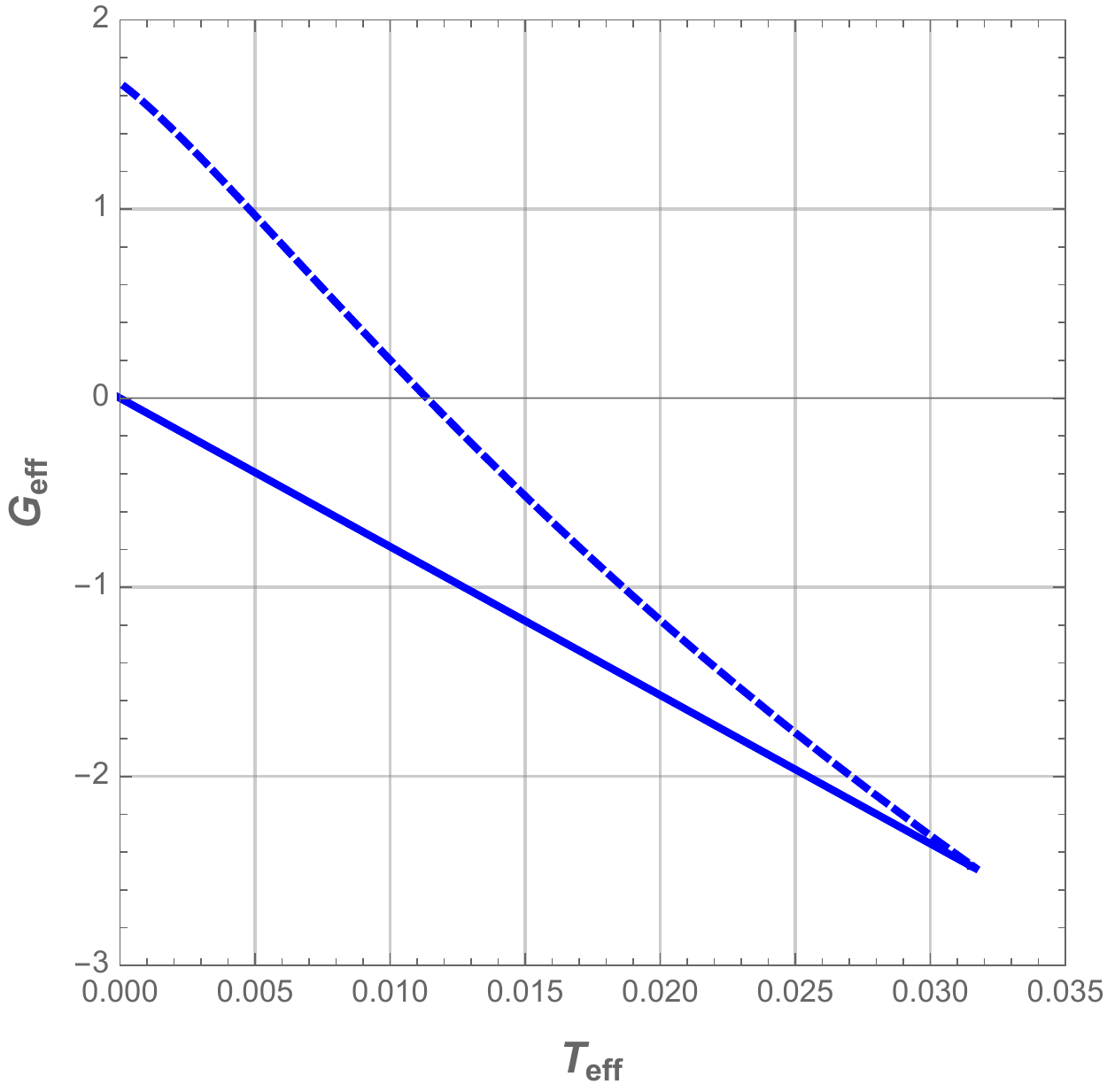}
		\caption{\footnotesize $a_\rho = 10000$.}
		\label{f27_2}
	\end{subfigure}
	
	\caption{\footnotesize  Effective free Gibbs energy $G_{eff}$ as a function of the effective temperature $T_{eff}$, with different values of the characteristic ratio $a_\rho$, and $\rho_c = 5$.}
	\label{f27}
\end{figure}

After that, we define the effective free Gibbs energy as $G_{eff} = M - T_{eff} S_{eff}$ and plot its variation as a function of the effective temperature in Fig.~\ref{f27}.
 For small values of the characteristic ratio $a_\rho$, the black hole undergoes a phase transition between a stable and unstable phases. Whereas, for large values of  $a_\rho$, the black hole is only stable  near the extreme case where $\rho_+ \to \rho_-$,  dubbed "cold black hole" \footnote{ For example, if $a_\rho = 10^3$, the black hole is stable  when $\frac{\rho_+}{\rho_-} < 1.04$. Hence, this phase can well be omitted  for $a_\rho \to \infty$,  since in this scenario, the  unstable phase dominates}.

\section{Conclusion}

The thermodynamics of asymptotically de Sitter black hole is considered as one of the main keys to our comprehension of Gravity. However, compared to anti-de sitter thermodynamics, it remains relatively unexplored. The present work considered such a topic: we have first extended the black hole solution of Einstein gravity coupled to a non-linear electromagnetic field and Yukawa potential initially treated in \cite{base} to the dS spacetime with a static and spherical symmetric ansatz. Then we have shown that such a solution presents three distinct horizons, namely the event, cosmological and inner horizons. We have employed several approaches to disclose the phase structure of Einstein nonlinear Maxwell Yukawa dS black hole in the extended phase space.

By following the strategy employed by Kubiznak et al. \cite{Kubiznak:2015bya} to derive several thermodynamic first laws (one for each horizon) and explore their thermodynamics as if they were independent thermodynamic systems, we were able to uncover the rich phase structure of de Einstein non-linear Maxwell Yukawa dS black hole solution. In addition to that, we have also proposed an analysis with a simultaneous treatment of all horizons. Afterward, by assuming that the event and cosmological horizons are not located far away, we have used an approach based on the effective thermodynamics of the black hole. We studied the internal correlation of these systems (cosmological and inner horizons), which enabled us to evaluate a single Gibbs free energy-like quantity $G$ that captures their individual thermodynamic behavior and contains information about possible phase transitions of the whole de Sitter system. In the end, we have extended the formalism of effective thermodynamics initially based on the correlation between two horizons to the correlation of three black hole horizons.

Last, from a close look at Eq.~\eqref{49} and Eq.~\eqref{57}, we determined that the effective temperatures behave like an association of ohmic resistances in an electrical circuit. This remarkable finding may open a new window on a possible analogy between both systems and  constitute an interesting field of future investigations.

\newpage
\bibliographystyle{unsrt}
\bibliography{biblio}

\begin{thebibliography}{10}

\bibitem{Hawking:1982dh}
S.~W. Hawking and Don~N. Page.
\newblock {Thermodynamics of Black Holes in anti-De Sitter Space}.
\newblock {\em Commun. Math. Phys.}, 87:577, 1983.

\bibitem{Hawking:1974sw}
S.~W. Hawking.
\newblock {Particle Creation by Black Holes}.
\newblock {\em Commun. Math. Phys.}, 43:199--220, 1975.
\newblock [,167(1975)].

\bibitem{Bardeen:1973gs}
James~M. Bardeen, B.~Carter, and S.~W. Hawking.
\newblock {The Four laws of black hole mechanics}.
\newblock {\em Commun. Math. Phys.}, 31:161--170, 1973.

\bibitem{Bekenstein:1972tm}
J.~D. Bekenstein.
\newblock {Black holes and the second law}.
\newblock {\em Lett. Nuovo Cim.}, 4:737--740, 1972.

\bibitem{Bekenstein:1973mi}
J.~D. Bekenstein.
\newblock {Extraction of energy and charge from a black hole}.
\newblock {\em Phys. Rev.}, D7:949--953, 1973.

\bibitem{Bekenstein:1974ax}
Jacob~D. Bekenstein.
\newblock {Generalized second law of thermodynamics in black hole physics}.
\newblock {\em Phys. Rev.}, D9:3292--3300, 1974.

\bibitem{Chamblin:1999tk}
Andrew Chamblin, Roberto Emparan, Clifford~V. Johnson, and Robert~C. Myers.
\newblock {Charged AdS black holes and catastrophic holography}.
\newblock {\em Phys. Rev.}, D60:064018, 1999.

\bibitem{KM}
David Kubiznak and Robert~B. Mann.
\newblock {P-V criticality of charged AdS black holes}.
\newblock {\em JHEP}, 07:033, 2012.

\bibitem{our}
A.~Belhaj, M.~Chabab, H.~El~Moumni, and M.~B. Sedra.
\newblock {On Thermodynamics of AdS Black Holes in Arbitrary Dimensions}.
\newblock {\em Chin. Phys. Lett.}, 29:100401, 2012.

\bibitem{Chabab:2018lzf}
M.~Chabab, H.~El~Moumni, S.~Iraoui, K.~Masmar, and S.~Zhizeh.
\newblock {Chaos in charged AdS black hole extended phase space}.
\newblock {\em Physics Letters}, B781:316--321, 2018.

\bibitem{Chabab:2019mlu}
M.~Chabab, H.~El~Moumni, S.~Iraoui, and K.~Masmar.
\newblock {Phase transitions and geothermodynamics of black holes in dRGT
  massive gravity}.
\newblock {\em Eur. Phys. J.}, C79(4):342, 2019.

\bibitem{our3}
A.~Belhaj, M.~Chabab, H.~EL~Moumni, K.~Masmar, and M.~B. Sedra.
\newblock {Ehrenfest scheme of higher dimensional AdS black holes in the
  third-order Lovelock-Born-Infeld gravity}.
\newblock {\em Int. J. Geom. Meth. Mod. Phys.}, 12(10):1550115, 2015.

\bibitem{Chabab:2017knz}
M.~Chabab, H.~El~Moumni, S.~Iraoui, and K.~Masmar.
\newblock {Phase Transition of Charged-AdS Black Holes and Quasinormal Modes :
  a Time Domain Analysis}.
\newblock {\em Astrophys. Space Sci.}, 362(10):192, 2017.

\bibitem{our8}
M.~Chabab, H.~El~Moumni, S.~Iraoui, and K.~Masmar.
\newblock {Behavior of quasinormal modes and high dimension RN-AdS black hole
  phase transition}.
\newblock {\em Eur. Phys. J.}, C76(12):676, 2016.

\bibitem{our6}
M.~Chabab, H.~El~Moumni, and K.~Masmar.
\newblock {On thermodynamics of charged AdS black holes in extended phases
  space via M2-branes background}.
\newblock {\em Eur. Phys. J.}, C76(6):304, 2016.

\bibitem{our7}
A.~Belhaj, M.~Chabab, H.~El~Moumni, K.~Masmar, M.~B. Sedra, and A.~Segui.
\newblock {On Heat Properties of AdS Black Holes in Higher Dimensions}.
\newblock {\em JHEP}, 05:149, 2015.

\bibitem{chin1}
B.~D. Koberlein and Ronald~L. Mallett.
\newblock {Charged, radiating black holes, inflation, and cosmic censorship}.
\newblock {\em Phys. Rev.}, D49:5111--5114, 1994.

\bibitem{chin2}
Rong-Gen Cai.
\newblock {Cardy-Verlinde formula and thermodynamics of black holes in de
  Sitter spaces}.
\newblock {\em Nucl. Phys.}, B628:375--386, 2002.

\bibitem{Urano:2009xn}
Miho Urano, Akira Tomimatsu, and Hiromi Saida.
\newblock {Mechanical First Law of Black Hole Spacetimes with Cosmological
  Constant and Its Application to Schwarzschild-de Sitter Spacetime}.
\newblock {\em Class. Quant. Grav.}, 26:105010, 2009.

\bibitem{Sekiwa:2006qj}
Yuichi Sekiwa.
\newblock {Thermodynamics of de Sitter black holes: Thermal cosmological
  constant}.
\newblock {\em Phys. Rev.}, D73:084009, 2006.

\bibitem{Dolan:2013ft}
Brian~P. Dolan, David Kastor, David Kubiznak, Robert~B. Mann, and Jennie
  Traschen.
\newblock {Thermodynamic Volumes and Isoperimetric Inequalities for de Sitter
  Black Holes}.
\newblock {\em Phys. Rev.}, D87(10):104017, 2013.

\bibitem{Li:2016zdi}
Huai-Fan Li, Meng-Sen Ma, Li-Chun Zhang, and Ren Zhao.
\newblock {Entropy of Kerr--de Sitter black hole}.
\newblock {\em Nucl. Phys.}, B920:211--220, 2017.

\bibitem{Zhao:2014raa}
Hui-Hua Zhao, Li-Chun Zhang, Meng-Sen Ma, and Ren Zhao.
\newblock {P-V criticality of higher dimensional charged topological dilaton de
  Sitter black holes}.
\newblock {\em Phys. Rev.}, D90(6):064018, 2014.

\bibitem{Zhang:2014jfa}
Li-Chun Zhang, Meng-Sen Ma, Hui-Hua Zhao, and Ren Zhao.
\newblock {Thermodynamics of phase transition in higher-dimensional
  Reissner-Nordstr{\"o}m-de Sitter black hole}.
\newblock {\em Eur. Phys. J.}, C74(9):3052, 2014.

\bibitem{Zhang:2016yek}
L.~C. Zhang and R.~Zhao.
\newblock {The critical phenomena of Schwarzschild-de Sitter black hole}.
\newblock {\em EPL}, 113(1):10008, 2016.

\bibitem{Simovic:2019zgb}
Fil Simovic and Robert~B. Mann.
\newblock {Critical Phenomena of Born-Infeld-de Sitter Black Holes in
  Cavities}.
\newblock {\em JHEP}, 05:136, 2019.

\bibitem{Mbarek:2018bau}
Saoussen Mbarek and Robert~B. Mann.
\newblock {Reverse Hawking-Page Phase Transition in de Sitter Black Holes}.
\newblock {\em JHEP}, 02:103, 2019.

\bibitem{Zhang:2016nws}
Li-Chun Zhang, Ren Zhao, and Meng-Sen Ma.
\newblock {Entropy of Reissner--Nordstr{\"o}m--de Sitter black hole}.
\newblock {\em Phys. Lett.}, B761:74--76, 2016.

\bibitem{Liu:2019qxt}
Fang Liu and Li-Chun Zhang.
\newblock {On thermodynamics of RN-dS black hole surrounded by the
  quintessence}.
\newblock {\em Chin. J. Phys.}, 2019.

\bibitem{Ma:2019pya}
Yubo Ma, Yang Zhang, Ren Zhao, Shuo Cao, Tonghua Liu, Shubiao Geng, Yuting Liu,
  and Yumei Huang.
\newblock {Phase transitions and entropy force of charged de Sitter black holes
  with cloud of string and quintessence}.
\newblock 2019.

\bibitem{Kubiznak:2015bya}
David Kubiznak and Fil Simovic.
\newblock {Thermodynamics of horizons: de Sitter black holes and reentrant
  phase transitions}.
\newblock {\em Class. Quant. Grav.}, 33(24):245001, 2016.

\bibitem{Sheykhi}
Ahmad Sheykhi.
\newblock {Higher-dimensional charged $f(R)$ black holes}.
\newblock {\em Phys. Rev.}, D86:024013, 2012.

\bibitem{ref1}
Max Born.
\newblock {Quantum theory of the electromagnetic field}.
\newblock {\em Proc. Roy. Soc. Lond.}, A143(849):410--437, 1934.

\bibitem{ref2}
J.~F. Plebanski and M.~Przanowski.
\newblock {Duality transformations in electrodynamics}.
\newblock {\em Int. J. Theor. Phys.}, 33:1535--1551, 1994.

\bibitem{ref3}
Nathan Seiberg and Edward Witten.
\newblock {String theory and noncommutative geometry}.
\newblock {\em JHEP}, 09:032, 1999.

\bibitem{ref4}
Ofer Aharony, Steven~S. Gubser, Juan~Martin Maldacena, Hirosi Ooguri, and Yaron
  Oz.
\newblock {Large N field theories, string theory and gravity}.
\newblock {\em Phys. Rept.}, 323:183--386, 2000.

\bibitem{ref5}
Ricardo Garcia-Salcedo and Nora Breton.
\newblock {Born-Infeld cosmologies}.
\newblock {\em Int. J. Mod. Phys.}, A15:4341--4354, 2000.

\bibitem{ref6}
M.~Novello, Santiago~Esteban Perez~Bergliaffa, and J.~Salim.
\newblock {Non-linear electrodynamics and the acceleration of the universe}.
\newblock {\em Phys. Rev.}, D69:127301, 2004.

\bibitem{ref7}
Eloy Ayon-Beato and Alberto Garcia.
\newblock {The Bardeen model as a nonlinear magnetic monopole}.
\newblock {\em Phys. Lett.}, B493:149--152, 2000.

\bibitem{ref8}
Eloy Ayon-Beato and Alberto Garcia.
\newblock {Four parametric regular black hole solution}.
\newblock {\em Gen. Rel. Grav.}, 37:635, 2005.

\bibitem{ref9}
K.~A. Bronnikov.
\newblock {Comment on `Regular black hole in general relativity coupled to
  nonlinear electrodynamics'}.
\newblock {\em Phys. Rev. Lett.}, 85:4641, 2000.

\bibitem{ref10}
Kirill~A. Bronnikov.
\newblock {Regular magnetic black holes and monopoles from nonlinear
  electrodynamics}.
\newblock {\em Phys. Rev.}, D63:044005, 2001.

\bibitem{ref11}
Irina~G. Dymnikova.
\newblock {The algebraic structure of a cosmological term in spherically
  symmetric solutions}.
\newblock {\em Phys. Lett.}, B472:33--38, 2000.

\bibitem{ref12}
Irina Dymnikova.
\newblock {Regular electrically charged structures in nonlinear electrodynamics
  coupled to general relativity}.
\newblock {\em Class. Quant. Grav.}, 21:4417--4429, 2004.

\bibitem{fan}
Zhong-Ying Fan and Xiaobao Wang.
\newblock {Construction of Regular Black Holes in General Relativity}.
\newblock {\em Phys. Rev.}, D94(12):124027, 2016.

\bibitem{Hassaine:2007py}
Mokhtar Hassaine and Cristian Martinez.
\newblock {Higher-dimensional black holes with a conformally invariant Maxwell
  source}.
\newblock {\em Phys. Rev.}, D75:027502, 2007.

\bibitem{Hassaine:2008pw}
Mokhtar Hassaine and Cristian Martinez.
\newblock {Higher-dimensional charged black holes solutions with a nonlinear
  electrodynamics source}.
\newblock {\em Class. Quant. Grav.}, 25:195023, 2008.

\bibitem{Kruglov:2014iwa}
S.~I. Kruglov.
\newblock {Nonlinear arcsin-electrodynamics}.
\newblock {\em Annalen Phys.}, 527:397--401, 2015.

\bibitem{Soleng:1995kn}
Harald~H. Soleng.
\newblock {Charged black points in general relativity coupled to the
  logarithmic U(1) gauge theory}.
\newblock {\em Phys. Rev.}, D52:6178--6181, 1995.

\bibitem{Hendi:2012zz}
S.~H. Hendi.
\newblock {Asymptotic charged BTZ black hole solutions}.
\newblock {\em JHEP}, 03:065, 2012.

\bibitem{Hendi:2013mka}
S.~H. Hendi and A.~Sheykhi.
\newblock {Charged rotating black string in gravitating nonlinear
  electromagnetic fields}.
\newblock {\em Phys. Rev.}, D88(4):044044, 2013.

\bibitem{Fradkin:1985qd}
E.~S. Fradkin and Arkady~A. Tseytlin.
\newblock {Nonlinear Electrodynamics from Quantized Strings}.
\newblock {\em Phys. Lett.}, 163B:123--130, 1985.

\bibitem{Karlhede:1987bg}
A.~Karlhede, U.~Lindstrom, M.~Rocek, and G.~Theodoridis.
\newblock {Supersymmetric Nonlinear Maxwell Theories and the String Effective
  Action}.
\newblock {\em Nucl. Phys.}, B294:498--504, 1987.

\bibitem{Hamada:1987ph}
Ken-ji Hamada, Jiro Kodaira, and Juichi Saito.
\newblock {HETEROTIC STRING IN BACKGROUND GAUGE FIELDS}.
\newblock {\em Nucl. Phys.}, B297:637--652, 1988.

\bibitem{Tseytlin:1997csa}
Arkady~A. Tseytlin.
\newblock {On nonAbelian generalization of Born-Infeld action in string
  theory}.
\newblock {\em Nucl. Phys.}, B501:41--52, 1997.

\bibitem{base}
S.~Habib Mazharimousavi and Mustafa Halilsoy.
\newblock {Einstein-nonlinear Maxwell--Yukawa black hole}.
\newblock {\em Int. J. Mod. Phys.}, D28(09):1950120, 2019.

\bibitem{Yukawa:1935xg}
Hideki Yukawa.
\newblock {On the Interaction of Elementary Particles I}.
\newblock {\em Proc. Phys. Math. Soc. Jap.}, 17:48--57, 1935.
\newblock [Prog. Theor. Phys. Suppl.1,1(1935)].

\bibitem{Birrell:1982ix}
N.~D. Birrell and P.~C.~W. Davies.
\newblock {\em {Quantum Fields in Curved Space}}.
\newblock Cambridge Monographs on Mathematical Physics. Cambridge Univ. Press,
  Cambridge, UK, 1984.

\bibitem{DeMartino:2018yqf}
Ivan De~Martino, Ruth Lazkoz, and Mariafelicia De~Laurentis.
\newblock {Analysis of the Yukawa gravitational potential in $f(R)$ gravity I:
  semiclassical periastron advance}.
\newblock {\em Phys. Rev.}, D97(10):104067, 2018.

\bibitem{DeLaurentis:2018ahr}
Mariafelicia De~Laurentis, Ivan De~Martino, and Ruth Lazkoz.
\newblock {Analysis of the Yukawa gravitational potential in $f(R)$ gravity II:
  relativistic periastron advance}.
\newblock {\em Phys. Rev.}, D97(10):104068, 2018.

\bibitem{Ribas:2016ulz}
Marlos~O. Ribas, Fernando~P. Devecchi, and Gilberto~M. Kremer.
\newblock {Cosmological model with fermion and tachyon fields interacting via
  Yukawa-type potential}.
\newblock {\em Mod. Phys. Lett.}, A31(06):1650039, 2016.

\bibitem{Berezhiani:2009kv}
Zurab Berezhiani, Fabrizio Nesti, Luigi Pilo, and Nicola Rossi.
\newblock {Gravity Modification with Yukawa-type Potential: Dark Matter and
  Mirror Gravity}.
\newblock {\em JHEP}, 07:083, 2009.

\bibitem{bookintegral}
Milton Abramowitz and Irene~A. Stegun.
\newblock {\em Handbook of Mathematical Functions with Formulas, Graphs, and
  Mathematical Tables}, volume~55 of {\em National Bureau of Standards Applied
  Mathematics Series}.
\newblock Tenth Printing, 1972.

\bibitem{nam82}
G.~W. Gibbons and S.~W. Hawking.
\newblock {Cosmological Event Horizons, Thermodynamics, and Particle Creation}.
\newblock {\em Phys. Rev.}, D15:2738--2751, 1977.

\bibitem{nam83}
G.~W. Gibbons and S.~W. Hawking.
\newblock {Action Integrals and Partition Functions in Quantum Gravity}.
\newblock {\em Phys. Rev.}, D15:2752--2756, 1977.

\bibitem{Ma:2013aqa}
Meng-Sen Ma, Hui-Hua Zhao, Li-Chun Zhang, and Ren Zhao.
\newblock {Existence condition and phase transition of
  Reissner-Nordstr{\"o}m-de Sitter black hole}.
\newblock {\em Int. J. Mod. Phys.}, A29:1450050, 2014.

\bibitem{Ma:2014hna}
Meng-Sen Ma, Li-Chun Zhang, Hui-Hua Zhao, and Ren Zhao.
\newblock {Phase transition of the higher dimensional charged Gauss-Bonnet
  black hole in de Sitter spacetime}.
\newblock {\em Adv. High Energy Phys.}, 2015:134815, 2015.

\bibitem{Guo:2015waa}
Xiongying Guo, Huaifan Li, Lichun Zhang, and Ren Zhao.
\newblock {Thermodynamics and phase transition in the Kerr--de Sitter black
  hole}.
\newblock {\em Phys. Rev.}, D91(8):084009, 2015.

\bibitem{Guo:2016eie}
Xiongying Guo, Huaifan Li, Lichun Zhang, and Ren Zhao.
\newblock {The critical phenomena of charged rotating de Sitter black holes}.
\newblock {\em Class. Quant. Grav.}, 33(13):135004, 2016.

\bibitem{Bhattacharya:2015mja}
Sourav Bhattacharya.
\newblock {A note on entropy of de Sitter black holes}.
\newblock {\em Eur. Phys. J.}, C76(3):112, 2016.

\bibitem{Kubiznak:2016qmn}
David Kubiznak, Robert~B. Mann, and Mae Teo.
\newblock {Black hole chemistry: thermodynamics with Lambda}.
\newblock {\em Class. Quant. Grav.}, 34(6):063001, 2017.

\bibitem{Abrahams1979}
E.~Abrahams, P.~W. Anderson, D.~C. Licciardello, and T.~V. Ramakrishnan.
\newblock Scaling theory of localization: Absence of quantum diffusion in two
  dimensions.
\newblock {\em Phys. Rev. Lett.}, 42:673--676, Mar 1979.

\bibitem{Cherroret2014}
Nicolas Cherroret, Beno\^{\i}t Vermersch, Jean~Claude Garreau, and Dominique
  Delande.
\newblock How nonlinear interactions challenge the three-dimensional anderson
  transition.
\newblock {\em Phys. Rev. Lett.}, 112:170603, May 2014.

\end{thebibliography}
\end{document}